\documentclass[twocolumn,prl,reprint]{revtex4}
\usepackage[latin9]{inputenc}
\usepackage{color}
\usepackage{amsmath}
\usepackage{amssymb}
\usepackage{graphicx}
\usepackage{esint}
\PassOptionsToPackage{version=3}{mhchem}
\usepackage{mhchem}

\makeatletter

\@ifundefined{textcolor}{}
{%
 \definecolor{BLACK}{gray}{0}
 \definecolor{WHITE}{gray}{1}
 \definecolor{RED}{rgb}{1,0,0}
 \definecolor{GREEN}{rgb}{0,1,0}
 \definecolor{BLUE}{rgb}{0,0,1}
 \definecolor{CYAN}{cmyk}{1,0,0,0}
 \definecolor{MAGENTA}{cmyk}{0,1,0,0}
 \definecolor{YELLOW}{cmyk}{0,0,1,0}
}

\makeatother

\begin{document}

\title{Exchange-Correlation Energy from Pairing Matrix Fluctuation and the
Particle-Particle Random Phase Approximation}

\author{Helen van Aggelen}

\affiliation{Ghent University, Department of Inorganic and Physical Chemistry,
9000 Ghent, Belgium}

\affiliation{Duke University, Department of Chemistry, NC 27708, U.S. }

\author{Yang Yang }

\affiliation{Duke University, Department of Chemistry, NC 27708, U.S. }

\author{Weitao Yang }

\affiliation{Duke University, Department of Chemistry and Department of Physics,
NC 27708, U.S. }


\date{\today}

\begin{abstract}
We formulate an adiabatic connection for the exchange-correlation
energy in terms of pairing matrix fluctuation. This connection opens
new channels for density functional approximations based on pairing
interactions. Even the simplest approximation to the pairing matrix
fluctuation, the particle-particle Random Phase Approximation (pp-RPA),
has some highly desirable properties. It has no delocalization error
with a nearly linear energy behavior for systems with fractional charges,
describes van der Waals interactions similarly and thermodynamic properties
significantly better than particle-hole RPA, and eliminates static
correlation error for single-bond systems. Most significantly, the
pp-RPA is the first known functional that has an explicit and closed-form
dependence on the occupied and unoccupied orbitals and captures the
energy derivative discontinuity in strongly correlated systems. These
findings illlustrate the potential of including pairing interactions
within a density functional framework.
\end{abstract}

\maketitle

The desire for systematic progress in Density Functional Approximations
(DFA) has drawn attention to functionals rooted in many-body perturbation
theory \cite{Cohen2012289,Burke2012,Onida2002601}, the most popular
of which is the Random Phase Approximation (RPA). The RPA originated
in nuclear many-body theory in the 1950s \cite{Bohm1951625,Ring1980}
but recently found new applications formulated as a DFA
of occupied and virtual orbitals \cite{Fuchs2002235109}. The DFA
perspective is justified by the adiabatic connection fluctuation dissipation
(ACFD) theorem\cite{Langreth19772884}, which establishes a fundamental
connection between DFT and many-body perturbation theory. It formulates
the exchange-correlation energy in terms of the polarization propagator,
for which the RPA provides an approximation. The RPA overcomes some
persistent problems of classical DFA functionals. In contrast to most
classical DFA functionals, it describes static correlation correctly
and thus dissociates, for instance, \ce{H2} correctly\cite{Hesselmann2011093001};
it captures long-range interactions adequately and is applicable to
systems with vanishing gap \cite{Eshuis2012}. These desirable features
have been the incentive to construct more efficient algorithms, such
that large-scale applications are now feasible \cite{Ren20127447}.
Nonetheless, the RPA still faces some major theoretical challenges:
it violates the Pauli principle, which leads to a large delocalization
error, as demonstrated in the dissociation of \ce{H2+} and other molecules\cite{Mori-Sanchez2012}. The Second Order Screened
Exchange (SOSEX) \cite{Freeman19775512} corrects this problem \cite{Gruneis2009154115},
but fails in cases of static correlation such as dissociating \ce{H2}.

All of the RPA-related DFA methods are based on particle-hole (ph)
interactions \cite{Eshuis2012,Ren20127447,Toulouse2010,Scheffler2011153003}.
However, the second-order Green's function naturally leads to another
channel of interactions: particle-particle (pp) and hole-hole (hh)
interactions \cite{Blaizot1986}. The present work establishes an
adiabatic connection\cite{Gunnarson19764274,Langreth19751425} for
the exchange-correlation energy in terms of the dynamic paring matrix
\textcolor{black}{fluctuation or particle-particle Green function}, parallel
to the ACFD theorem in terms of the density fluctuation or polarization
propagator. Like the ACFD theorem, it is in principle exact, but requires
the particle-particle Green function as a function of the interaction
strength. The pp-RPA, a Random Phase Approximation in the pp and hh
correlation channels, provides an approximation to the $\lambda$-
dependence of the Green function that leads to a simple closed expression
for the exchange-correlation energy. In this paper we therefore explore
the pp-RPA as a DFA functional, based on the adiabatic connection
we formulate, to illustrate the potential of using pairing interactions
in DFA. Despite its close relationship to the ph-RPA, particle-particle
interactions have received limited attention only in spectral calculations
\cite{Romaniello2012}, but not as a DFA for ground state energies.
The theoretical framework underlying the pp-RPA is very similar to
that of ph-RPA, but its features as a DFA functional are quite different,
as we will illustrate with applications to molecular dissociation
and thermodynamical properties.

\label{theoretical}

The exact exchange-correlation energy in KS-DFT can be related to
paring matrix fluctuation $\bar{\mathbf{K}}(E)$ (or the particle-particle
Green function $\mathbf{K}(E)$) in many-body perturbation theory via
the adiabatic connection. The pairing matrix fluctuation $\mathbf{\bar{K}}(t-t')$
describes the response of the pairing matrix $\kappa_{ij}(t)=\langle\Psi_{0}^{N}\vert a_{H_{i}}(t)a_{H_{j}}(t)\vert\Psi_{0}^{N}\rangle$
to a perturbation in the form of a pairing field, $\hat{F}(t')=f_{kl}a_{H_{l}}^{\dagger}(t')a_{H_{k}}^{\dagger}(t')\theta(t')$.
In the energy domain, $\mathbf{\bar{K}}(E)$ has the form

\begin{align*}
\bar{K}(E)_{ijkl} & =\sum_{n}\frac{\langle\Psi_{0}^{N}\vert a_{i}a_{j}\vert\Psi_{n}^{N+2}\rangle\langle\Psi_{n}^{N+2}\vert a_{l}^{+}a_{k}^{+}\vert\Psi_{0}^{N}\rangle}{E-\omega_{n}^{N+2}+i\eta}\\
 & -\sum_{n}\frac{\langle\Psi_{0}^{N}\vert a_{l}^{+}a_{k}^{+}\vert\Psi_{n}^{N-2}\rangle\langle\Psi_{n}^{N-2}\vert a_{i}a_{j}\vert\Psi_{0}^{N}\rangle}{E-\omega_{n}^{N-2}+i\eta}
\end{align*}
and therefore contains information on the 2-electron addition and
removal energies $\omega_{n}^{N+2}$ and $\omega_{n}^{N-2}$ and the
corresponding transition amplitudes. Moreover, these response functions
not only provide information on the $N\pm2$ electron excited states,
they also indirectly determine ground state properties. The ground
state correlation energy can be formulated in terms of the pairing
matrix fluctuation (or, equivalently, the pp-Green function) through the
adiabatic connection:

\begin{align}
E^{c} & =\frac{1}{2\pi i}\int_{0}^{1}d\lambda\int_{-i\infty}^{+i\infty}dE\int d\mathbf{x}d\mathbf{x}'\frac{\bar{K}^{\lambda}(\mathbf{x},\mathbf{x}',E)-\bar{K}^{0}(\mathbf{x},\mathbf{x}',E)}{\vert\mathbf{r}-\mathbf{r}'\vert}.\label{eq:Ec_K}
\end{align}
Since the exchange energy is the exact exchange, we focus on the correlation
energy. Further background and full derivations are presented in sections
1A-1C of the supplementary material, ref. (\cite{supp}). This formula
can be considered the pairing interaction counterpart of the ACFD
theorem. Like the ACFD theorem, formula (\ref{eq:Ec_K}) is in principle
exact, but requires an approximation to compute the pairing matrix
fluctuation $\mathbf{\bar{K}}^{\lambda}$. The simplest approximation
to the pairing matrix fluctuation is the particle-particle RPA. In
this work, we will focus on the particle-particle RPA to illustrate
the potential of including pairing interactions in a DFT framework. 

The pp-RPA approximates the dynamic pairing matrix fluctuation $\mathbf{\bar{K}}^{\lambda}$
in terms of its non-interacting counterpart $\mathbf{\bar{K}}^{0}$

\begin{align*}
\mathbf{\bar{K}}^{\lambda} & =\mathbf{\bar{K}}^{0}+\lambda\mathbf{\bar{K}}^{0}\mathbf{V}\mathbf{\bar{K}}^{\lambda},
\end{align*}
where the Coulomb interaction is $V_{abcd}=\langle ab\Vert cd\rangle=\langle ab\vert cd\rangle-\langle ba\vert cd\rangle,$
and $\langle ab\vert cd\rangle=\int\phi_{a}^*(\mathbf{x})\phi_{b}^*(\mathbf{x}')\frac{1}{\vert\mathbf{r}-\mathbf{r}'\vert}\phi_{c}(\mathbf{x})\phi_{d}(\mathbf{x}')d\mathbf{x}d\mathbf{x}'$.
Under this approximation, the integration over the interaction strength
$\lambda$ in Eq. (\ref{eq:Ec_K}) can be carried out analytically.
The resulting expression for the correlation energy in terms of the
non-interacting Green function $\mathbf{K}^{0}$ is equivalent to the
sum of all ladder diagrams in the context of many-body perturbation
theory \cite{Blaizot1986}

\begin{eqnarray}
E^{c} & = & \frac{-1}{2\pi i}\ \sum_{n=2}\frac{1}{n}\int_{-i\infty}^{+i\infty}\mathrm{tr}\ [\mathbf{\bar{K}}^{0}(E))\mathbf{V}]^{n}\ dE\nonumber \\
 & = & \frac{1}{2\pi i}\ \int_{-i\infty}^{+i\infty}\mathrm{tr}\ [\ln(\mathbf{I}-\mathbf{\bar{K}}^{0}(E)\mathbf{V})+\mathbf{\bar{K}}^{0}(E)\mathbf{V}]\ dE\label{eq:Ec_ppRPA_intform}
\end{eqnarray}

The pairing matrix fluctuation $\bar{\mathbf{K}}(E)$ is antisymmetrical
under particle exchange, so Eq. (\ref{eq:Ec_K})-(\ref{eq:Ec_ppRPA_intform})
are formulated in an antisymmetrical basis, which includes only ordered
two-particle indices. While the correlation energy can be computed
directly from Eq. (\ref{eq:Ec_ppRPA_intform}), it can also be cast
in terms of the solution to a generalized eigenvalue problem (see
Eq. (11) of ref. (\cite{supp})), with the same formal $O(N^{6})$
scaling as the ph-RPA eigenvalue problem:
\begin{align}
\sum_{c<d}\left(\langle ab\Vert cd\rangle+\delta_{ac}\delta_{bd}\omega^0_{ab}\right)X_{cd}^{n}+\sum_{i<j}\langle ab\Vert ij\rangle Y_{ij}^{n} & =\omega_{n}X_{ab}^{n} \nonumber \\
\sum_{a<b}\langle ij\Vert ab\rangle X_{ab}^{n}+\sum_{k<l}\left(\langle ij\Vert kl\rangle-\delta_{ik}\delta_{jl}\omega^0_{ij}\right)Y_{kl}^{n} & =-\omega_{n}Y_{ij}^{n} \label{ppRPA_matrixform}
\end{align}
where $\omega^0_{ab}=\epsilon_a +\epsilon_b -2\nu$ and $\nu$ is the chemical potential.
This eigenvalue problem is then solved for the pp-RPA eigenvectors
$\mathbf{X^{n}},\mathbf{Y}^{\mathbf{n}}$ and their corresponding
eigenvalues $\omega_{n}$. The generalized eigenvalues $\omega_{n}$
have a clear physical meaning: they are either positive 2-electron
addition energies, $\omega_{n}^{N+2}=E_{n}^{N+2}-E_{0}^{N}-2\nu$,
or negative 2-electron removal energies, $\omega_{n}^{N-2}=E_{0}^{N}-E_{n}^{N-2}-2\nu$.
The eigenvectors are the corresponding amplitudes, $X_{ab}^{n}=\langle\Psi_{0}^{N}\vert a_{a}a_{b}\vert\Psi_{n}^{N+2}\rangle$
and $Y_{ij}^{n}=\langle\Psi_{0}^{N}\vert a_{i}a_{j}\vert\Psi_{n}^{N+2}\rangle$
when $\omega_{n}>0$; $X_{ab}^{n}=\langle\Psi_{0}^{N}\vert a_{b}^{+}a_{a}^{+}\vert\Psi_{n}^{N-2}\rangle$
and $Y_{ij}^{n}=\langle\Psi_{0}^{N}\vert a_{j}^{+}a_{i}^{+}\vert\Psi_{n}^{N-2}\rangle$
when $\omega_{n}<0$.

The pp-RPA correlation energy can be reformulated in terms of the
solution to this generalized eigenvalue system (see section 1C of
ref. (\cite{supp}) ):

\begin{eqnarray}
E^{c} & = & \sum_{n}\omega_{n}^{N+2}-\sum_{a<b}\left( \epsilon_{a}+\epsilon_{b}-2\nu+\langle ab\Vert ab\rangle \right) \label{eq:Ec_matrixform}
\end{eqnarray}
where the summation over $n$ runs over all 2-electron addition energies. Since Eq. (\ref{ppRPA_matrixform}) depends only
on the orthonormal set of orbitals $\{\phi_{i}\}$ and their occupations
$n_{i}$, the correlation energy can be viewed as a functional $E[\{\phi_{i}\},n_{i}]$.
The total pp-RPA energy expression combines the HF-energy functional
with the pp-RPA correlation energy, Eq. (\ref{eq:Ec_matrixform}).

The density functional perspective on the pp-RPA raises some prominent
questions: how does the pp-RPA behave for systems with fractional
spins or charges, which present a major challenge for DFA? \cite{Cohen2012289,Cohen20084}.
Most approximate density functionals give physically incorrect properties
for systems that arise from an ensemble, such as molecule fragments
with fractional electron numbers or spins. Such systems naturally
arise for instance as the dissociation products of a molecule. While
the molecule as a whole has integer electron number and (half) integer
spin, each of its dissociation products may have a fractional electron
number or spin. The exact conditions on density functionals for fractional
charges \cite{Perdew19821691,Cohen2008}, fractional spins
\cite{Cohen2008a}, and their combination\cite{Mori-Sanchez2009}
are now known.

The performance of density functionals for systems with fractional
occupation number has therefore become an important criterion in the
development of DFA. The behavior of the pp-RPA for such systems can
be quantified by taking the fractional orbital occupations into account
explicitly in the pp-RPA equations (section 1E of ref. (\cite{supp})), following previous work extending other DFAs to fractionals \cite{Mori-Sanchez2012,Cohen2008a}.

\label{computational}

We computed the KS reference wavefunction with Gaussian03 \cite{Frisch2004}
for the systems with integer electron number and with the QM4D package
for systems with fractional electron number or spin \cite{QM4D}.
For the subsequent pp-RPA calculation, we used our implementation,
which diagonalizes the pp-RPA matrix. Since the diagonalization is
computationally expensive, we used a cc-pVDZ basis set for all calculations,
except for the Ar and Ne atoms, for which we used an aug-cc-pVDZ (FC)
basis set. For the calculations on thermodynamical properties, we
used a cc-pVTZ basis set limited to F-functions because the pp-RPA energy converges slowly with the basis set size
(Fig. 13 of ref. (\cite{supp})) and geometries from
the G2 test set \cite{Curtiss19971063}. Accurate potential energy
functions for the dimers of the noble gases have been taken from the
work of Ogilvie et al. \cite{Ogilvie1992277,Ogilvie1993313} and the
MRCI potential energy function for the \ce{N2} in the cc-pVDZ basis
set has been taken from previous work \cite{Aggelen2010}.

\label{results}

\begin{figure}
\centering \includegraphics[width=0.45\textwidth]{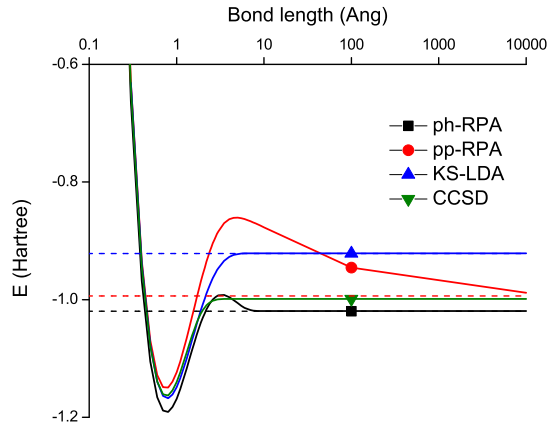}
\caption{The pp-RPA(LDA) energy for the \ce{H2} molecule approaches the correct
value in the dissociation limit, but has an unphysical 'bump', much
more so than ph-RPA(LDA). The dashed lines indicate the dissociation limits from the fractional analysis of the H atom.}
\label{fig:HH_LDA} 
\end{figure}

\begin{figure}
\centering \includegraphics[width=0.45\textwidth]{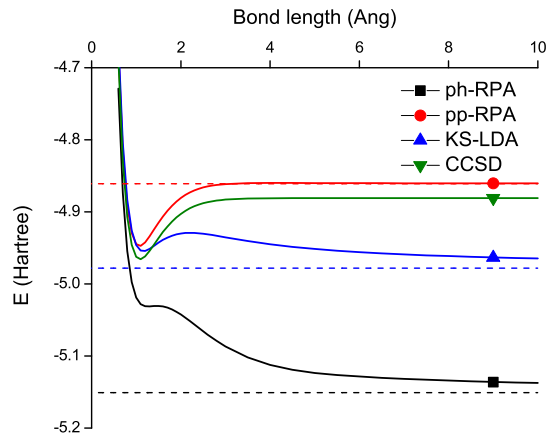}
\caption{The pp-RPA(LDA) also gives a correct energy profile for \ce{He2+}, in
contrast to ph-RPA(LDA). The dashed lines indicate the dissociation limits from the fractional analysis of the He atom.}
\label{fig:HeHe+_LDA} 
\end{figure}

The pp-RPA has negligible delocalization error and static correlation
error and thus produces the correct dissociation limit for \ce{H2}
and \ce{H2+}. The \ce{H2} and \ce{H2+} molecules are paradigmatic
examples of challenges for DFA \cite{Cohen20084}, because few DFA
functionals give the correct dissociation limit for both \ce{H2}
and \ce{H2+}. The ph-RPA dissociates \ce{H2} correctly, but
produces a huge delocalization error for \ce{H2+} \cite{Mori-Sanchez2012}.
The pp-RPA, however, gives the correct dissociation limit for \ce{H2}
and \ce{H2+}, although the potential energy curve of \ce{H2}
has an unphysical local maximum around 10\ \AA{}\ (Figs. \ref{fig:HH_LDA}
and Fig. 2 of ref. (\cite{supp})). While the dissociation of \ce{H2+}
is described correctly by construction in pp-RPA -- the pp-RPA energy
reduces to the HF functional for a one-electron system -- it also
gives a good dissociation profile for \ce{He2+}, for instance (Fig.
\ref{fig:HeHe+_LDA}). Other RPA methods have been constructed to
dissociate these positively charged molecules correctly, such as ph-RPA+SOSEX,
which a posteriori corrects for neglecting antisymmetry in the ph-RPA.
However, RPA+SOSEX gives a much too high dissociation limit for \ce{H2}\cite{Gruneis2009154115}.

The pp-RPA satisfies the Hydrogen Test Set (\cite{Cohen2012289}):
it has no delocalization error for \ce{H2+} and almost no static
correlation error for \ce{H2} because it has a nearly physically
correct energy profile for the H atom with fractional charges and
fractional spins. Describing both cases correctly requires
that the functional has constant energy for all spin projections between
0 and 1 \cite{Mori-Sanchez2009,Cohen2008a}, and that it has a linear
energy profile for electron numbers between 0 and 1 \cite{Perdew19821691}.
Most DFA functionals do not have these features, which results in
static correlation errors and/or delocalization errors. The ph-RPA,
for instance, has a nearly constant energy for different spin projections
in the H atom, but has a significant delocalization error for fractional
electron numbers \cite{Mori-Sanchez2012} (Figs. 4 and 5 of ref.
(\cite{supp})). It thus describes the dissociation of \ce{H2}
correctly but gives a much too low dissociation limit for \ce{H2+}.
The pp-RPA not only has a nearly constant energy for different spin
projections of the H atom but also has a nearly linear energy between
electron numbers of 0 and 1 (Fig. 4 of ref. (\cite{supp})). These
properties ensure that it gives the right dissociation limit for \ce{H2}
and \ce{H2+}.

\begin{figure}
\centering \includegraphics[width=0.45\textwidth]{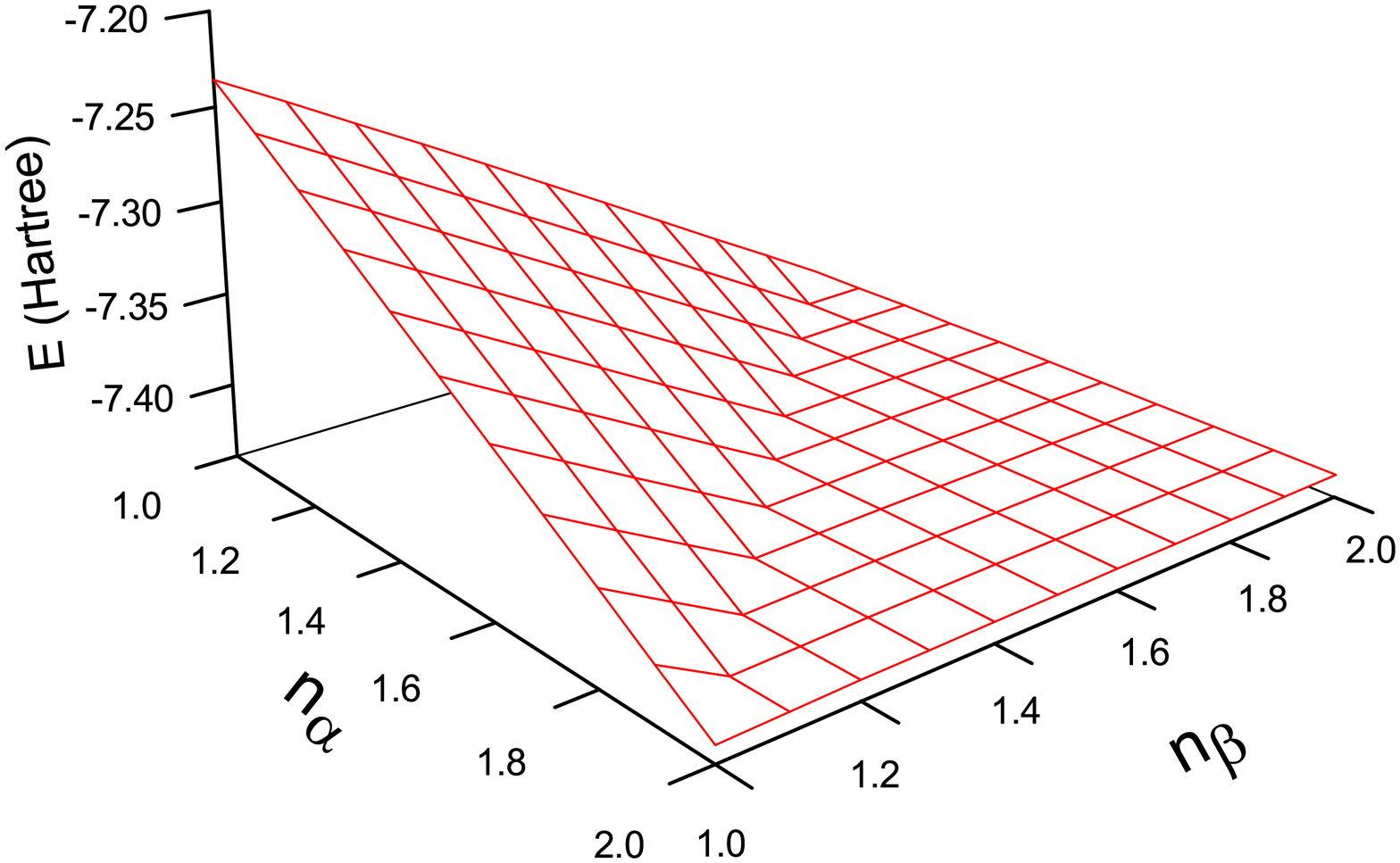}
\caption{The pp-RPA(LDA) energy for the Li atom is a nearly constant function of
the fractional spin projection and a nearly linear function of the
fractional electron number. Like the exact functional, its derivative
has a discontinuity at N=3.}
\label{fig:Li_ppRPA_LDA} 
\end{figure}

Most significantly, the pp-RPA captures the energy derivative discontinuity
for strongly correlated systems (SCS) at integer electron numbers.
Traditional DFA functionals have a smooth dependence on the occupied
orbitals and cannot capture the required derivative discontinuities
for SCS at integer electron number \cite{Mori-Sanchez2009,Cohen2008a}.
Even the ph-RPA energy, which is a functional of the occupied and
the unoccupied orbitals, does not have a derivative discontinuity
at integer electron numbers for SCS. (Figs. 4 and 5 of ref. (\cite{supp})).\cite{Mori-Sanchez2012}
However, the pp-RPA adequately captures the energy derivative discontinuity
and satisfies the flat-plane condition\cite{Mori-Sanchez2009} , as
Fig. 4 of ref. (\cite{supp}) and Fig. \ref{fig:Li_ppRPA_LDA}
illustrate for the H atom and Li atom.

The pp-RPA describes the ionization energies exceptionally well, although
in the present basis set the sign of some of the very small electron
affinities is wrong. Finite-difference calculations on the pp-RPA
chemical potential for a set of second-period atoms (table II of ref.
(\cite{supp})) demonstrate the superiority of the pp-RPA over the
ph-RPA.

The pp-RPA has almost no static correlation error for single-bond
systems, and gives the proper dissociation limit for ethane, for instance
(Fig. 7 of ref. (\cite{supp})). However, it has a substantial error for 
the dissociation of \ce{N2} (Fig. 8 of ref. (\cite{supp})). Breaking multiple bonds like those
in \ce{N2} within a singlet description is problematic for pp-RPA
because \ce{N2} dissociates into two spin-unpolarized spherical
N atoms, which have equal fractional numbers of alpha and beta electrons
distributed evenly over the three p-orbitals, and pp-RPA assigns much
too low energy to these spin-unpolarized spherical atoms (Fig. 9
of ref. (\cite{supp})).

\begin{figure}
\centering \includegraphics[width=0.45\textwidth]{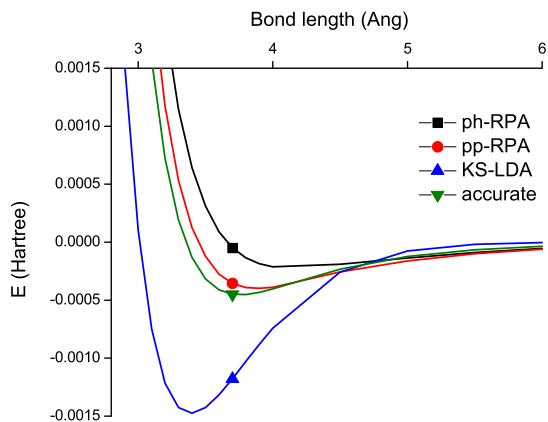}
\caption{The ph-RPA(LDA) and pp-RPA(LDA) both describe the van der Waals interactions
in the Ar dimer well.}
\label{fig:ArAr_LDA} 
\end{figure}

The pp-RPA describes van der Waals interactions to a very good extent,
similar to ph-RPA and ph-RPA+SOSEX \cite{Gruneis2009154115,Irelan201194105}.
One of the main strengths of ph-RPA is its ability to capture non-covalent
long-range interactions smoothly and seamlessly. Although the nature
of the interactions is different in pp-RPA from that in ph-RPA, pp-RPA
also captures the van der Waals interactions in \ce{Ar2} and \ce{NeAr}
well (Figs. \ref{fig:ArAr_LDA} and Fig. 11 of ref. (\cite{supp})).

\begin{figure}
\includegraphics[width=0.45\textwidth]{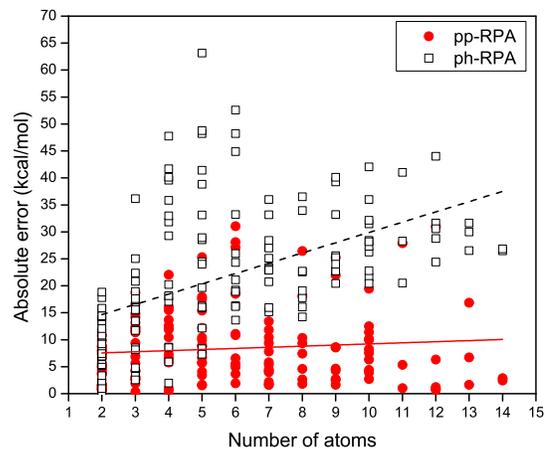}
\caption{The ph-RPA(PBE) enthalpies of formation for the molecules in the G2/97
database show a steadily increasing error with the number of atoms, with a MAE of 22.7 kcal/mol
whereas the pp-RPA(PBE) enthalpies show nearly constant errors with
the number of atoms, with a MAE of 8.3 kcal/mol. }
\label{fig:G2set} 
\end{figure}

The pp-RPA performs much better than ph-RPA on the heats of formation
and atomization energies for a set of small molecules. The mean absolute errors (MAE) on the
heats of formation computed for a set of small molecules is 5.8 kcal/mol
for the pp-RPA and 10.4 kcal/mol -- in good agreement with the results
by Ren et al. \cite{Ren20127447} -- for the ph-RPA (table III of
ref. (\cite{supp})). The 4.6 kcal/mol difference shows that the accuracy
of the heats of formation computed with pp-RPA is better than that
of ph-RPA. Furthermore, a test on the whole G2 set shows that the errors in the 
ph-RPA heats of formation increase steadily with the number of atoms in the molecules, 
whereas the errors in the pp-RPA
heats of formation remain nearly constant (Fig. \ref{fig:G2set}).

Finally, a perturbation theory analysis (section 1D of ref. (\cite{supp}))
shows that pp-PRA has the correct second-order energy, in contrast
to the ph-RPA, which contains only the direct terms of the second-order
energy.

To summarize, we have shown that the exact exchange-correlation energy
can be expressed in terms of the dynamic paring matrix fluctuation
via the adiabatic connection and illustrated the potential of this
approach with the pp-RPA. The pp-RPA is a remarkable DFA, because
it is the first functiona\textcolor{black}{l that has an explicit
and closed-form dependence on the occupied and virtual orbitals and
captures the derivative discontinuity }of the energy at integer electron
numbers for the whole range of spin polarizations in\textcolor{black}{{}
strongly correlated systems.}

The pp-RPA meets the flat-plane energy requirement for systems with
fractional charges and spins \cite{Mori-Sanchez2009}. This flat-plane
energy behavior has been actively pursued in recent years, with limited
success up to now \cite{johnson2011081103}. It was shown that explicit,
differentiable functionals of the density or density matrix cannot
capture it \cite{Cohen2008a,Mori-Sanchez2012}. Even the inclusion
of virtual orbitals in the ph-RPA does not prove helpful \cite{Mori-Sanchez2012}.
The discontinuity in the pp-RPA energy as shown presently proves that
this goal can be achieved in closed form with a functional that depends
on both the occupied and unoccupied orbitals, or on the Green's function
of the non-interacting KS or GKS reference system, highlighting the
path forward for development of functionals for strongly correlated
systems.

\begin{acknowledgments}
Support from FWO-Flanders (Scientific Research Fund Flanders) (H.v.A),
the Office of Naval Research (N00014-09- 0576) and the National Science
Foundation (CHE-09-11119) (W.Y.) is appreciated.
\end{acknowledgments}
\clearpage{}

\expandafter\ifx\csname natexlab\endcsname\relax\global\long\def\natexlab#1{#1}
\fi \expandafter\ifx\csname bibnamefont\endcsname\relax \global\long\def\bibnamefont#1{#1}
\fi \expandafter\ifx\csname bibfnamefont\endcsname\relax \global\long\def\bibfnamefont#1{#1}
\fi \expandafter\ifx\csname citenamefont\endcsname\relax \global\long\def\citenamefont#1{#1}
\fi \expandafter\ifx\csname url\endcsname\relax \global\long\def\url#1{\texttt{#1}}
\fi \expandafter\ifx\csname urlprefix\endcsname\relax\global\long\def\urlprefix{URL }
\fi \providecommand{\bibinfo}[2]{#2} \providecommand{\eprint}[2][]{\url{#2}}

\end{document}


Supporting information of

\title{Exchange-Correlation Energy from Pairing Matrix Fluctuation and the
Particle-Particle Random Phase Approximation}


\author{Helen van Aggelen}

\affiliation{Ghent University, Department of Inorganic and Physical Chemistry,
9000 Ghent, Belgium}

\affiliation{Duke University, Department of Chemistry, NC 27708, U.S. }

\author{Yang Yang }

\affiliation{Duke University, Department of Chemistry, NC 27708, U.S. }

\author{Weitao Yang }

\affiliation{Duke University, Department of Chemistry and Department of Physics, NC 27708, U.S. }

\date{\today}

\maketitle

\section{Theory in Detail}

\subsection{The paring matrix fluctuation, particle-particle Green function,
and the particle-particle Random Phase Approximation\label{sub:background_ppRPA}}

In the absence of a pairing field, the pairing matrix

\begin{align*}
\kappa_{ij}(t)=\langle\Psi_{0}^{N}\vert a_{H_{i}}(t)a_{H_{j}}(t)\vert\Psi_{0}^{N}\rangle\\
\end{align*}
where $\vert\Psi_{0}^{N}\rangle$ is the $N$-electron ground state,
is identically zero. The operators $a_{H_{i}}^{\dagger}(t)$ are the
creation operators in the Heisenberg picture, $a_{H_{i}}^{\dagger}(t)=e^{\frac{i}{\hbar}(\hat{H}-\nu\hat{N})}a_{i}^{\dagger}e^{\frac{-i}{\hbar}(\hat{H}-\nu\hat{N})}$
and the term $-\nu\hat{N}$, with $\nu$ the chemical potential, is
added to the Hamiltonian such that the $N$-electron state is the
minimum under the total Hamiltonian $\hat{H}-\nu\hat{N}$ when the
particle number is allowed to change. Under a perturbation $\hat{F}(t)$
in the form of a pairing field, $\hat{F}(t')=\sum_{kl} f_{kl}a_{H_{l}}^{\dagger}(t')a_{H_{k}}^{\dagger}(t')\theta(t')$,
the retarded Green function $\mathbf{\bar{K}}^{R}$ 
\begin{equation}
\bar{K}_{ijkl}^{R}(t-t')=\frac{-i}{\hbar}\theta(t-t')\langle\Psi_{0}^{N}\vert[a_{H_{i}}(t)a_{H_{j}}(t),a_{H_{l}}^{\dagger}(t')a_{H_{k}}^{\dagger}(t')]\vert\Psi_{0}^{N}\rangle,\label{eq:Green_K_retarded-1}
\end{equation}
describes the linear change in the paring matrix $\langle\Psi_{0}^{N}\vert a_{H_{i}}(t)a_{H_{j}}(t)\vert\Psi_{0}^{N}\rangle$:

\begin{align*}
\kappa_{ij}(t) & =\frac{-i}{\hbar}\int_{0}^{t}\langle\Psi_{0}^{N}\vert[a_{H_{i}}(t)a_{H_{j}}(t),\hat{F}(t')]\vert\Psi_{0}^{N}\rangle dt'\\
 & =\sum_{kl}\bar{K}^{R}(t-t')_{ijkl}f_{kl}
\end{align*}
Since the paring matrix $\langle\Psi_{0}^{N}\vert a_{H_{i}}(t)a_{H_{j}}(t)\vert\Psi_{0}^{N}\rangle=\langle\Psi_{0}^{N}\vert a_{{i}}a_{{j}}\vert\Psi_{0}^{N}\rangle=0$
in the absence of the pairing field, the retarded Green function is
identical to the dynamic pairing matrix fluctuation, $\mathbf{\bar{K}}(t-t')$

\[
\bar{K}_{ijkl}(t-t')=\frac{-i}{\hbar}\theta(t-t')\langle\Psi_{0}^{N}\vert[\left(a_{H_{i}}(t)a_{H_{j}}(t)-\langle \Psi_{0}^{N}\vert  a_{{i}}a_{{j}}\vert \Psi_{0}^{N}\rangle\right),\left(a_{H_{l}}^{\dagger}(t')a_{H_{k}}^{\dagger}(t')-\langle \Psi_{0}^{N} \vert  a_{{l}}^{\dagger}a_{{k}}^{\dagger}\vert \Psi_{0}^{N} \rangle\right)]\vert\Psi_{0}^{N}\rangle,
\]
The particle-particle Green function $\mathbf{K}(t-t')$, defined
as \cite{Blaizot1986}
\begin{equation}
K_{ijkl}(t-t')=\frac{-i}{\hbar}\langle\Psi_{0}^{N}\vert\mathcal{T}[a_{H_{i}}(t)a_{H_{j}}(t)a_{H_{l}}^{\dagger}(t')a_{H_{k}}^{\dagger}(t')]\vert\Psi_{0}^{N}\rangle\label{eq:Green_K}
\end{equation}
where $\mathcal{T}$ is the time-ordering operator, is a closely related quantity. The dynamic paring matrix fluctuation
$\mathbf{\bar{K}}(t-t')$ and the pp-Green function $\mathbf{K}(t-t')$
contain information on the same physical properties, namely 2-electron
removal and addition energies and their corresponding transition amplitudes.
This becomes apparent from their Fourier Transform 
\begin{eqnarray*}
K_{ijkl}(E) & = & \int_{-\infty}^{+\infty}e^{\frac{i}{\hbar}E(t-t')}K_{ijkl}(t-t')d(t-t')\\
 & = & \frac{-i}{\hbar}\sum_{n}\int_{-\infty}^{\infty}e^{\frac{i}{\hbar}(E_{0}^{N}-E_{n}^{n+2}+2\nu+E)(t-t')}\theta(t-t')d(t-t')\langle\Psi_{0}^{N}\vert a_{i}a_{j}\vert\Psi_{n}^{N+2}\rangle\langle\Psi_{n}^{N+2}\vert a_{l}^{\dagger}a_{k}^{\dagger}\vert\Psi_{0}^{N}\rangle\\
 &  & -\frac{i}{\hbar}\sum_{n}\int_{-\infty}^{\infty}e^{\frac{i}{\hbar}(E_{0}^{N}-E_{n}^{N-2}-2\nu-E)(t'-t)}\theta(t'-t)d(t-t')\langle\Psi_{0}^{N}\vert a_{l}^{\dagger}a_{k}^{\dagger}\vert\Psi_{n}^{N-2}\rangle\langle\Psi_{n}^{N-2}\vert a_{i}a_{j}\vert\Psi_{0}^{N}\rangle.\\
\end{eqnarray*}
where the last line invokes the completeness of the $N-2$ and $N+2$ electron wavefunction basis.
At this point, it is convenient to introduce a short-hand notation
for the transition pairing matrix elements

\begin{align}
\chi_{ij}^{n,N-2} & =\langle\Psi_{n}^{N-2}\vert a_{i}a_{j}\vert\Psi_{0}^{N}\rangle\label{eq:amplitudes}\\
\chi_{ij}^{n,N+2} & =\langle\Psi_{0}^{N}\vert a_{i}a_{j}\vert\Psi_{n}^{N+2}\rangle\nonumber 
\end{align}
and the transition energies

\begin{align}
\omega_{n}^{N-2} & =E_{0}^{N}-E_{n}^{N-2}-2\nu\\
\omega_{n}^{N+2} & =E_{n}^{N+2}-E_{0}^{N}-2\nu.
\end{align}
For a physical system, the energy decreases monotonically with the
number of electrons, so the term $-2\nu$ makes it possible to distinguish
the 2-electron removal energies from the 2-electron addition energies
by their sign: the 2-electron removal energies $\omega_{n}^{N-2}=E_{0}^{N}-E_{n}^{N-2}-2\nu$
are negative and the 2-electron addition energies $\omega_{n}^{N+2}=E_{n}^{N+2}-E_{0}^{N}-2\nu$
are positive. The particle-particle Green function expressed in the
energy domain is then

\begin{eqnarray}
K_{ijkl}(E) & = & \frac{-i}{\hbar}\sum_{n}\int_{-\infty}^{\infty}e^{\frac{i}{\hbar}(-\omega_{n}^{N+2}+E)(t-t')}\theta(t-t')d(t-t')\chi_{ij}^{n,N+2}\left(\chi_{kl}^{n,N+2}\right)^{*}\nonumber \\
 &  & -\frac{i}{\hbar}\sum_{n}\int_{-\infty}^{\infty}e^{\frac{i}{\hbar}(\omega_{n}^{N-2}-E)(t'-t)}\theta(t'-t)d(t-t')\left(\chi_{kl}^{n,N-2}\right){}^{*}\chi_{ij}^{n,N-2}\nonumber \\
 & = & \sum_{n}\frac{\chi_{ij}^{n,N+2}(\chi_{kl}^{n,N+2})^{*}}{E-\omega_{n}^{N+2}+i\eta}\nonumber \\
 &  & -\sum_{n}\frac{(\chi_{kl}^{n,N-2})^{*}\chi_{ij}^{n,N-2}}{E-\omega_{n}^{N-2}-i\eta}.\label{eq:pp_GreenFunction}
\end{eqnarray}
Similarly, the dynamic paring matrix fluctuation and the retarded
particle-particle Green function in energy domain are

\begin{eqnarray*}
\bar{K}_{ijkl}(E)=\bar{K}_{ijkl}^{R}(E) & = & \sum_{n}\frac{\chi_{ij}^{n,N+2}(\chi_{kl}^{n,N+2})^{*}}{E-\omega_{n}^{N+2}+i\eta}-\sum_{n}\frac{(\chi_{kl}^{n,N-2})^{*}\chi_{ij}^{n,N-2}}{E-\omega_{n}^{N-2}+i\eta}
\end{eqnarray*}
This form of the particle-particle Green function and the dynamic
pairing matrix fluctuation reveals their most interesting properties:
they contain information on the vectors $\mathbf{\chi}^{\mathbf{n},N-2}$
and $\mathbf{\chi}^{\mathbf{n},N+2}$ with the amplitudes defined
in (\ref{eq:amplitudes}) and the 2-electron removal and addition
energies, $\omega_{n}^{N-2}$ and $\omega_{n}^{N+2}$ . Since the
particle-particle Green function and the dynamic pairing matrix fluctuation
contain the same physical information, the following derivations can
be expressed equivalently in terms of the dynamic pairing matrix fluctuation.
While we feel that the dynamic pairing matrix fluctuation has a more straightforward
interpretation as the response to a pairing perturbation than the
pp-Green function, the majority of the literature on many-body perturbation
theory uses the language of Green functions. We will therefore adopt the
Green function formalism in the following derivations as well.

There are several ways to derive the pp-RPA equations, which are similar
in nature to their ph-RPA counterparts \cite{Blaizot1986,Ring1980}.
In the same way the particle-hole Green function can be approximated by
an infinite series in terms of the non-interacting Green function in the
ph-RPA, the particle-particle Green function $\mathbf{K}(E)$ can be approximated in terms
of the non-interacting Green function $\mathbf{K}^{0}(E)$ by 
\begin{equation}
\mathbf{K}(E)=\mathbf{K}^{0}(E)+\mathbf{K}^{0}(E)\mathbf{VK}(E), \label{dyson}
\end{equation}
an equivalent form of which can be found in Ref. (\cite{Blaizot1986}).
In Eq. (\ref{dyson}) all quantities, including the two-electron integrals

\begin{align*}
V_{ijkl} & =\langle ij\vert\vert kl\rangle\\
 & =\langle ij\vert kl\rangle-\langle ji\vert kl\rangle\\
 & =\int\frac{\phi_{i}^*(\mathbf{x}_{1})\phi_{j}^*(\mathbf{x}_{2})(1-\hat{P}_{12})\phi_{k}(\mathbf{x}_{1})\phi_{l}(\mathbf{x}_{2})}{\vert\mathbf{r_{1}}-\mathbf{r_{2}}\vert}d\mathbf{x}_{1}d\mathbf{x}_{2},
\end{align*}
where $\mathbf{x}$ represents the one-electron spatial vector and
spin coordinate, are expressed in an antisymmetrized basis, so only
matrix indices $ab$ with $a<b$ and $ij$ with $i<j$ need to be
considered. All matrix operations, such as the trace operation and
matrix multiplication, are defined accordingly. The non-interacting
particle-particle Green function, expressed in an antisymmetrical basis,
is the particle-particle Green function in the non-interacting limit,

\begin{align*}
K_{ijkl}^{0}(t-t') & =\frac{-i}{\hbar}\langle\Phi_{0}^{N}\vert\mathcal{T}[a_{I_{i}}(t)a_{I_{j}}(t)a_{I_{l}}^{\dagger}(t')a_{I_{k}}^{\dagger}(t')]\vert\Phi_{0}^{N}\rangle\\
 & =\frac{-i}{\hbar}(\delta_{jl}\delta_{ik}-\delta_{il}\delta_{jk})\left(e^{-\frac{i}{\hbar}(\epsilon_{i}+\epsilon_{j}-2\nu)(t-t')}\theta(i-F)\theta(j-F)\theta(t-t')+e^{\frac{i}{\hbar}(\epsilon_{i}+\epsilon_{j}+2\nu)(t'-t)}\theta(F-i)\theta(F-j)\theta(t'-t)\right)\\
 & =\frac{-i}{\hbar}(\delta_{jl}\delta_{ik}-\delta_{il}\delta_{jk})e^{-\frac{i}{\hbar}(\epsilon_{i}+\epsilon_{j}-2\nu)(t-t')}\left(\theta(i-F)\theta(j-F)\theta(t-t')+\theta(F-i)\theta(F-j)\theta(t'-t)\right),
\end{align*}
where $\vert\Phi_{0}^{N}\rangle$ is the $N$-electron non-interacting
reference state and the operators $a_{I_{i}}^{\dagger}(t)$ are the
creation operators in the interaction picture, $a_{I_{i}}^{\dagger}(t)=e^{\frac{i}{\hbar}(\hat{H_{0}}-\nu\hat{N})}a_{i}^{\dagger}e^{\frac{-i}{\hbar}(\hat{H_{0}}-\nu\hat{N})}$
with $\hat{H}_{0}$ the non-interacting (one-electron) Hamiltonian.
Note that the non-interacting particle-particle Green function can also
be written in terms of the non-interacting one-particle Green function
$\mathbf{G}^{0}$, 
\begin{align*}
G_{ij}^{0}(t-t') & =\frac{-i}{\hbar}\langle\Phi_{0}^{N}\vert\mathcal{T}[a_{I_{i}}(t)a_{I_{j}}^{\dagger}(t')]\vert\Phi_{0}^{N}\rangle\\
 & =\frac{-i}{\hbar}\delta_{ij}e^{\frac{-i}{\hbar}(\epsilon_{i}-\nu)(t-t')}\Big(\theta(i-F)\theta(t-t')-\theta(F-i)\theta(t'-t)\Big),
\end{align*}
namely

\begin{align*}
K_{ijkl}^{0}(t-t') & =\frac{-\hbar}{i}(\delta_{ik}\delta_{jl}-\delta_{il}\delta_{jk})G_{ik}^{0}(t-t')G_{jl}^{0}(t-t')\\
 & =\frac{-\hbar}{i}\left(G_{ik}^{0}(t-t')G_{jl}^{0}(t-t')-G_{il}^{0}(t-t')G_{jk}^{0}(t-t')\right).\\
\end{align*}
The Fourier Transform of the non-interacting particle-particle Green function
is 
\begin{eqnarray}
K_{ijkl}^{0}(E) & = & (\delta_{ik}\delta_{jl}-\delta_{il}\delta_{jk})\frac{-\hbar}{i}\int_{-\infty}^{+\infty}e^{\frac{i}{\hbar}Et}G_{ik}^{0}(t)G_{jl}^{0}(t)dt\nonumber \\
 & = & (\delta_{ik}\delta_{jl}-\delta_{il}\delta_{jk})\frac{-i}{\hbar}\int_{-\infty}^{+\infty}e^{\frac{-i}{\hbar}(\epsilon_{i}+\epsilon_{j}-2\nu-E)t}\Big(\theta(i-F)\theta(j-F)\theta(t)+\theta(F-i)\theta(F-j)\theta(-t)\Big)dt\\
 & = & (\delta_{ik}\delta_{jl}-\delta_{il}\delta_{jk})\left[\frac{\theta(i-F)\theta(j-F)}{E-(\epsilon_{i}+\epsilon_{j}-2\nu)+i\eta}-\frac{\theta(F-i)\theta(F-j)}{E-(\epsilon_{i}+\epsilon_{j}-2\nu)-i\eta}\right]\label{pp0}
\end{eqnarray}
where $\left\{ \epsilon_{i}\right\} $ are the orbital energies of
the non-interacting reference system. 

Eq. (\ref{dyson}) can be solved by multiplying each side of the equation
by $(E-\omega_{n}^{N-2})$ and subsequently taking the limit $E\rightarrow\omega_{n}^{N-2}$
\[
\underbrace{lim}_{E\rightarrow\omega_{n}^{N-2}}(E-\omega_{n}^{N-2})K(E)_{ijkl}=\underbrace{lim}_{E\rightarrow\omega_{n}^{N-2}}(E-\omega_{n}^{N-2})\Big(K^{0}(E)_{ijkl}+\sum_{m<n,o<p}K^{0}(E)_{ijmn}V_{mnop}K(E)_{opkl}\Big).
\]
This will separate out one single term on both sides of the equation:
the term that has $(E-\omega_{n}^{N-2})$ in the denominator.

\begin{align*}
(\chi_{kl}^{n,N-2})^{*}\chi_{ij}^{n,N-2} & =\sum_{m<n,o<p}K^{0}(\omega_{n}^{N-2})_{ijmn}V_{mnop}(\chi_{kl}^{n,N-2})^{*}\chi_{op}^{n,N-2}.
\end{align*}
The factor $(\chi_{kl}^{n,N-2})^{*}$ that appears on both sides of
the equation can then be canceled out

\begin{align}
\chi_{ij}^{n,N-2} & =\sum_{m<n,o<p}K^{0}(\omega_{n}^{N-2})_{ijmn}V_{mnop}\chi_{op}^{n,N-2}\nonumber \\
 & =\sum_{o<p}\left(\frac{\theta(i-F)\theta(j-F)}{\omega_{n}^{N-2}-(\epsilon_{i}+\epsilon_{j}-2\nu)+i\eta}-\frac{\theta(F-i)\theta(F-j)}{\omega_{n}^{N-2}-(\epsilon_{i}+\epsilon_{j}-2\nu)-i\eta}\right)V_{ijop}\chi_{op}^{n,N-2}.\label{eq:dyson_N-2}
\end{align}
This leads to a set of equations for the pp-indices $ab$ and a set
of equations for the hh-indices $hi$

\begin{align*}
\chi_{ab}^{n,N-2} & =\frac{1}{\omega_{n}^{N-2}-(\epsilon_{a}+\epsilon_{b}-2\nu)}\left(\sum_{c<d}^{N_{p}}V_{abcd}\chi_{cd}^{n,N-2}+\sum_{h<i}^{N_{h}}V_{abhi}\chi_{hi}^{n,N-2}\right)\\
\chi_{hi}^{n,N-2} & =\frac{-1}{\omega_{n}^{N-2}-(\epsilon_{h}+\epsilon_{i}-2\nu)}\left(\sum_{c<d}^{N_{p}}V_{hicd}\chi_{cd}^{n,N-2}+\sum_{h<i}^{N_{h}}V_{hijk}\chi_{jk}^{n,N-2}\right).
\end{align*}
which can be rearranged to reveal a generalized eigenvalue problem
in the eigenvalues $\omega_{n}$ and the eigenvectors $\chi^{n}$
\begin{align*}
\sum_{c<d}\chi_{cd}^{n,N-2}\left(V_{abcd}+\delta_{ac}\delta_{bd}(\epsilon_{a}+\epsilon_{b}-2\nu)\right)+\sum_{h<i}\chi_{hi}^{n,N-2}V_{abhi} & =\chi_{ab}^{n,N-2}\omega_{n}^{N-2}\\
-\sum_{c<d}\chi_{cd}^{n,N-2}V_{hicd}-\sum_{j<k}\chi_{jk}^{n,N-2}\left(V_{hijk}-\delta_{jh}\delta_{ik}(\epsilon_{h}+\epsilon_{i}-2\nu)\right) & =\chi_{hi}^{n,N-2}\omega_{n}^{N-2},
\end{align*}
where $a,b,c,d$ are particle indices, $h,i,j,k$ are hole indices
and $m,n,o,p$ are general indices. This can be written in matrix
form by defining $\mathbf{\mathbf{\mathbf{\chi}}^{n}}\equiv\begin{pmatrix}\mathbf{X^{n}}\\
\mathbf{Y^{n}}
\end{pmatrix}$, where $\mathbf{X}^{n}$ contains the elements of the vector $\chi^{n}$
with pp-labels and the vector $\mathbf{Y}^{n}$ contains the elements with
hh-labels,

\begin{align}
\begin{pmatrix}\mathbf{A} & \mathbf{B}\\
\mathbf{B^{\dagger}} & \mathbf{C}
\end{pmatrix}\begin{pmatrix}\mathbf{X^{n}}\\
\mathbf{Y^{n}}
\end{pmatrix} & =\omega_{n}\begin{pmatrix}\mathbf{1} & \mathbf{0}\\
\mathbf{0} & \mathbf{-1}
\end{pmatrix}\begin{pmatrix}\mathbf{X^{n}}\\
\mathbf{Y^{n}}
\end{pmatrix}\label{eq:pp-RPA_eigen}
\end{align}
with 
\begin{align}
 & A_{abcd}=\langle ab\Vert cd\rangle+\delta_{ac}\delta_{bd}(\epsilon_{a}+\epsilon_{b}-2\nu)\nonumber \\
 & B_{abij}=\langle ab\Vert ij\rangle\nonumber \\
 & C_{ijkl}=\langle ij\Vert kl\rangle-\delta_{ik}\delta_{jl}(\epsilon_{i}+\epsilon_{j}-2\nu).\label{app:ppRPA_matrixform}
\end{align}
\textcolor{black}{In our implementation, we have used $\nu=\frac{\epsilon_{HOMO}+\epsilon_{LUMO}}{2}$
, which corresponds to the average chemical potential for the physical
system under the non-interacting KS or generalized KS DFA \cite{Cohen2008}.
The constant $\nu$ does not affect the correlation energy; it only
ensures that the pp-RPA matrix on the left hand side of Eq. (\ref{eq:pp-RPA_eigen})
is positive semidefinite. This implies that the 2-electron removal
energies are negative and the two-electron addition energies are positive,
which makes it easier to separate them among the entire set of eigenvalues
$\omega_{n}$. }Since the pp-RPA matrix is expressed in an anti-symmetric
basis, only ordered pp-indices $ab$ with $a<b$ and hh-indices $hi$
with $h<i$ are included. The dimension of the $A$ and $C$ matrix
is therefore the number of ordered pp and hh pairs:

\begin{align*}
dim(\mathbf{A}) & =\frac{1}{2}N_{p}(N_{p}-1)\\
dim(\mathbf{C}) & =\frac{1}{2}N_{h}(N_{h}-1)
\end{align*}
where $N_{p}$ and $N_{h}$ are the number of particles (unoccupied
orbitals) and holes (occupied orbitals) respectively. Since in general,
$N_{p}>N_{h}$, the dimension of the pp-RPA matrix is $O(N_{p}^{2})$,
so a straightforward diaonalization of the pp-RPA matrix leads to
an $O(N_{p}^{6})$ scaling. Eq. (\ref{dyson}) can be rearranged for
the $N+2$ electron states in a similar manner, by multiplying by
$(E-\omega_{n}^{N+2})$ and taking the limit $E\rightarrow\omega_{n}^{N+2}$.
This leads to the same set of equations for the 2-electron addition
energies;

\begin{align*}
\chi_{ij}^{n,N+2} & =\sum_{m<n,o<p}K^{0}(\omega_{n}^{N+2})_{ijmn}V_{mnop}\chi_{op}^{n,N+2}\\
\chi_{ij}^{n,N+2} & =\sum_{o<p}\left(\frac{\theta(i-F)\theta(j-F)}{\omega_{n}^{N+2}-(\epsilon_{i}+\epsilon_{j}-2\nu)+i\eta}-\frac{\theta(F-i)\theta(F-j)}{\omega_{n}^{N+2}-(\epsilon_{i}+\epsilon_{j}-2\nu)-i\eta}\right)V_{ijop}\chi_{op}^{n,N+2},
\end{align*}
which has the exact same form as Eq. (\ref{eq:dyson_N-2}) for the
2-electron removal energies. The eigenvectors $\mathbf{X}^{\mathbf{n}}$
and $\mathbf{Y}^{\mathbf{n}}$ that satisfy Eq. (\ref{app:ppRPA_matrixform})
may thus involve either the $N+2$ electron states or $N-2$ electron
states. The generalized eigenvalues $\omega_{n}$ are either positive
2-electron addition energies, $\omega_{n}^{N+2}=E_{n}^{N+2}-E_{0}^{N}-2\nu$,
or negative 2-electron removal energies, $\omega_{n}^{N-2}=E_{0}^{N}-E_{n}^{N-2}-2\nu$.

\subsection{Exchange-correlation energy from dynamic pairing matrix fluctuations}

In this section, we develop an exact expression for the exchange-correlation
energy in terms of dynamic pairing matrix fluctuations via the adiabatic
connection \cite{Langreth19751425,Langreth19772884,Gunnarson19764274}.
The result is the dynamic pairing matrix\textcolor{black}{{} fluctuation}
counterpart of the well-known adiabatic-connection fluctuation-dissipation
(ACFD)\cite{Langreth19772884,Callen195134} theorem which expresses
the exchange-correlation energy in terms of dynamic density fluctuations. Just 
like the ACFD theorem, it formulates the exact correlation energy in terms of dynamic
fluctuations; it only considers different correlation
channels: the dynamic pairing matrix fluctuation involves the pp- and
hh-correlation channels, while the dynamic density fluctuation involves the
ph-correlation channel. These two different types of correlation
channels are closely related to the division of the second order density
matrix space into P-, Q- and G-matrices \cite{Mazziotti201262}. The
energy can be expressed in either one of these matrices, which naturally
leads to equivalent formulations for the exchange-correlation energy
in terms of dynamic pairing matrix fluctuations and dynamic density
fluctuations via the adiabatic connection. The resulting adiabatic-connection
formulae are in principle exact. In section \ref{sub:Ec_ppRPA}, we
show that the approximate exchange-correlation energy that follows
from the pp-RPA is equivalent to the summation of ladder diagrams
in many body perturbation theory.

The adiabatic connection considers a non-interacting reference system,
described by the Hamiltonian

\begin{align*}
\hat{H}_{0} & =\hat{h}+\hat{u},\\
\end{align*}
where $\hat{h}$ is the core Hamiltonian and $\hat{u}$ is the --
local or non-local, and possibly spin-dependent -- one-body operator
that defines the non-interacting system. The adiabatic connection
then defines a path from the non-interacting model to the fully interacting
system, parametrized by the interaction strength $\lambda$:

\begin{align*}
\hat{H}_{\lambda} & =\hat{H}_{0}+\lambda(\hat{V}-\hat{u}_{\lambda}).\\
\end{align*}
The operator $\hat{u}_{\lambda}$ is restricted to satisfy $\hat{u}_{1}=\hat{u}$
such that $\hat{H}_{1}$ is the Hamiltonian for the fully interacting
system. The Hellmann-Feynman theorem

\[
\frac{\partial E}{\partial\lambda}=\langle\Psi^{\lambda}\vert\frac{\partial\hat{H}_{\lambda}}{\partial\lambda}\vert\Psi^{\lambda}\rangle 
\]
then formulates the correlation energy $E^{1}-E^{0}$ as an integration
along the adiabatic connection path

\begin{align*}
E^{1}-E^{0} & =\int_{0}^{1}\langle\Psi^{\lambda}\vert\frac{\partial\hat{H}_{\lambda}}{\partial\lambda}\vert\Psi^{\lambda}\rangle d\lambda\\
 & =\int_{0}^{1}\langle\Psi^{\lambda}\vert\hat{V}-\hat{u}_{\lambda}-\lambda\frac{\partial\hat{u}_{\lambda}}{\partial\lambda}\vert\Psi^{\lambda}\rangle d\lambda.
\end{align*}
Since $\hat{V}$ is a two-body operator and $\hat{u}_{\lambda}$ is
a one-body operator, this can be written more compactly in terms of
the second-order density matrix $\Gamma^{\lambda}$ and the first-order
density matrix $\gamma^{\lambda}$ for the system with interaction
strength $\lambda$:

\begin{align*}
E^{1}-E^{0} & =\mathrm{tr}\ \int_{0}^{1}\mathbf{V\Gamma^{\lambda}}d\lambda-\mathrm{tr}\ \int_{0}^{1}\mathbf{u_{\lambda}\gamma^{\lambda}}d\lambda-\mathrm{tr}\ \int_{0}^{1}\lambda\mathbf{\frac{\partial u_{\lambda}}{\partial\lambda}\gamma^{\lambda}}d\lambda.
\end{align*}
Given that $E^{0}=\mathrm{tr}\ \mathbf{h\gamma^{0}}+\mathrm{tr}\ \mathbf{u\gamma^{0}}$,
the energy for the fully interacting system is

\begin{align*}
E^{1} & =\mathrm{tr}\ \mathbf{h\gamma^{0}}+\mathrm{tr}\ \int_{0}^{1}\mathbf{V\Gamma^{\lambda}}d\lambda-\mathrm{tr}\ \int_{0}^{1}(\mathbf{u_{\lambda}\gamma^{\lambda}}-\mathbf{u\gamma^{0}})d\lambda-\mathrm{tr}\ \int_{0}^{1}\lambda\frac{\partial\mathbf{u_{\lambda}}}{\partial\lambda}\mathbf{\gamma^{\lambda}}d\lambda.
\end{align*}
Relative to the Hartree-Fock/Exact Exchange energy functional, $E^{HF}=\mathrm{tr}\ \mathbf{h\gamma^{0}}+\mathrm{tr}\ \mathbf{V}\mathbf{\Gamma}^{\mathbf{0}}$,
the correlation energy functional $E^{c}\equiv E^{1}-E^{HF}$ is then

\begin{align*}
E^{c} & =\mathrm{tr}\ \int_{0}^{1}\mathbf{V}\mathbf{(\Gamma^{\lambda}}-\mathbf{\Gamma}^{\mathbf{0}})d\lambda-\mathrm{tr}\ \int_{0}^{1}(\mathbf{u_{\lambda}\gamma^{\lambda}}-\mathbf{u\gamma^{0}})d\lambda-\mathrm{tr}\ \int_{0}^{1}\lambda\frac{\partial\mathbf{u}_{\mathbf{\lambda}}}{\partial\lambda}\mathbf{\gamma^{\lambda}}d\lambda
\end{align*}

The two-body part of the energy can be written equivalently in terms
of the second-order density matrix, the Q-matrix or the G-matrix,
\textcolor{black}{defined by }

\begin{align*}
\Gamma_{ijkl} & =\langle\Psi\vert a_{k}^{+}a_{l}^{+}a_{j}a_{i}\vert\Psi\rangle\\
Q_{ijkl} & =\langle\Psi\vert a_{k}a_{l}a_{j}^{+}a_{i}^{+}\vert\Psi\rangle\\
G_{ijkl} & =\langle\Psi\vert a_{k}^{+}a_{l}a_{j}^{+}a_{i}\vert\Psi\rangle,
\end{align*}
because the anti-commutation properties of the creation and annihilation
operators define maps between the second-order density matrix, the
Q-matrix and the G-matrix:

\begin{align*}
\Gamma_{ijkl} & =Q_{lkji}+(\delta\wedge\gamma)_{ijkl}-(\delta\wedge\delta)_{ijkl}\\
\Gamma_{ijkl} & =-G_{ilkj}+\delta_{jl}\gamma_{ik}=G_{jlki}-\delta_{il}\gamma_{jk},
\end{align*}
where $\wedge$ denotes the wedge product, which includes all unique
anti-symmetrical product terms, $(\delta\wedge\gamma)_{ijkl}=\delta_{ik}\gamma_{jl}+\delta_{jl}\gamma_{ik}-\delta_{il}\gamma_{jk}-\delta_{jk}\gamma_{il}$
and $(\delta\wedge\delta)_{ijkl}=\delta_{ik}\delta_{jl}-\delta_{il}\delta_{jk}.$
This results in three equivalent expressions for the correlation energy

\begin{align}
E^{c} & =\mathrm{tr}\ \int_{0}^{1}\mathbf{V}\mathbf{(\Gamma^{\lambda}}-\mathbf{\Gamma}^{\mathbf{0}})d\lambda-\mathrm{tr}\ \int_{0}^{1}(\mathbf{u_{\lambda}\gamma^{\lambda}}-\mathbf{u\gamma^{0}})d\lambda-\mathrm{tr}\ \int_{0}^{1}\lambda\frac{\partial\mathbf{u}_{\mathbf{\lambda}}}{\partial\lambda}\mathbf{\gamma^{\lambda}}d\lambda,\label{eq:Ec_p}
\end{align}

\begin{align}
E^{c} & =\mathrm{tr}\ \int_{0}^{1}\mathbf{V}\mathbf{(Q^{\lambda}}-\mathbf{Q}^{\mathbf{0}})d\lambda+\mathrm{tr}\ \int_{0}^{1}\mathbf{V}\mathbf{(\delta\wedge(\gamma^{\lambda}}-\mathbf{\gamma}^{\mathbf{0}})d\lambda-\mathrm{tr}\ \int_{0}^{1}(\mathbf{u_{\lambda}\gamma^{\lambda}}-\mathbf{u\gamma^{0}})d\lambda-\mathrm{tr}\ \int_{0}^{1}\lambda\frac{\partial\mathbf{u}_{\mathbf{\lambda}}}{\partial\lambda}\mathbf{\gamma^{\lambda}}d\lambda,\label{eq:Ec_q}
\end{align}
and

\begin{align}
E^{c} & =\mathrm{tr}\ \int_{0}^{1}\mathbf{\tilde{V}}\mathbf{(G^{\lambda}}-\mathbf{G}^{\mathbf{0}})d\lambda-\sum_{ijk}\int_{0}^{1}\langle ij\vert ki \rangle (\gamma_{jk}^{\lambda}-\gamma_{jk}^{0})d\lambda-\mathrm{tr}\ \int_{0}^{1}(\mathbf{u_{\lambda}\gamma^{\lambda}}-\mathbf{u\gamma^{0}})d\lambda-\mathrm{tr}\ \int_{0}^{1}\lambda\frac{\partial\mathbf{u}_{\mathbf{\lambda}}}{\partial\lambda}\mathbf{\gamma^{\lambda}}d\lambda.\label{eq:Ec_g}
\end{align}
In Eq. (\ref{eq:Ec_g}), \textbf{$\tilde{\mathbf{V}}$ }is a rearranged
form of the two-electron integral matrix that pairs up indices associated
to the same electron, $\tilde{V}_{ijkl}=\langle il\vert jk\rangle$.
Equations (\ref{eq:Ec_p}-\ref{eq:Ec_g}) are general expressions
for the correlation energy functional, valid for any adiabatic connection
path.

In the context of KS-DFT, these formulae can be simplified by assuming
that the potential $\hat{u}_{\lambda}=\hat{u}_{\lambda}(\mathbf{x})$
is local and chosing a constant-density adiabatic connection path,
such that the spin density remains constant: $\rho^{\lambda}(\mathbf{x})=\rho^{0}(\mathbf{x})=\rho(\mathbf{x}).$
The terms $\mathrm{tr}\ \int_{0}^{1}(\mathbf{u_{\lambda}\gamma^{\lambda}}-\mathbf{u\gamma^{0}})d\lambda$
can then be expressed in terms of the density $\rho^{\lambda}=\rho$
instead of the density matrix $\gamma^{\lambda}$

\begin{align*}
\mathrm{tr}\ \int_{0}^{1}(\mathbf{u_{\lambda}\gamma^{\lambda}}-\mathbf{u\gamma^{0}})d\lambda & =\mathrm{tr}\ \int_{0}^{1}(\mathbf{u_{\lambda}\rho}-\mathbf{u\rho})d\lambda
\end{align*}
and the last term $\mathrm{tr}\ \int_{0}^{1}\lambda\frac{\partial\mathbf{u}_{\mathbf{\lambda}}}{\partial\lambda}\mathbf{\mathbf{\gamma^{\lambda}}}d\lambda$
can be simplified through partial integration

\begin{align*}
\mathrm{tr}\ \int_{0}^{1}\lambda\frac{\partial\mathbf{u}_{\mathbf{\lambda}}}{\partial\lambda}\mathbf{\mathbf{\gamma^{\lambda}}}d\lambda & =\mathrm{tr}\ \int_{0}^{1}\lambda\frac{\partial\mathbf{u}_{\mathbf{\lambda}}}{\partial\lambda}d\lambda\mathbf{\mathbf{\rho}}\\
 & =\mathrm{tr}\ [\lambda\mathbf{u_{\lambda}}]_{0}^{1}\mathbf{\rho}-\mathrm{tr}\ \int_{0}^{1}\mathbf{u}_{\mathbf{\lambda}}d\lambda\mathbf{\rho}\\
 & =\mathrm{tr}\ \mathbf{u\rho}-\mathrm{tr}\ \int_{0}^{1}\mathbf{u}_{\mathbf{\lambda}}d\lambda\mathbf{\rho}
\end{align*}
All terms involving $\hat{u}_{\lambda}$ cancel out:

\begin{align*}
-\mathrm{tr}\ \int_{0}^{1}(\mathbf{u_{\lambda}\gamma^{\lambda}}-\mathbf{u\gamma^{0}})d\lambda-\mathrm{tr}\ \int_{0}^{1}\lambda\frac{\partial\mathbf{u}_{\mathbf{\lambda}}}{\partial\lambda}\mathbf{\mathbf{\gamma^{\lambda}}}d\lambda & =-\mathrm{tr}\ \int_{0}^{1}(\mathbf{u_{\lambda}}-\mathbf{u})d\lambda\mathbf{\rho}-\mathrm{tr}\ \mathbf{u\rho}+\mathrm{tr}\ \int_{0}^{1}\mathbf{u}_{\mathbf{\lambda}}d\lambda\mathbf{\rho}\\
 & =0.
\end{align*}
Furthermore, the terms $\mathrm{tr}\ \int_{0}^{1}\mathbf{V}\mathbf{(\delta\wedge(\gamma^{\lambda}}-\mathbf{\gamma}^{\mathbf{0}})d\lambda$
and $\sum_{ijk}\int_{0}^{1}\langle ij\vert ki\rangle(\gamma_{jk}^{\lambda}-\gamma_{jk}^{0})d\lambda$
vanish\textcolor{black}{{} because of the following:}

\begin{align*}
\sum_{ijk}\langle ij\vert ik\rangle(\gamma_{jk}^{\lambda}-\gamma_{jk}^{0}) & =\int\frac{\sum_{i}\phi^*_{i}(\mathbf{x}')\phi_{i}(\mathbf{x'})\sum_{jk}\phi^*_{j}(\mathbf{x})\phi_{k}(\mathbf{x})(\gamma_{jk}^{\lambda}-\gamma_{jk}^{0})}{\mathbf{\vert r-r'\vert}}d\mathbf{x}d\mathbf{x}'\\
 & =\int\delta(0)\frac{\gamma^{\lambda}(\mathbf{x},\mathbf{x})-\gamma^{0}(\mathbf{x},\mathbf{x})}{\mathbf{\vert r-r'\vert}}d\mathbf{x}d\mathbf{x}'\\
 & =\int\delta(0)\frac{\rho^{\lambda}(\mathbf{x})-\rho^{0}(\mathbf{x})}{\mathbf{\vert r-r'\vert}}d\mathbf{x}d\mathbf{x}'\\
 & =0\\
\sum_{ijk}\langle ij\vert ki\rangle(\gamma_{jk}^{\lambda}-\gamma_{jk}^{0}) & =\int\frac{\sum_{i}\phi^*_{i}(\mathbf{x})\phi_{i}(\mathbf{x'})\sum_{jk}\phi_{j}^*(\mathbf{x'})\phi_{k}(\mathbf{x})(\gamma_{jk}^{\lambda}-\gamma_{jk}^{0})}{\mathbf{\vert r-r'\vert}}d\mathbf{x}d\mathbf{x}'\\
 & =\int\delta(\mathbf{x}-\mathbf{x}')\frac{\gamma^{\lambda}(\mathbf{x}',\mathbf{x})-\gamma^{0}(\mathbf{x'},\mathbf{x})}{\vert\mathbf{r}-\mathbf{r}'\vert}d\mathbf{x}d\mathbf{x}'\\
 & =\int\delta(\mathbf{x}-\mathbf{x}')\frac{\rho^{\lambda}(\mathbf{x})-\rho^{0}(\mathbf{x})}{\vert\mathbf{r}-\mathbf{r}'\vert}d\mathbf{x}d\mathbf{x}'\\
 & =0.
\end{align*}
Thus for a local potential $\hat{u}_{\lambda}(r)$ the adiabatic connection
along the constant-density path leads to the equivalent formulae

\begin{align*}
E^{c} & =\mathrm{tr}\ \int_{0}^{1}\mathbf{V}\mathbf{(\Gamma^{\lambda}}-\mathbf{\Gamma}^{\mathbf{0}})d\lambda
\end{align*}

\begin{align*}
E^{c} & =\mathrm{tr}\ \int_{0}^{1}\mathbf{V}\mathbf{(Q^{\lambda}}-\mathbf{Q}^{\mathbf{0}})d\lambda
\end{align*}

\begin{align*}
E^{c} & =\mathrm{tr}\ \int_{0}^{1}\mathbf{\tilde{V}}\mathbf{(G^{\lambda}}-\mathbf{G}^{\mathbf{0}})d\lambda.
\end{align*}

The correlation energy can then be expressed in terms of dynamic fluctuations:
the P- and Q-matrix can be written in terms of the pairing matrix fluctuation
and the G-matrix in terms of  the density matrix fluctuation. The second-order
density matrix can be related to the transition paring matrix elements
$\chi_{ij}^{n,N-2}=\langle\Psi_{n}^{N-2}\vert a_{i}a_{j}\vert\Psi_{0}^{N}\rangle$
through the completeness of the $N-2$ electron wavefunction basis,
\begin{align}
\Gamma_{ijkl} & =\langle\Psi_{0}^{N}\vert a_{k}^{+}a_{l}^{+}a_{j}a_{i}\vert\Psi_{0}^{N}\rangle\nonumber \\
 & =\sum_n\langle\Psi_{0}^{N}\vert a_{k}^{+}a_{l}^{+}\vert\Psi_{n}^{N-2}\rangle\langle\Psi_{n}^{N-2}\vert a_{j}a_{i}\vert\Psi_{0}^{N}\rangle\nonumber \\
 & =\sum_{n}\chi_{ji}^{n,N-2}(\chi_{lk}^{n,N-2})^{*},\label{eq:Pmatrix}
\end{align}
and the Q-matrix can be related to the transition pairing matrix elements
$\chi_{ij}^{n,N+2}=\langle\Psi_{n}^{N}\vert a_{i}a_{j}\vert\Psi_{0}^{N+2}\rangle$
through the completeness of the $N+2$ electron wavefunction basis,

\begin{align}
Q_{ijkl} & =\langle\Psi_{0}^{N}\vert a_{k}a_{l}a_{j}^{+}a_{i}^{+}\vert\Psi_{0}^{N}\rangle\nonumber \\
 & =\sum_n \langle\Psi_{0}^{N}\vert a_{k}a_{l}\vert\Psi_{n}^{N+2}\rangle\langle\Psi_{n}^{N+2}\vert a_{j}^{+}a_{i}^{+}\vert\Psi_{0}^{N}\rangle\nonumber \\
 & =\sum_{n}\chi_{kl}^{n,N+2}(\chi_{ij}^{n,N+2})^{*},\label{eq:Qmatrix}
\end{align}
and the G-matrix can be written in terms of the transition density
matrix elements $\chi_{ij}^{n,N}\equiv\langle\Psi_{n}^{N}\vert a_{j}^{+}a_{i}\vert\Psi_{0}^{N}\rangle$
through the completeness of the $N$-electron wavefunction basis

\begin{align}
G_{ijkl} & =\langle\Psi_{0}^{N}\vert a_{k}^{+}a_{l}a_{j}^{+}a_{i}\vert\Psi_{0}^{N}\rangle\nonumber \\
 & =\sum_{n}\langle\Psi_{0}^{N}\vert a_{k}^{+}a_{l}\vert\Psi_{n}^{N}\rangle\langle\Psi_{n}^{N}\vert a_{j}^{+}a_{i}\vert\Psi_{0}^{N}\rangle\nonumber \\
 & =\sum_{n\neq0}\chi_{ij}^{n,N}(\chi_{kl}^{n,N})^{*}+\gamma_{ij}\gamma_{kl}.\label{eq:Gmatrix}
\end{align}
The exact correlation energy can thus be expressed in terms of transition
pairing matrix elements,

\begin{align}
E^{c} & =\sum_{n}\sum_{ijkl}\int_{0}^{1}\left((\chi_{\lambda}^{n,N-2})_{ji}(\chi_{\lambda}^{n,N-2})_{lk}^{*}-(\chi_{0}^{n,N-2})_{ji}(\chi_{0}^{n,N-2})_{lk}^{*}\right)V_{ijkl}d\lambda\nonumber \\
 & =\sum_{n}\int_{0}^{1}\int d\mathbf{x}d\mathbf{x}'\frac{\chi_{\lambda}^{n,N-2}(\mathbf{x},\mathbf{x}')\chi_{\lambda}^{n,N-2}(\mathbf{x},\mathbf{x}')^{*}-\chi_{0}^{n,N-2}(\mathbf{x},\mathbf{x}')\chi_{0}^{n,N-2}(\mathbf{x},\mathbf{x}')^{*}}{\vert\mathbf{r}-\mathbf{r}'\vert}d\lambda,\label{eq:Ec_pairing_P}
\end{align}
and

\begin{align}
E^{c} & =\sum_{n}\sum_{ijkl}\int_{0}^{1}\left((\chi_{\lambda}^{n,N+2})_{ij}^{*}(\chi_{\lambda}^{n,N+2})_{kl}-(\chi_{0}^{n,N+2})_{ij}^{*}(\chi_{0}^{n,N+2})_{kl}\right)V_{ijkl}d\lambda\nonumber \\
 & =\sum_{n}\int_{0}^{1}\int d\mathbf{x}d\mathbf{x}'\frac{\chi_{\lambda}^{n,N+2}(\mathbf{x},\mathbf{x}')^{*}\chi_{\lambda}^{n,N+2}(\mathbf{x},\mathbf{x}')-\chi_{0}^{n,N+2}(\mathbf{x},\mathbf{x}')^{*}\chi_{0}^{n,N+2}(\mathbf{x},\mathbf{x}')}{\vert\mathbf{r}-\mathbf{r}'\vert}d\lambda,\label{eq:Ec_pairing_Q}
\end{align}
or in terms of transition density matrix elements,

\begin{align}
E^{c} & =\sum_{n\neq0}\sum_{ijkl}\int_{0}^{1}\left((\chi_{\lambda}^{n,N})_{ij}(\chi_{\lambda}^{n,N})_{kl}^{*}-(\chi_{0}^{n,N})_{ij}(\chi_{0}^{n,N})_{kl}^{*}\right)\tilde{V}_{ijkl}d\lambda\nonumber \\
 & =\sum_{n\neq0}\int_{0}^{1}\int d\mathbf{x}d\mathbf{x}'\frac{\chi_{\lambda}^{n,N}(\mathbf{x})\chi_{\lambda}^{n,N}(\mathbf{x}')^{*}-\chi_{0}^{n,N}(\mathbf{x})\chi_{0}^{n,N}(\mathbf{x}')^{*}}{\vert\mathbf{r}-\mathbf{r}'\vert}d\lambda.\label{eq:Ec_trans_dens}
\end{align}
Note that the ground-state density matrix elements in Eq. (\ref{eq:Gmatrix})
do not contribute along the constant-density adiabatic-connection
path.

Equation (\ref{eq:Ec_trans_dens}) for the correlation energy in terms
of transition density matrix elements has been exploited in the context
of ph-RPA, because the transition density matrix elements involved
can be extracted from the polarization propagator $\mathbf{\Pi}$,
define\textcolor{black}{d as \cite{Blaizot1986}}

\begin{align*}
\Pi(E)_{ijkl} & =\sum_{n\neq0}\frac{\langle\Psi_{0}^{N}\vert a_{k}^{+}a_{l}\vert\Psi_{n}^{N}\rangle\langle\Psi_{n}^{N}\vert a_{j}^{+}a_{i}\vert\Psi_{0}^{N}\rangle}{E-\omega_{n}^{N}+i\eta}-\sum_{n\neq0}\frac{\langle\Psi_{0}^{N}\vert a_{j}^{+}a_{i}\vert\Psi_{n}^{N}\rangle\langle\Psi_{n}^{N}\vert a_{k}^{+}a_{l}\vert\Psi_{0}^{N}\rangle}{E+\omega_{n}^{N}-i\eta}\\
 & =\sum_{n\neq0}\frac{\langle\Psi_{0}^{N}\vert a_{k}^{+}a_{l}\vert\Psi_{n}^{N}\rangle\langle\Psi_{n}^{N}\vert a_{j}^{+}a_{i}\vert\Psi_{0}^{N}\rangle}{E-\omega_{n}^{N}+i\eta}-\sum_{n\neq0}\frac{\langle\Psi_{0}^{N}\vert a_{j}^{+}a_{i}\vert\Psi_{n}^{N}\rangle\langle\Psi_{n}^{N}\vert a_{k}^{+}a_{l}\vert\Psi_{0}^{N}\rangle}{E+\omega_{n}^{N}-i\eta}\\
 & =\sum_{n\neq0}\frac{(\chi_{kl}^{n,N})^{*}\chi_{ij}^{n,N}}{E-\omega_{n}^{N}+i\eta}-\sum_{n\neq0}\frac{(\chi_{ji}^{n,N})^{*}\chi_{lk}^{n,N}}{E+\omega_{n}^{N}-i\eta}.
\end{align*}
Integrating over a semi-circular path in the positive real plane gives

\begin{align}
\frac{-1}{2\pi i}\int_{-i\infty}^{+i\infty}e^{-E\eta}\Pi(E)_{ijkl}dE & =\sum_{n\neq0}\chi_{ij}^{n,N}(\chi_{kl}^{n,N})^{*}\label{eq:Pi_int}\\
\nonumber 
\end{align}
while integrating over a semi-circular path in the negative real plane
gives

\begin{align*}
\frac{-1}{2\pi i}\int_{-i\infty}^{+i\infty}e^{E\eta}\Pi(E)_{ijkl}dE & =\sum_{n\neq0}\chi_{lk}^{n,N}(\chi_{ji}^{n,N})^{*}.\\
\end{align*}
Using Eqs. (\ref{eq:Ec_trans_dens}) and (\ref{eq:Pi_int}), the correlation
energy can be expressed in terms of the polarization propagator:

\begin{align}
E^{c} & =\sum_{ijkl}\tilde{V}_{ijkl}\sum_{n\neq0}\int_{0}^{1}(\chi_{\lambda}^{n,N})_{ij}(\chi_{\lambda}^{n,N})_{kl}^{*}d\lambda-(\chi_{0}^{n,N})_{ij}(\chi_{0}^{n,N})_{kl}^{*}\nonumber \\
 & =\frac{-1}{2\pi i}\int_{0}^{1}\int_{-i\infty}^{+i\infty}e^{-E\eta}\mathrm{tr}\ \mathbf{\tilde{V}}[\mathbf{\Pi^{\lambda}}(E)-\mathbf{\Pi^{0}}(E)]dEd\lambda\nonumber \\
 & =\frac{-1}{2\pi i}\int_{0}^{1}\int_{-i\infty}^{+i\infty}e^{-E\eta}\int d\mathbf{x}d\mathbf{x}'\int\frac{\Pi^{\lambda}(\mathbf{x},\mathbf{x}',E)-\Pi^{0}(\mathbf{x},\mathbf{x}',E)}{\vert\mathbf{r}-\mathbf{r}'\vert}dEd\lambda.\label{eq:Ec_G}
\end{align}
This result is in principle exact, but requires an expression for
$\Pi^{\lambda}(\mathbf{x},\mathbf{x}',E)$. The ph-RPA approximates
the polarization propagator for the interacting strength $\lambda$
by the Dyson-like equation $\mathbf{\Pi}^{\lambda}=\mathbf{\Pi}^{0}+\lambda\mathbf{\Pi}^{0}\tilde{\mathbf{V}}\mathbf{\Pi}^{\lambda}$,
which leads to the well-known energy expression for the RPA \cite{Langreth19772884,Bohm1951625}.

The correlation energy can also be expressed in terms of pairing matrix
fluctuations or the particle-particle Green function, based on Eqs. (\ref{eq:Pmatrix},\ref{eq:Qmatrix}).
The transition pairing matrix elements involved can be extracted from
the particle-particle Green function, Eq. (\ref{eq:pp_GreenFunction}):
integrating the particle-particle Green function over a semi-circular
path in the negative real plane gives

\begin{align}
\frac{-1}{2\pi i}\int_{-i\infty}^{+i\infty}e^{E\eta}K(E)_{ijkl}dE & =\sum_{n}(\chi_{lk}^{n,N-2})^{*}\chi_{ji}^{n,N-2}\label{eq:K_P}\\
\nonumber 
\end{align}
while closing the contour in the positive real plane gives

\begin{align}
\frac{-1}{2\pi i}\int_{-i\infty}^{+i\infty}e^{-E\eta}K(E)_{ijkl}dE & =\sum_{n}(\chi_{kl}^{n,N+2})^{*}\chi_{ij}^{n,N+2}.\label{eq:K_Q}\\
\nonumber 
\end{align}
Equations (\ref{eq:Pmatrix}) and (\ref{eq:K_P}) then lead to an
expression for the correlation energy in terms of the particle-particle
Green function, integrated over a contour in the negative real plane:

\begin{align}
E^{c} & =\sum_{ijkl}V_{ijkl}\sum_{n}\int_{0}^{1}(\chi_{\lambda}^{n,N-2})_{ij}(\chi_{\lambda}^{n,N-2})_{kl}^{*}d\lambda-(\chi_{0}^{n,N-2})_{ij}(\chi_{0}^{n,N-2})_{kl}^{*}\nonumber \\
 & =\frac{-1}{2\pi i}\int_{0}^{1}\int_{-i\infty}^{+i\infty}e^{E\eta}\mathrm{tr}\ \mathbf{V}[\mathbf{K^{\lambda}}(E)-\mathbf{K^{0}}(E)]dEd\lambda\nonumber \\
 & =\frac{-1}{2\pi i}\int_{0}^{1}\int_{-i\infty}^{+i\infty}e^{E\eta}\int d\mathbf{x}d\mathbf{x}'\frac{K^{\lambda}(\mathbf{x},\mathbf{x}',E)-K^{0}(\mathbf{x},\mathbf{x}',E)}{\vert\mathbf{r}-\mathbf{r}'\vert}dE\label{eq:Ec_P}
\end{align}
where

\begin{equation}
K^{\lambda}(\mathbf{x}_{1},\mathbf{x}_{2},E)=\frac{1}{2}\sum_{ijkl}K(E)_{ijkl}\phi_{i}(\mathbf{x}_{1})\phi_{j}(\mathbf{x}_{2})\phi_{k}^{*}(\mathbf{x}_{1})\phi_{l}^{*}(\mathbf{x}_{2})\label{eq:K_RealSpace}
\end{equation}
Equations (\ref{eq:Qmatrix}) and (\ref{eq:K_Q}) lead to the same
formula, integrated over a contour in the positive real plane:

\begin{align}
E^{c} & =\sum_{ijkl}V_{ijkl}\sum_{n}\int_{0}^{1}(\chi_{\lambda}^{n,N+2})_{ij}^{*}(\chi_{\lambda}^{n,N+2})_{kl}d\lambda-(\chi_{0}^{n,N+2})_{ij}^{*}(\chi_{0}^{n,N+2})_{kl}\nonumber \\
 & =\frac{-1}{2\pi i}\int_{0}^{1}\int_{-i\infty}^{+i\infty}e^{-E\eta}\mathrm{tr}\ \mathbf{V}[\mathbf{K^{\lambda}}(E)-\mathbf{K^{0}}(E)]dEd\lambda\nonumber \\
 & =\frac{-1}{2\pi i}\int_{0}^{1}\int_{-i\infty}^{+i\infty}e^{-E\eta}\int d\mathbf{x}d\mathbf{x}'\frac{\mathbf{K^{\lambda}}(\mathbf{x},\mathbf{x}',E)-\mathbf{K^{0}}(\mathbf{x},\mathbf{x}',E)}{\vert\mathbf{r}-\mathbf{r}'\vert}dEd\lambda.\label{eq:Ec_Q}
\end{align}
The equivalence of (\ref{eq:Ec_P}) and (\ref{eq:Ec_Q}) shows that
the integration path can be closed in either half plane. Although
the previous equations integrate the Green functions along the imaginary
axis, similar equations hold for integration along the real axis,
namely 
\begin{align}
E^{c} & =\frac{-1}{2\pi i}\int_{0}^{1}\int_{-\infty}^{+\infty}e^{-iE\eta}\mathrm{tr}\ \mathbf{\tilde{V}}[\mathbf{\Pi^{\lambda}}(E)-\mathbf{\Pi^{0}}(E)]dEd\lambda\nonumber \\
 & =\frac{-1}{2\pi i}\int_{0}^{1}\int_{-\infty}^{+\infty}e^{-iE\eta}\int d\mathbf{x}d\mathbf{x}'\int\frac{\Pi^{\lambda}(\mathbf{x},\mathbf{x}',E)-\Pi^{0}(\mathbf{x},\mathbf{x}',E)}{\vert\mathbf{r}-\mathbf{r}'\vert}dEd\lambda,\label{eq:Ec_G-1}
\end{align}
\begin{align}
E^{c} & =\frac{-1}{2\pi i}\int_{0}^{1}\int_{-\infty}^{+\infty}e^{iE\eta}\mathrm{tr}\ \mathbf{V}[\mathbf{K^{\lambda}}(E)-\mathbf{K^{0}}(E)]dEd\lambda\nonumber \\
 & =\frac{-1}{2\pi i}\int_{0}^{1}\int_{-\infty}^{+\infty}e^{iE\eta}\int d\mathbf{x}d\mathbf{x}'\frac{K^{\lambda}(\mathbf{x},\mathbf{x}',E)-K^{0}(\mathbf{x},\mathbf{x}',E)}{\vert\mathbf{r}-\mathbf{r}'\vert}dE,\label{eq:Ec_P-1}
\end{align}
and 
\begin{align}
E^{c} & =\frac{-1}{2\pi i}\int_{0}^{1}\int_{-\infty}^{+\infty}e^{-iE\eta}\mathrm{tr}\ \mathbf{V}[\mathbf{K^{\lambda}}(E)-\mathbf{K^{0}}(E)]dEd\lambda\nonumber \\
 & =\frac{-1}{2\pi i}\int_{0}^{1}\int_{-\infty}^{+\infty}e^{-iE\eta}\int d\mathbf{x}d\mathbf{x}'\frac{K^{\lambda}(\mathbf{x},\mathbf{x}',E)-K^{0}(\mathbf{x},\mathbf{x}',E)}{\vert\mathbf{r}-\mathbf{r}'\vert}dEd\lambda.\label{eq:Ec_Q-1}
\end{align}
From the numerical point of view, integration along the imaginary
axis is more convenient because it avoids the poles on the real axis.
The integration along the imaginary energy axis is also valid for
the retarded Green function or the paring matrix fluctuation, such
that Eq. (\ref{eq:Ec_G},\ref{eq:Ec_P} and \ref{eq:Ec_Q}) also apply
to the retarded Green function or the pairing matrix fluctuation.

\subsection{Exchange-correlation energy from the particle-particle RPA}

\label{sub:Ec_ppRPA}

Expressions (\ref{eq:Ec_P}) and (\ref{eq:Ec_Q}) for the correlation
energy in terms of the particle-particle Green function are in principle
exact, but require knowledge of the Green function $\mathbf{K}^{\lambda}(E)$
as a function of the interaction strength $\lambda$. The pp-RPA approximates
$\mathbf{K}^{\lambda}(E)$ through the Dyson-like equation

\begin{equation}
\mathbf{K}^{\lambda}(E)=\mathbf{K}^{0}(E)+\lambda\mathbf{K}^{0}(E)\mathbf{V}\mathbf{K}^{\lambda}(E)\label{eq:dyson_lambda}
\end{equation}
such that, based on Eq. (\ref{eq:Ec_P}), 
\begin{align}
E_{pp}^{c} & =\frac{-1}{2\pi i}\int_{0}^{1}\int_{-i\infty}^{+i\infty}\mathrm{tr}\ [\mathbf{K}^{\lambda}(E)\mathbf{V}-\mathbf{K}^{0}(E)\mathbf{V}]dEd\lambda\nonumber \\
 & =\frac{-1}{2\pi i}\int_{0}^{1}\int_{-i\infty}^{+i\infty}\Big(\lambda\mathrm{tr}\ [\mathbf{K}^{0}(E)\mathbf{V}\mathbf{K}^{0}(E)\mathbf{V}]+\lambda^{2}\mathrm{tr}\ [\mathbf{K}^{0}(E)\mathbf{V}\mathbf{K}^{0}(E)\mathbf{V}\mathbf{K}^{0}(E)\mathbf{V}]+\ldots\Big)dEd\lambda\nonumber \\
 & =\frac{-1}{2\pi i}\int_{0}^{1}\int_{-i\infty}^{+i\infty}\sum_{n=2}^{\infty}\lambda^{n-1}\mathrm{tr}\ [(\mathbf{K}^{0}\mathbf{V})^{n}]dEd\lambda\nonumber \\
 & =\frac{-1}{2\pi i}\int_{-i\infty}^{+i\infty}\Big[\sum_{n=2}^{\infty}\frac{1}{n}(\lambda)^{n}\mathrm{tr}\ [(\mathbf{K}^{0}\mathbf{V})^{n}]dE\Big]_{0}^{1}\nonumber \\
 & =-\frac{1}{2\pi i}\int_{-i\infty}^{+i\infty}\sum_{n=2}^{\infty}\frac{1}{n}\mathrm{tr}\ [(\mathbf{K}^{0}\mathbf{V})^{n}]dE\nonumber \\
 & =\frac{1}{2\pi i}\int_{-i\infty}^{+i\infty}\mathrm{tr}\ [\mathrm{ln}(\mathbf{I}-\mathbf{K}^{0}\mathbf{V})+\mathbf{K}^{0}\mathbf{V}]dE.\label{eq:Ec_Re_int}
\end{align}
Note that no convergence factors $e^{\pm E\eta}$ are needed here,
since the third line shows that no first-order poles are included.
This expression is consistent with the diagrammatic expansion of the
particle-particle Green function in many body perturbation theory. Similarly
to the ph-RPA, which approximates the ground-state correlation energy
by the sum of all ring diagrams, the pp-RPA approximates the correlation
energy by the sum of all ladder diagrams\cite{Blaizot1986}: 
\begin{eqnarray}
E_{Ladder}^{c} & = & \frac{-1}{2\pi i}\ \sum_{n=2}^{\infty}\frac{1}{n}\int_{-i\infty}^{+i\infty}\mathrm{tr}\ [\mathbf{K}^{0}(E))\mathbf{V}]^{n}\ dE\nonumber \\
 & = & \frac{-1}{2\pi i}\ \sum_{n=1}^{\infty}\frac{1}{n}\int_{-i\infty}^{+i\infty}\mathrm{tr}\ [\mathbf{K}^{0}(E))\mathbf{V}]^{n}\ dE+\frac{1}{2\pi i}\ \int_{-i\infty}^{+i\infty}\mathrm{tr}\ \mathbf{K}^{0}(E))\mathbf{V}\ dE\\
 & = & \frac{1}{2\pi i}\ \int_{-i\infty}^{+i\infty}\mathrm{tr}\ [\mathrm{ln}(\mathbf{I}-\mathbf{K}^{0}(E)\mathbf{V})+\mathbf{K}^{0}(E)\mathbf{V}]\ dE.\label{eq:Ec_Im_int}
\end{eqnarray}
This expression is equivalent to adiabatic connection result, Eq.
(\ref{eq:Ec_Re_int}). The pp-RPA equations have an equivalent real
space representation. To derive their real space counterpart, it is
convenient to rewrite the Dyson-like equation in terms of the two-electron
integrals that are not antisymmetrized

\begin{align*}
\frac{1}{2}K_{ijkl}^{\lambda} & =\frac{1}{2}K_{ijkl}^{0}+\lambda\sum_{mnop}\frac{1}{2}K_{ijmn}^{0}\langle mn\vert op\rangle\frac{1}{2}K_{opkl}^{\lambda}\\
\end{align*}
Because $v(\mathbf{x}_{1},\mathbf{x}_{2})=\frac{1}{\vert\mathbf{r_{1}}-\mathbf{r}_{\mathbf{2}}\vert}$
is diagonal the real space representation, the real-space equivalent
of Eq. (\ref{eq:dyson_lambda}) is a four-point equation

\begin{align*}
K^{\lambda}(\mathbf{x}_{1},\mathbf{x}_{2},\mathbf{x}_{1}',\mathbf{x}_{2}',E) & =K^{0}(\mathbf{x}_{1},\mathbf{x}_{2},\mathbf{x}_{1}',\mathbf{x}_{2}',E)+\lambda\int d\mathbf{x}_{1}"d\mathbf{x}_{2}"K^{0}(\mathbf{x}_{1},\mathbf{x}_{2},\mathbf{x}_{1}",\mathbf{x}_{2}",E)v(\mathbf{x}_{1}",\mathbf{x}_{2}")K^{\lambda}(\mathbf{x}_{1}",\mathbf{x}_{2}",\mathbf{x}_{1}',\mathbf{x}_{2}',E).\\
\end{align*}
This leads to the correlation energy expression

\begin{eqnarray}
E_{pp}^{c} & = & \frac{-1}{2\pi i}\int_{0}^{1}\int_{-i\infty}^{+i\infty}\lambda\int\int K^{0}(\mathbf{x}_{1},\mathbf{x}_{2},\mathbf{x}_{1}',\mathbf{x}_{2}',E)v(\mathbf{x}'_{1},\mathbf{x}'_{2})K^{0}(\mathbf{x}_{1}',\mathbf{x}_{2}',\mathbf{x}_{1},\mathbf{x}_{2},E)v(\mathbf{x}_{1},\mathbf{x}_{2})d\mathbf{x}_{1}d\mathbf{x}_{2}d\mathbf{x}_{1}'d\mathbf{x}_{2}'\ dEd\lambda\nonumber \\
 & - & \frac{1}{2\pi i}\int_{0}^{1}\int_{-i\infty}^{+i\infty}\lambda^{2}\int\int\int K^{0}(\mathbf{x}_{1},\mathbf{x}_{2},\mathbf{x}_{1}',\mathbf{x}_{2}',E)v(\mathbf{x}'_{1},\mathbf{x}'_{2})K^{0}(\mathbf{x}_{1}',\mathbf{x}_{2}',\mathbf{x}_{1}",\mathbf{x}_{2}",E)v(\mathbf{x}_{1}",\mathbf{x}_{2}")\nonumber \\
 &  & \times K^{0}(\mathbf{x}_{1}",\mathbf{x}_{2}",\mathbf{x}_{1},\mathbf{x}_{2},E)v(\mathbf{x}_{1},\mathbf{x}_{2})d\mathbf{x}_{1}d\mathbf{x}_{2}d\mathbf{x}_{1}'d\mathbf{x}_{2}'d\mathbf{x}_{1}"d\mathbf{x}_{2}"\ dEd\lambda\nonumber \\
 & - & \frac{1}{2\pi i}\int_{0}^{1}\int_{-i\infty}^{+i\infty}\lambda^{3}\int\int\int\int\ldots\nonumber \\
 & - & \ldots\nonumber \\
 & = & \frac{-1}{2\pi i}\int_{-i\infty}^{+i\infty}\frac{1}{2}\int\int K^{0}(\mathbf{x}_{1},\mathbf{x}_{2},\mathbf{x}_{1}',\mathbf{x}_{2}',E)v(\mathbf{x}'_{1},\mathbf{x}'_{2})K^{0}(\mathbf{x}_{1}',\mathbf{x}_{2}',\mathbf{x}_{1},\mathbf{x}_{2},E)v(\mathbf{x}_{1},\mathbf{x}_{2})d\mathbf{x}_{1}d\mathbf{x}_{2}d\mathbf{x}_{1}'d\mathbf{x}_{2}'\ dE\nonumber \\
 & - & \frac{1}{2\pi i}\int_{-i\infty}^{+i\infty}\frac{1}{3}\int\int\int K^{0}(\mathbf{x}_{1},\mathbf{x}_{2},\mathbf{x}_{1}',\mathbf{x}_{2}',E)v(\mathbf{x}'_{1},\mathbf{x}'_{2})K^{0}(\mathbf{x}_{1}',\mathbf{x}_{2}',\mathbf{x}_{1}",\mathbf{x}_{2}",E)v(\mathbf{x}_{1}",\mathbf{x}_{2}")\nonumber \\
 &  & \times K^{0}(\mathbf{x}_{1}",\mathbf{x}_{2}",\mathbf{x}_{1},\mathbf{x}_{2},E)v(\mathbf{x}_{1},\mathbf{x}_{2})d\mathbf{x}_{1}d\mathbf{x}_{2}d\mathbf{x}_{1}'d\mathbf{x}_{2}'d\mathbf{x}_{1}"d\mathbf{x}_{2}"\ dE\nonumber \\
 & - & \frac{1}{2\pi i}\ \int_{-i\infty}^{+i\infty}\frac{1}{4}\int\int\int\int\ldots\nonumber \\
 & - & \ldots\nonumber \\
 & = & \frac{1}{2\pi i}\ \int_{-i\infty}^{+i\infty}\mathrm{tr}\left(\mathrm{ln}(\mathbf{I}-\mathbf{S})+\mathbf{S}\right)dE\label{eq:Ec_Im_int_realspace}
\end{eqnarray}
where $\mathbf{S}$ is a matrix represented in real space with its
elements 
\[
S(\mathbf{x}_{1},\mathbf{x}_{2},\mathbf{x}_{1}',\mathbf{x}_{2}',E)=K^{0}(\mathbf{x}_{1},\mathbf{x}_{2},\mathbf{x}_{1}',\mathbf{x}_{2}',E)v(\mathbf{x}'_{1},\mathbf{x}'_{2})
\]

The correlation energy can be computed directly from Eq. (\ref{eq:Ec_Im_int})
or (\ref{eq:Ec_Im_int_realspace}) through numerical integration,
since the non-interacting pp-function $\mathbf{K}^{0}$ has a simple,
known structure (Eq. (\ref{pp0})), but it can also be reformulated
in terms of the eigenvalues of equation (\ref{app:ppRPA_matrixform})\cite{Blaizot1986}:

\begin{eqnarray}
E_{pp}^{c} & = & \frac{1}{2\pi i}\ \int_{-i\infty}^{+i\infty}\mathrm{tr}\ [\mathrm{ln}(\mathbf{I}-\mathbf{K}^{0}(E)\mathbf{V})+\mathbf{K}^{0}(E)\mathbf{V}]\ dE\nonumber \\
 & = & \sum_{n}^{N_{pp}}\omega_{n}^{N+2}-\mathrm{tr}\ \mathbf{A}\label{Ec}\\
 & = & -\sum_{n}^{N_{hh}}\omega_{n}^{N-2}-\mathrm{tr}\ \mathbf{C}\\
 & = & \frac{1}{2}\sum_{n}^{N_{pp}}\omega_{n}^{N+2}-\frac{1}{2}\mathrm{tr}\ \mathbf{A}-\frac{1}{2}\sum_{n}^{N_{hh}}\omega_{n}^{N-2}-\frac{1}{2}\mathrm{tr}\ \mathbf{C}.
\end{eqnarray}
In order to show how the expression Eq. (\ref{eq:Ec_Im_int}), or
equivalently Eq. (\ref{eq:Ec_Re_int}), reduces to the three equivalent
expressions in terms of the eigenvalues $\omega_{n}^{N+2}$ or $\omega_{n}^{N-2}$,
we will consider the integrals of the two terms, $\mathrm{tr}\ [\ln(\mathbf{I}-\mathbf{K}^{0}(E)\mathbf{V})]$
and $\mathrm{tr}\ [\mathbf{K}^{0}(E)\mathbf{V}]$, separately. First
of all,

$ $ 
\begin{align*}
 & \frac{1}{2\pi i}\ \int_{-i\infty}^{+i\infty}\mathrm{tr}\ \mathbf{K}^{0}(E)\mathbf{V}\ dE\\
 & =\frac{1}{2\pi i}\ \int_{-i\infty}^{+i\infty}\sum_{a<b}^{N_{p}}V_{abab}\frac{1}{E-(\epsilon_{a}+\epsilon_{b}-2\nu)+i\eta}-\sum_{h<i}^{N_{h}}V_{hihi}\frac{1}{E-(\epsilon_{h}+\epsilon_{i}-2\nu)-i\eta}\ dE.
\end{align*}
Integrating this over a semi-circle in the positive real plane --
a negatively oriented curve -- gives

\begin{align*}
 & \frac{1}{2\pi i}\ \int_{-i\infty}^{+i\infty}\mathrm{tr}\ \mathbf{K}^{0}(E)\mathbf{V}\ dE\\
 & =-\sum_{a<b}^{N_{p}}V_{abab},
\end{align*}
whereas integrating this over a semi-circle in the negative real plane
-- a positively oriented curve -- gives

\begin{align*}
 & \frac{1}{2\pi i}\ \int_{-i\infty}^{+i\infty}\mathrm{tr}\ \mathbf{K}^{0}(E)\mathbf{V}\ dE\\
 & =-\sum_{h<i}^{N_{h}}V_{hihi}.
\end{align*}
The remaining integral of $\mathrm{tr}\ [\ln(\mathbf{I}-\mathbf{K}^{0}(E)\mathbf{V})]$
can be evaluated using partial integration.

\begin{align}
\frac{1}{2\pi i}\ \int_{-i\infty}^{+i\infty}\mathrm{tr}\ [\mathrm{ln}(\mathbf{I}-\mathbf{K}^{0}(E)\mathbf{V})] & dE=\frac{1}{2\pi i}\ \left[E\ \mathrm{tr}\ \mathrm{ln}(\mathbf{I}-\mathbf{K}^{0}(E)\mathbf{V})\right]_{-i\infty}^{+i\infty}-\frac{1}{2\pi i}\ \int_{-i\infty}^{+i\infty}E\ \mathrm{tr}\ [\frac{\partial}{\partial E}\mathrm{ln}(\mathbf{I}-\mathbf{K}^{0}(E)\mathbf{V})]dE\nonumber \\
 & \quad=-\frac{1}{2\pi i}\ \int_{-i\infty}^{+i\infty}E\ \mathrm{tr}\ [\frac{\partial}{\partial E}\mathrm{ln}(\mathbf{I}-\mathbf{K}^{0}(E)\mathbf{V})]dE.\label{eq:Ec_logterm}
\end{align}
In order to tackle the integrand, the identity $\mathbf{I}-\mathbf{K}^{0}\mathbf{V}=\mathbf{K}^{0}\mathbf{K}^{-1}$,
which follows simply from Eq. (\ref{dyson}), can be applied:

\begin{align}
\frac{\partial}{\partial E}\mathrm{ln}(\mathbf{I}-\mathbf{K}^{0}(E)\mathbf{V}) & =\frac{\partial}{\partial E}\mathrm{ln}\ \mathbf{K}^{0}\mathbf{K}^{-1}\nonumber \\
 & =\mathbf{K}(\mathbf{K}^{0})^{-1}\left(\frac{\partial\mathbf{K}^{0}}{\partial E}\mathbf{K}^{-1}+\mathbf{K}^{0}\frac{\partial\mathbf{K}^{-1}}{\partial E}\right)\nonumber \\
 & =\mathbf{K}(\mathbf{K}^{0})^{-1}\left(\frac{\partial\mathbf{K}^{0}}{\partial E}\mathbf{K}^{-1}+\mathbf{K}^{0}\frac{\partial(\mathbf{K}^{0})^{-1}}{\partial E}\right).\label{eq:deriv_logterm}
\end{align}
In the last line, the relationship $\mathbf{K}^{-1}=(\mathbf{K}^{0})^{-1}-\mathbf{V}$, which implies that $\frac{\partial\mathbf{K}^{-1}}{\partial E}=\frac{\partial(\mathbf{K}^{0})^{-1}}{\partial E}$, has been used. The integral then becomes

\begin{align}
-\frac{1}{2\pi i}\ \int_{-i\infty}^{+i\infty}E\ \mathrm{tr}\ [\frac{\partial}{\partial E}\mathrm{ln}(\mathbf{I}-\mathbf{K}^{0}(E)\mathbf{V})]dE & =-\frac{1}{2\pi i}\ \int_{-i\infty}^{+i\infty}E\ \mathrm{tr}\ \left[\mathbf{K}(\mathbf{K}^{0})^{-1}\left(\frac{\partial\mathbf{K}^{0}}{\partial E}\mathbf{K}^{-1}+\mathbf{K}^{0}\frac{\partial(\mathbf{K}^{0})^{-1}}{\partial E}\right)\right]dE\nonumber \\
 & =-\frac{1}{2\pi i}\ \int_{-i\infty}^{+i\infty}E\ \mathrm{tr}\ \left[(\mathbf{K}^{0})^{-1}\frac{\partial\mathbf{K}^{0}}{\partial E}+\mathbf{K}\frac{\partial(\mathbf{K}^{0})^{-1}}{\partial E}\right]dE.\label{eq:int_logterm}
\end{align}
The terms needed to compute the integrand are

\begin{align}
\left(\frac{\partial K^{0}}{\partial E}\right)_{ijkl} & =-(\delta_{ik}\delta_{jl}-\delta_{il}\delta_{jk})\left[\frac{\theta(i-F)\theta(j-F)}{\left(E-(\epsilon_{i}+\epsilon_{j}-2\nu)+i\eta\right)^{2}}-\frac{\theta(F-i)\theta(F-j)}{\left(E-(\epsilon_{i}+\epsilon_{j}-2\nu)-i\eta\right)^{2}}\right]\nonumber \\
\left(K^{0}\right)_{ijkl}^{-1} & =(\delta_{ik}\delta_{jl}-\delta_{il}\delta_{jk})\left[\theta(i-F)\theta(j-F)\left(E-(\epsilon_{i}+\epsilon_{j}-2\nu)+i\eta\right)-\theta(F-i)\theta(F-j)\left(E-(\epsilon_{i}+\epsilon_{j}-2\nu)-i\eta\right)\right]\nonumber \\
\frac{\partial\left(K^{0}\right)_{ijkl}^{-1}}{\partial E} & =(\delta_{ik}\delta_{jl}-\delta_{il}\delta_{jk})\left[\theta(i-F)\theta(j-F)-\theta(F-i)\theta(F-j)\right].\label{eq:deriv_P0}
\end{align}
With the aid of expressions (\ref{eq:Ec_logterm}), (\ref{eq:deriv_logterm})
and (\ref{eq:int_logterm}), the first part of the integral (\ref{eq:Ec_logterm})
becomes

\begin{align*}
-\frac{1}{2\pi i}\ \int_{-i\infty}^{+i\infty}E\ \mathrm{tr}\ \left[(\mathbf{K}^{0})^{-1}\frac{\partial\mathbf{K}^{0}}{\partial E}\right]dE & =-\frac{1}{2\pi i}\ \int_{-i\infty}^{+i\infty}E\ \sum_{i<j}^{K}\ \left[-\frac{\theta(i-F)\theta(j-F)}{E-(\epsilon_{i}+\epsilon_{j}-2\nu)+i\eta}-\frac{\theta(F-i)\theta(F-j)}{E-(\epsilon_{i}+\epsilon_{j}-2\nu)-i\eta}\right]dE.\\
\end{align*}
Integration over a semi-circular path in the positive real plane gives

\begin{align*}
-\frac{1}{2\pi i}\ \int_{-i\infty}^{+i\infty}E\ \sum_{i<j}^{K}\ \left[-\frac{\theta(i-F)\theta(j-F)}{E-(\epsilon_{i}+\epsilon_{j}-2\nu)+i\eta}-\frac{\theta(F-i)\theta(F-j)}{E-(\epsilon_{i}+\epsilon_{j}-2\nu)-i\eta}\right]dE & =-\sum_{a<b}^{N_{p}}(\epsilon_{b}+\epsilon_{a}-2\nu),\\
\end{align*}
whereas integration over a semi-circular path in the negative real
plane gives

\begin{align*}
-\frac{1}{2\pi i}\ \int_{-i\infty}^{+i\infty}E\ \sum_{i<j}^{K}\ \left[-\frac{\theta(i-F)\theta(j-F)}{E-(\epsilon_{i}+\epsilon_{j}-2\nu)+i\eta}-\frac{\theta(F-i)\theta(F-j)}{E-(\epsilon_{i}+\epsilon_{j}-2\nu)-i\eta}\right]dE & =\sum_{h<i}^{N_{h}}(\epsilon_{h}+\epsilon_{i}-2\nu).\\
\end{align*}
The second part of the integral (\ref{eq:Ec_logterm}) becomes

\begin{align*}
-\frac{1}{2\pi i}\ \int_{-i\infty}^{+i\infty}E\ \mathrm{tr}\ \left[\mathbf{K}\frac{\partial(\mathbf{K}^{0})^{-1}}{\partial E}\right]dE & =-\frac{1}{2\pi i}\ \int_{-i\infty}^{+i\infty}E\ \left(\sum_{a<b}^{N_{p}}K(E)_{abab}-\sum_{h<i}^{N_{h}}K(E)_{hihi}\right)\ dE.\\
\end{align*}
Integration over a semi-circular path in the positive real plane gives

\begin{align*}
-\frac{1}{2\pi i}\ \int_{-i\infty}^{+i\infty}E\ \mathrm{tr}\ \left[\mathbf{K}\frac{\partial(\mathbf{K}^{0})^{-1}}{\partial E}\right]dE & =\sum_{n}\omega_{n}^{N+2}\left(\sum_{a<b}^{N_{p}}\chi_{ab}^{n,N+2}\left(\chi_{ab}^{n,N+2}\right)^{*}-\sum_{h<i}^{N_{h}}\chi_{hi}^{n,N+2}\left(\chi_{hi}^{n,N+2}\right)^{*}\right)\\
 & =\sum_{n}\omega_{n}^{N+2}
\end{align*}
and integration over a semi-circular path in the negative real plane
gives

\begin{align*}
-\frac{1}{2\pi i}\ \int_{-i\infty}^{+i\infty}E\ \mathrm{tr}\ \left[\mathbf{K}\frac{\partial(\mathbf{K}^{0})^{-1}}{\partial E}\right]dE & =-\sum_{n}\omega_{n}^{N-2}\left(-\sum_{a<b}^{N_{p}}\chi_{ab}^{n,N-2}\left(\chi_{ab}^{n,N-2}\right)^{*}+\sum_{h<i}^{N_{h}}\chi_{hi}^{n,N-2}\left(\chi_{hi}^{n,N-2}\right)^{*}\right)\\
 & =-\sum_{n}\omega_{n}^{N-2}.
\end{align*}
where we have used the normalization conditions, Eqs. (\ref{eq:orthonormality_N+2}-\ref{eq:orthonomrality_N-2}).

To summarize, by closing a semi-circular path in the positive real
plane, we find

$ $ 
\begin{align*}
E_{pp}^{c} & =\sum_{n}^{N_{pp}}\omega_{n}^{N+2}-\sum_{a<b}^{N_{p}}(\epsilon_{b}+\epsilon_{a}-2\nu)-\sum_{a<b}^{N_{p}}V_{abab}\\
 & =\sum_{n}^{N_{pp}}\omega_{n}^{N+2}-\mathrm{tr}\ \mathbf{A}
\end{align*}
and by closing a semi-circular path in the negative real plane,

\begin{align*}
E_{pp}^{c} & =-\sum_{n}^{N_{hh}}\omega_{n}^{N-2}+\sum_{h<i}^{N_{h}}(\epsilon_{h}+\epsilon_{i}-2\nu)-\sum_{h<i}^{N_{h}}V_{hihi}\\
 & =-\sum_{n}^{N_{hh}}\omega_{n}^{N-2}-\mathrm{tr}\ \mathbf{C}.
\end{align*}
The two expressions for the correlation energy are equivalent, which
follows from the orthonormality and completeness of the pp-RPA eigenvector
basis. At this point, it is convenient to introduce a simplified notation
for the pp-RPA matrix,

\begin{eqnarray*}
\mathbf{R} & = & \begin{pmatrix}\mathbf{A} & \mathbf{B}\\
\mathbf{B^{\dagger}} & \mathbf{C}
\end{pmatrix}\\
\end{eqnarray*}
and for its eigenvectors,

\begin{eqnarray*}
\mathbf{\chi^{n}} & = & \begin{pmatrix}\mathbf{\mathbf{X}^{n}}\\
\mathbf{\mathbf{Y}}^{\mathbf{n}}
\end{pmatrix}.\\
\end{eqnarray*}
The norm matrix can be denoted as $\mathbf{M}=\begin{pmatrix}\mathbf{1} & \mathbf{0}\\
\mathbf{0} & \mathbf{-1}
\end{pmatrix}$ so that the pp-RPA equations take the form

\begin{eqnarray}
\mathbf{R}\mathbf{\chi^{n}} & = & \omega_{n}\mathbf{M}\mathbf{\chi^{n}},\label{eq:notation}
\end{eqnarray}
for both the 2-electron addition and the 2-electron removal. The orthonormality
and completeness of the eigenvector basis can then be expressed as

\begin{align}
\left(\mathbf{\chi}^{\mathbf{n},N+2}\right)^{\dagger}\mathbf{M}\mathbf{\chi}^{\mathbf{m},N+2} & =\delta_{mn}\label{eq:orthonormality_N+2}\\
\left(\mathbf{\chi}^{\mathbf{n},N-2}\right)^{\dagger}\mathbf{M}\mathbf{\chi}^{\mathbf{m},N-2} & =-\delta_{mn}\label{eq:orthonomrality_N-2}\\
\sum_{n}^{N_{pp}}\mathbf{\chi}^{\mathbf{n},N+2}\left(\mathbf{\chi}^{n,N+2}\right)^{\dagger}-\sum_{n}^{N_{hh}}\mathbf{\chi}^{\mathbf{n},N-2}\left(\mathbf{\chi}^{\mathbf{n},N-2}\right)^{\dagger} & =\mathbf{M}\nonumber 
\end{align}
The pp-RPA equations imply that 
\begin{align*}
\ \sum_{n}^{N_{pp}}\left(\mathbf{\chi}^{\mathbf{n},N+2}\right){}^{\dagger}\begin{pmatrix}\mathbf{A} & \mathbf{B}\\
\mathbf{B^{\dagger}} & \mathbf{C}
\end{pmatrix}\mathbf{\chi}^{\mathbf{n},N+2} & =\sum_{n}^{N_{pp}}\omega_{n}^{N+2}\left(\mathbf{\chi}^{\mathbf{n},N+2}\right)^{\dagger}\mathbf{M}\mathbf{\chi}^{\mathbf{n},N+2}\\
-\sum_{n}^{N_{hh}}\left(\mathbf{\chi}^{\mathbf{n},N-2}\right)^{\dagger}\begin{pmatrix}\mathbf{A} & \mathbf{B}\\
\mathbf{B^{\dagger}} & \mathbf{C}
\end{pmatrix}\mathbf{\chi}^{\mathbf{n},N-2} & =-\sum_{n}^{N_{hh}}\omega_{n}^{N-2}\begin{pmatrix}\mathbf{\chi}^{\mathbf{n},N-2}\end{pmatrix}^{\dagger}\mathbf{M}\mathbf{\chi}^{\mathbf{n},N-2}.
\end{align*}
This, together with the normalization and completeness of the eigenvectors,
and Eq.(\ref{eq:orthonormality_N+2}), leads to the following relation
between the $N-2$ electron quantities and $N+2$ electron quantities
\begin{equation}
\mathrm{tr}\ \mathbf{A}-\mathrm{tr}\ \mathbf{C}=\sum_{n}^{N_{pp}}\omega_{n}^{N+2}+\sum_{n}^{N_{hh}}\omega_{n}^{N-2}.\label{eq:relation_N-2_N+2}
\end{equation}

The correlation energy can be viewed as a functional $E[\{\phi_{i}\},n_{i}]$
because equation (\ref{app:ppRPA_matrixform}) depends only on the
orthonormal set of orbitals $\{\phi_{i}\}$ and their occupations
$n_{i}$. The total pp-RPA energy expression combines the HF-energy
functional with the pp-RPA correlation energy: 
\begin{align*}
E^{pp}[\{\phi_{i}\},n_{i}] & =E^{HF}[\{\phi_{i}\},n_{i}]+E_{pp}^{c}[\{\phi_{i}\},n_{i}]\\
 & =\sum_{i}h_{ii}n_{i}+\frac{1}{2}\sum_{ij}\langle ij\vert\vert ij\rangle n_{i}n_{j}+E_{pp}^{c}[\{\phi_{i}\},n_{i}]
\end{align*}
with $\mathbf{h}$ the core Hamiltonian matrix.

\subsection{Perturbation analysis of the pp-RPA energy \label{sub:Perturbation_analysis}}

In the context of many-body perturbation theory, the pp-RPA energy
arises as the sum of all ladder diagrams up to infinite order \cite{Blaizot1986}:
\begin{eqnarray}
E_{pp}^{c} & = & \frac{-1}{2\pi i}\ \sum_{n=2}^{\infty}\frac{1}{n}\int_{-i\infty}^{+i\infty}\mathrm{tr}\ [\mathbf{K}^{0}(E))\mathbf{V}]^{n}\ dE\nonumber \\
 & = & \frac{-1}{2\pi i}\ \sum_{n=1}^{\infty}\frac{1}{n}\int_{-i\infty}^{+i\infty}\mathrm{tr}\ [\mathbf{K}^{0}(E))\mathbf{V}]^{n}\ dE+\frac{1}{2\pi i}\ \int_{-i\infty}^{+i\infty}\mathrm{tr}\ \mathbf{K}^{0}(E))\mathbf{V}\ dE\\
 & = & \frac{1}{2\pi i}\ \int_{-i\infty}^{+i\infty}\mathrm{tr}\ [\mathrm{ln}(\mathbf{I}-\mathbf{K}^{0}(E)\mathbf{V})+\mathbf{K}^{0}(E)\mathbf{V}]\ dE.
\end{eqnarray}
In contrast, the ph-RPA energy originates from the summation of all
ring diagrams \cite{Blaizot1986}:

\begin{eqnarray}
E_{ph}^{c} & = & \frac{1}{2\pi i}\ \sum_{n=2}^{\infty}\frac{-1}{2n}\int_{-i\infty}^{+i\infty}\mathrm{tr}\ [\mathbf{\Pi}^{0}(E))\mathbf{\tilde{\mathbf{V}}}]^{n}\ dE\nonumber \\
 & = & \frac{1}{2\pi i}\ \sum_{n=1}^{\infty}\frac{-1}{2n}\int_{-i\infty}^{+i\infty}\mathrm{tr}\ [\mathbf{\Pi}^{0}(E))\tilde{\mathbf{V}}]^{n}\ dE+\frac{1}{4\pi i}\ \int_{-i\infty}^{+i\infty}\mathrm{tr}\ \mathbf{\Pi}^{0}(E))\tilde{\mathbf{V}}\ dE\\
 & = & \frac{1}{4\pi i}\ \int_{-i\infty}^{+i\infty}\mathrm{tr}\ [\mathrm{ln}(\mathbf{I}-\mathbf{\Pi}^{0}(E)\tilde{\mathbf{V}})+\mathbf{\Pi}^{0}(E)\tilde{\mathbf{V}}]\ dE
\end{eqnarray}
where $\tilde{V}_{ahib}=\langle ab\vert hi\rangle$ does not include
exchange. The ph-RPAX uses antisymmetrized two-electron integrals
and the corersponding correlation energy can be derived from the adiabatic
connection to be \cite{Eshuis2012}:

\begin{eqnarray}
E_{phX}^{c} & = & \frac{1}{4\pi i}\ \sum_{n=2}^{\infty}\frac{-1}{2n}\int_{-i\infty}^{+i\infty}\mathrm{tr}\ [\mathbf{\Pi}^{0}(E))\mathbf{\tilde{\mathbf{V}}}]^{n}\ dE\nonumber \\
 & = & \frac{1}{4\pi i}\ \sum_{n=1}^{\infty}\frac{-1}{2n}\int_{-i\infty}^{+i\infty}\mathrm{tr}\ [\mathbf{\Pi}^{0}(E))\tilde{\mathbf{V}}]^{n}\ dE+\frac{1}{4\pi i}\ \int_{-i\infty}^{+i\infty}\mathrm{tr}\ \mathbf{\Pi}^{0}(E))\tilde{\mathbf{V}}\ dE\\
 & = & \frac{1}{8\pi i}\ \int_{-i\infty}^{+i\infty}\mathrm{tr}\ [\mathrm{ln}(\mathbf{I}-\mathbf{\Pi}^{0}(E)\tilde{\mathbf{V}})+\mathbf{\Pi}^{0}(E)\tilde{\mathbf{V}}]\ dE
\end{eqnarray}
where $\bar{V}_{ahib}=\langle ab\Vert hi\rangle$ now includes exchange.

The pp-RPA energy is correct through second order:

\begin{eqnarray*}
E_{pp}^{(2)} & = & -\frac{1}{2}\frac{1}{2\pi i}\int_{-i\infty}^{+i\infty}tr\ [\mathbf{\mathbf{K}}^{0}(E)\mathbf{V}]^{2}dE\\
 & = & -\frac{1}{2}\frac{1}{2\pi i}\int_{-i\infty}^{+i\infty}\sum_{a<b,c<d}\frac{V_{abcd}V_{cdab}}{(E-(\epsilon_{a}+\epsilon_{b}))(E-(\epsilon_{c}+\epsilon_{d}))}+\sum_{h<i,j<k}\frac{V_{hijk}V_{jkhi}}{(E-(\epsilon_{h}+\epsilon_{i}))(E-(\epsilon_{j}+\epsilon_{k}))}\\
 &  & -2\sum_{a<b,h<i}\frac{V_{abhi}V_{hiab}}{(E-(\epsilon_{a}+\epsilon_{b}))(E-(\epsilon_{h}+\epsilon_{i}))}dE\\
 & = & -\sum_{a<b,h<i}\frac{V_{abhi}V_{hiab}}{\epsilon_{a}+\epsilon_{b}-\epsilon_{h}-\epsilon_{i}}\\
 & = & -\frac{1}{4}\sum_{abhi}\frac{\vert\langle hi\Vert ab\rangle\vert^{2}}{\epsilon_{a}+\epsilon_{b}-\epsilon_{h}-\epsilon_{i}}
\end{eqnarray*}
where only the third term in the second line makes a non-zero contribution.
This expression includes all possible second-order diagrams, and is
hence exact. The ph-RPAX has the same second-order energy contribution,

\begin{eqnarray*}
E_{phX}^{(2)} & = & -\frac{1}{4}\frac{1}{4\pi i}\int_{-i\infty}^{+i\infty}tr\ [\mathbf{\Pi}^{0}(E)\mathbf{\bar{V}}]^{2}dE\\
 & = & -\frac{1}{4}\frac{1}{4\pi i}\int_{-i\infty}^{+i\infty}\sum_{abqi}\frac{\bar{V}_{ahbi}\bar{V}_{biah}}{(E-(\epsilon_{a}-\epsilon_{h}))(E-(\epsilon_{b}-\epsilon_{i}))}+\sum_{hijk}\frac{\bar{V}_{haib}\bar{V}_{ibha}}{(E-(\epsilon_{h}-\epsilon_{a}))(E-(\epsilon_{i}-\epsilon_{b}))}\\
 &  & -2\sum_{pqhi}\frac{\bar{V}_{ahib}\bar{V}_{ibah}}{(E-(\epsilon_{a}-\epsilon_{h}))(E-(\epsilon_{i}-\epsilon_{b}))}dE\\
 & = & -\frac{1}{4}\sum_{abhi}\frac{\bar{V}_{ahib}\bar{V}{}_{ibah}}{\epsilon_{a}-\epsilon_{h}-\epsilon_{i}+\epsilon_{b}}\\
 & = & -\frac{1}{4}\sum_{abhi}\frac{\vert\langle hi\Vert ab\rangle\vert^{2}}{\epsilon_{a}+\epsilon_{b}-\epsilon_{h}-\epsilon_{i}}\\
\end{eqnarray*}
but an inherent drawback of the ph-RPAX is its sensitivity to instabilities
in the non-interacting reference state: when the non-interacting reference
state is unstable with respect to orbital rotations, the ph-RPAX breaks
down and produces imaginary eigenvalues \cite{Ring1980}. For this
reason, molecular calculations are done almost exclusively using the
`direct' ph-RPA\cite{Furche2001,Eshuis2012,Ren20127447}, which does
not suffer from such instabilities. The ph-RPA, however, does not
have the correct second-order energy expression because it does not
consider antisymmetrized two-electron integrals:

\begin{eqnarray*}
E_{ph}^{(2)} & = & -\frac{1}{4}\frac{1}{2\pi i}\int_{-i\infty}^{+i\infty}tr\ [\mathbf{\Pi}^{0}(E)\tilde{\mathbf{V}}]^{2}dE\\
 & = & -\frac{1}{4}\frac{1}{2\pi i}\int_{-i\infty}^{+i\infty}\sum_{abqi}\frac{\tilde{V}_{ahbi}\tilde{V}_{biah}}{(E-(\epsilon_{a}-\epsilon_{h}))(E-(\epsilon_{b}-\epsilon_{i}))}+\sum_{hijk}\frac{\tilde{V}_{haib}\tilde{V}_{ibha}}{(E-(\epsilon_{h}-\epsilon_{a}))(E-(\epsilon_{i}-\epsilon_{b}))}\\
 &  & -2\sum_{pqhi}\frac{\tilde{V}_{ahib}\tilde{V}_{ibah}}{(E-(\epsilon_{a}-\epsilon_{h}))(E-(\epsilon_{i}-\epsilon_{b}))}dE\\
 & = & -\frac{1}{2}\sum_{abhi}\frac{\tilde{V}_{ahib}\tilde{V}{}_{ibah}}{\epsilon_{a}-\epsilon_{h}-\epsilon_{i}+\epsilon_{b}}\\
 & = & -\frac{1}{2}\sum_{abhi}\frac{\vert\langle hi\vert ab\rangle\vert^{2}}{\epsilon_{a}+\epsilon_{b}-\epsilon_{h}-\epsilon_{i}}\\
\end{eqnarray*}
Only the last term in the second line does not vanish upon integration.

\subsection{The particle-particle RPA for systems with fractional electron number\label{sub:fracppRPA}}

While equation (\ref{app:ppRPA_matrixform}) describes the pp-RPA
for systems with integer electron number, the behavior of the pp-RPA
for systems with fractional electron number or spin can be quantified
by taking the fractional orbital occupations into account explicitly
in the pp-RPA equations (\ref{app:ppRPA_matrixform}) 
\begin{align}
A_{abcd}= & \sqrt{(1-n_{a})(1-n_{b})}\langle ab\Vert cd\rangle\sqrt{(1-n_{c})(1-n_{d})}\nonumber \\
 & +\delta_{ac}\delta_{bd}(\epsilon_{a}+\epsilon_{b}-2\nu)\nonumber \\
B_{abij}= & \sqrt{(1-n_{a})(1-n_{b})}\langle ab\Vert ij\rangle\sqrt{n_{i}n_{j}}\nonumber \\
C_{ijkl}= & \sqrt{n_{i}n_{j}}\langle ij\Vert kl\rangle\sqrt{n_{k}n_{l}}-\delta_{ij}\delta_{kl}(\epsilon_{i}+\epsilon_{j}-2\nu).\label{app:ppRPA_fracmatrixform}
\end{align}
This extension to fractional occupation number follows the same approach
as the one taken in previous work by Cohen, Mori-Sanchez and Yang
\cite{Cohen2009786,Mori-Sanchez2012} and is explained in more detail
in Ref. \cite{Yang2013}. When all orbital occupation numbers are
integer these equations reduce to the usual pp-RPA equations.

\clearpage{}

\section{Additional figures and tables}

We computed the KS reference wavefunctions with Gaussian03 \cite{Frisch2004}
for the systems with integer electron number and with the QM4D package
for systems with fractional electron number or spin \cite{QM4D}.
For the subsequent pp-RPA calculation, we used our implementation,
which diagonalizes the pp-RPA matrix. Since the diagonalization is
computationally expensive, we used a cc-pVDZ basis set for all calculations,
except for the Ar and Ne atoms, for which we used an aug-cc-pVDZ (FC)
basis set. For the calculations on thermodynamical properties, we
used a cc-pVTZ basis set limited to F-functions because the pp-RPA energy converges slowly with the basis set size
(Fig. 13) and geometries from
the G2 test set \cite{Curtiss19971063}. Accurate potential energy
functions for the dimers of the noble gases have been taken from the
work of Ogilvie et al. \cite{Ogilvie1992277,Ogilvie1993313} and the
MRCI potential energy function for the \ce{N2} in the cc-pVDZ basis
set has been taken from previous work \cite{Aggelen2010}.

\begin{figure*}[h]
\begin{minipage}[c]{0.45\linewidth}%
\centering \includegraphics[width=0.99\textwidth]{H2LDA.eps} %
\end{minipage}%
\begin{minipage}[c]{0.45\linewidth}%
\centering \includegraphics[width=0.99\textwidth]{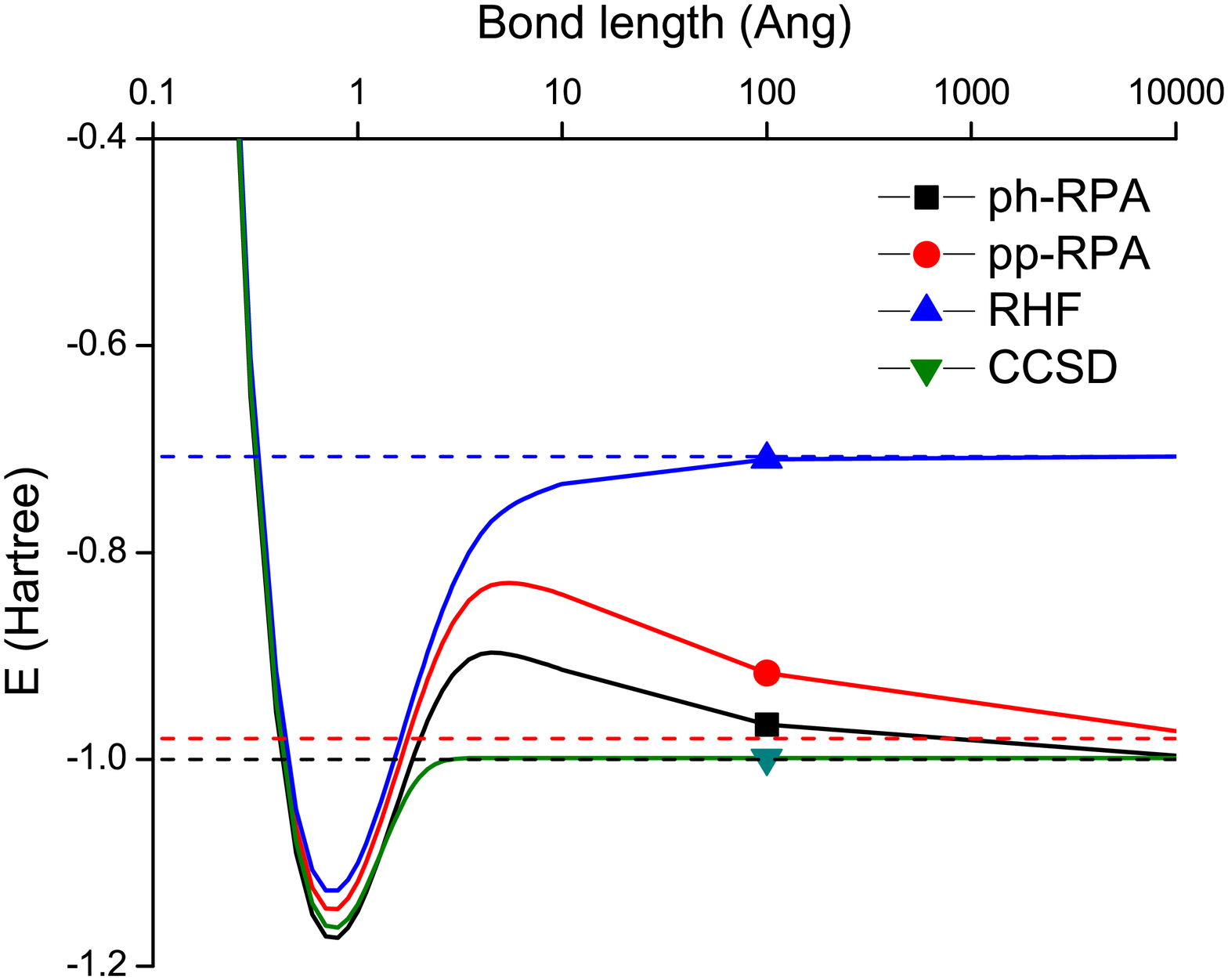} %
\end{minipage}\caption{The pp-RPA energy (left: restricted LDA reference, right: restricted HF reference) for the
\ce{H2} molecule approaches the correct value in the dissociation
limit, but has an unphysical 'bump', much more so than ph-RPA. The dashed lines indicate the dissociation limit from the fractional analysis of the H atom.}
\end{figure*}

\begin{figure*}[h]
\begin{minipage}[c]{0.45\linewidth}%
\centering \includegraphics[width=0.99\textwidth]{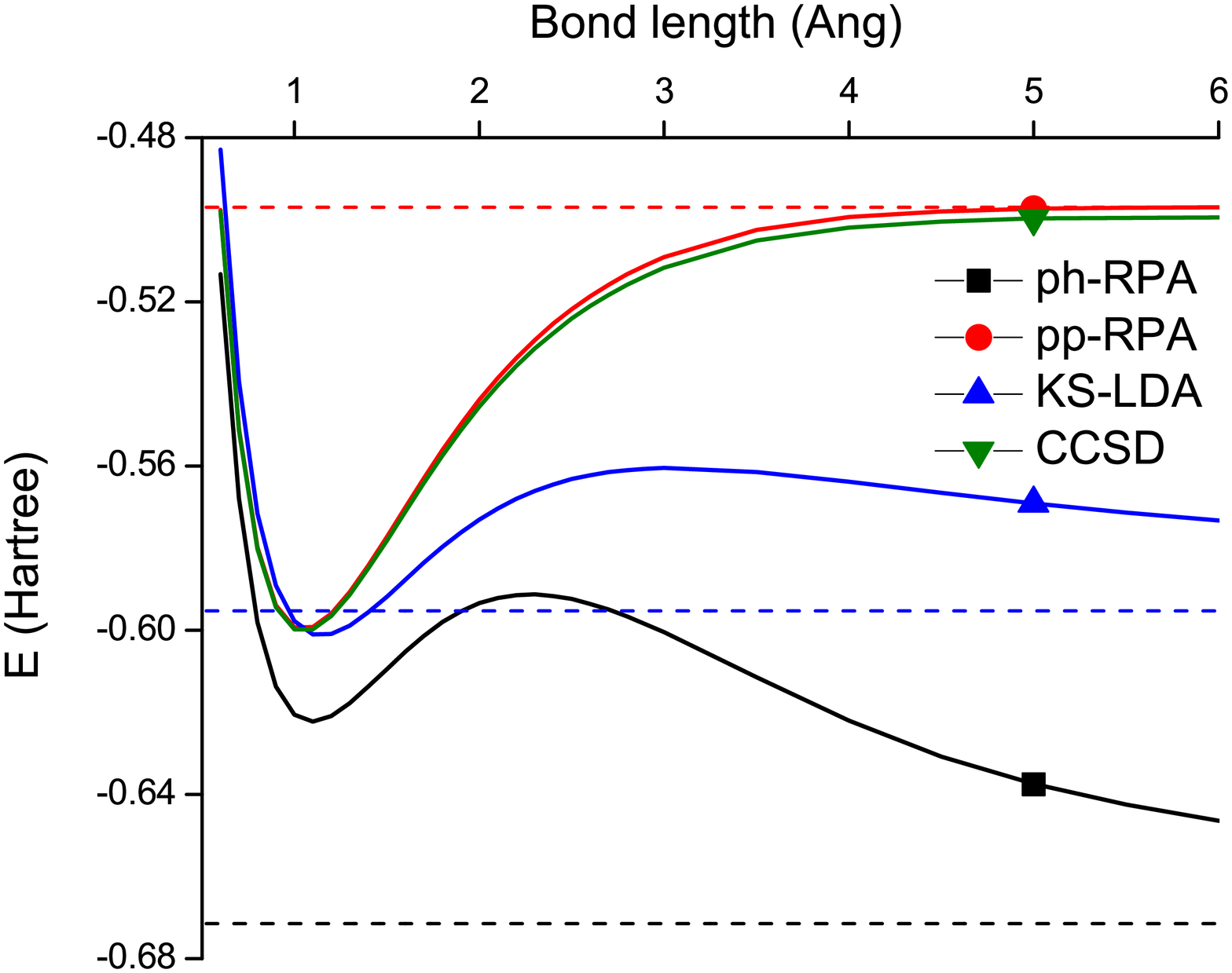} %
\end{minipage}%
\begin{minipage}[c]{0.45\linewidth}%
\includegraphics[width=0.99\textwidth]{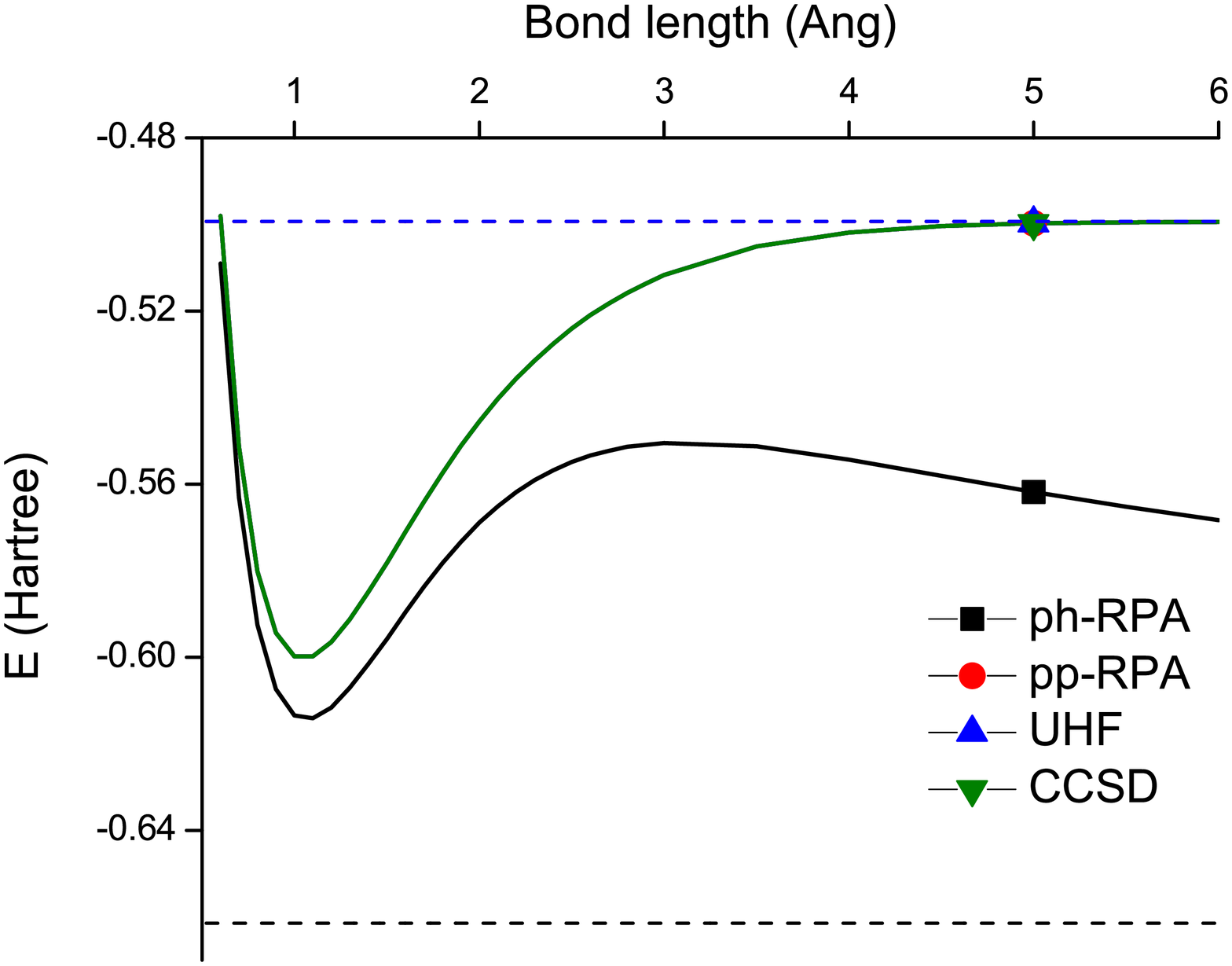} %
\end{minipage}\caption{In contrast to the ph-RPA, the pp-RPA dissociates \ce{H2+} correctly
(left: LDA reference, right: HF reference). The dashed lines indicate the dissociation limit from the fractional analysis of the H atom.}

\label{fig:HH+_LDA} 
\end{figure*}

\begin{figure*}
\begin{minipage}[c]{0.45\linewidth}%
\centering \includegraphics[width=0.99\textwidth]{He2+LDA.eps} %
\end{minipage}%
\begin{minipage}[c]{0.45\linewidth}%
\includegraphics[width=0.99\textwidth]{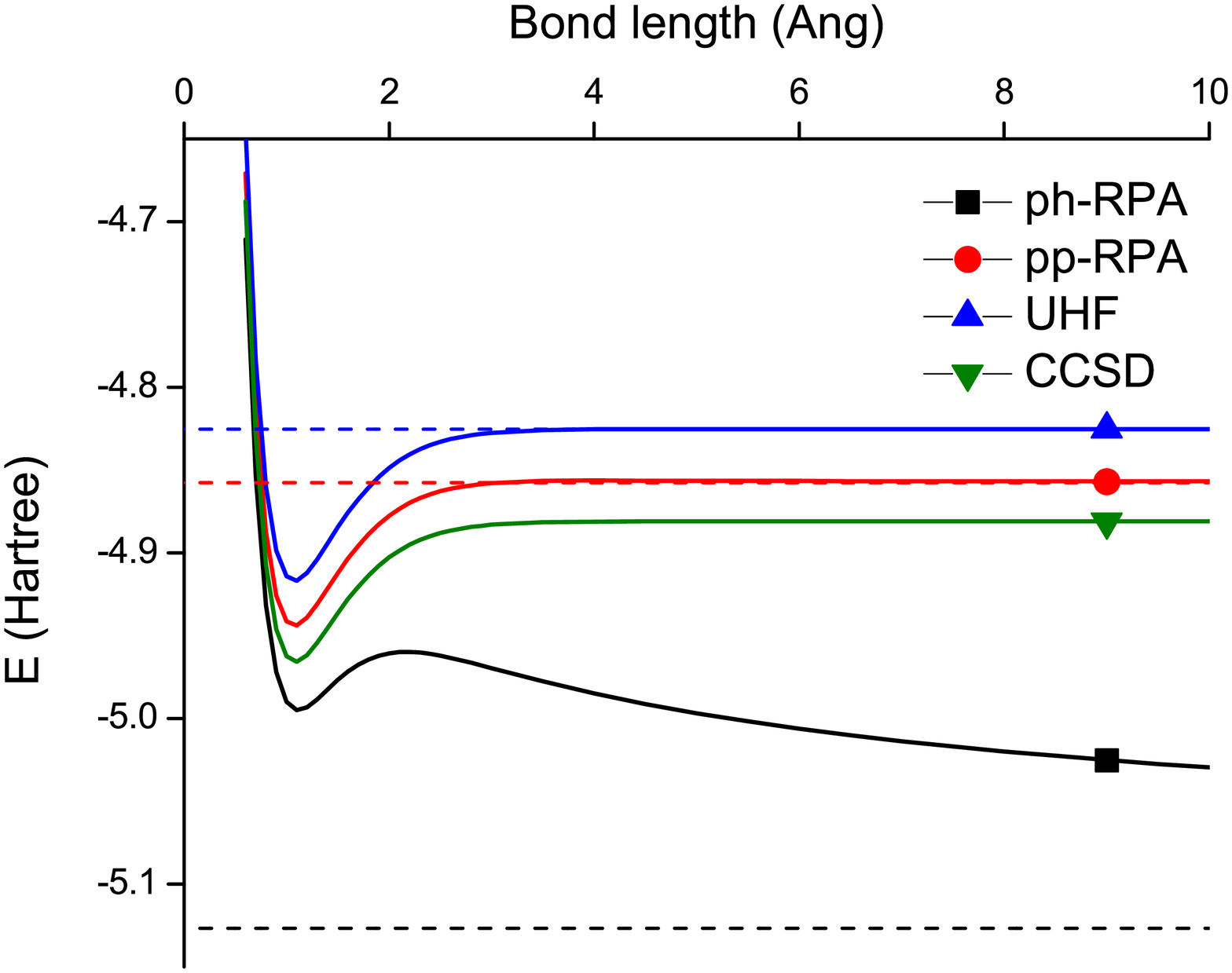} %
\end{minipage}\caption{The pp-RPA also gives a correct energy profile for \ce{He2+}, in
contrast to the ph-RPA (left: LDA reference, right: HF reference). The dashed lines indicate the dissociation limit from the fractional analysis of the He atom. }
\end{figure*}

\begin{figure*}
\begin{minipage}[c]{0.45\linewidth}%
\includegraphics[width=0.99\textwidth]{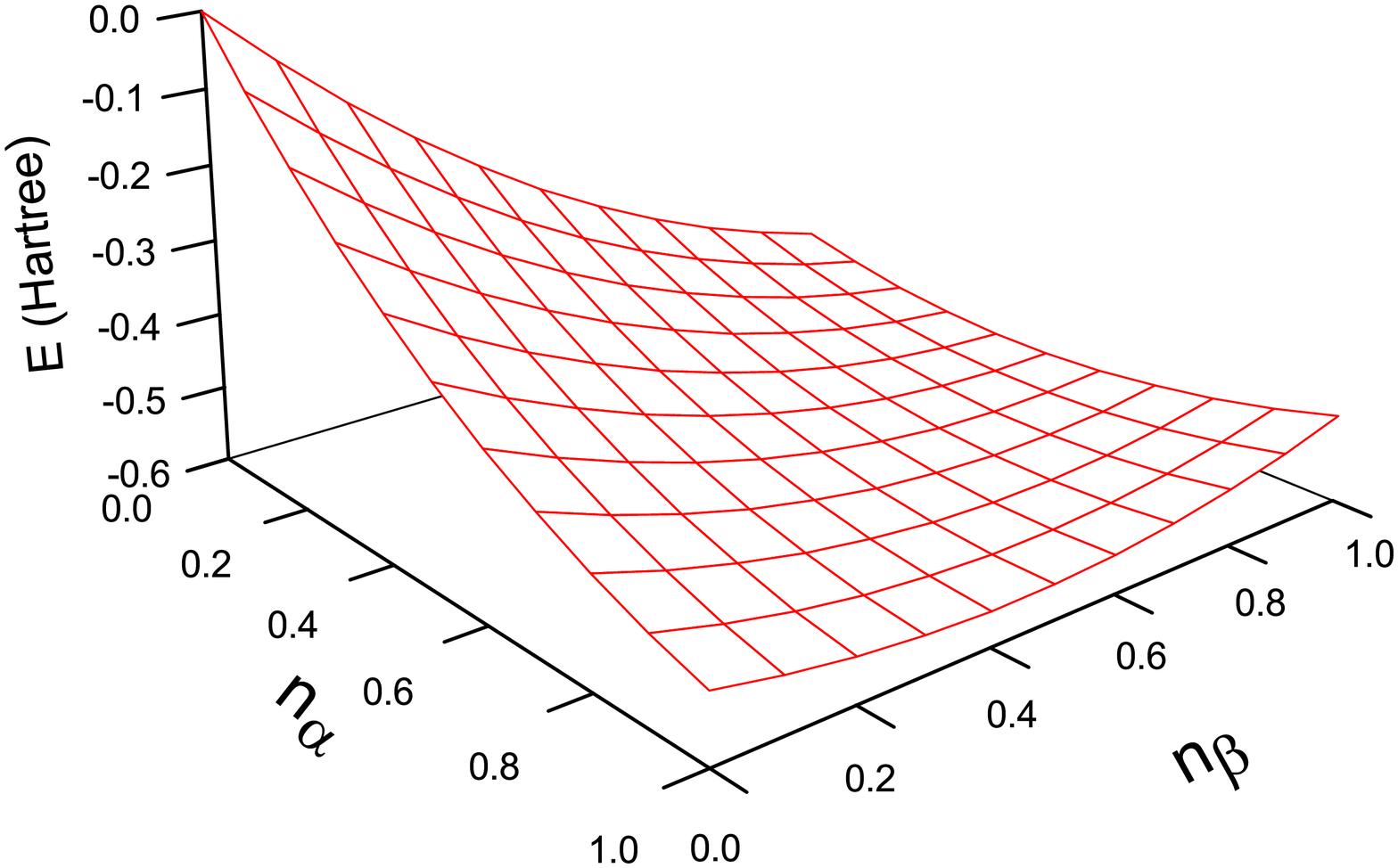} %
\end{minipage}%
\begin{minipage}[c]{0.45\linewidth}%
\includegraphics[width=0.99\textwidth]{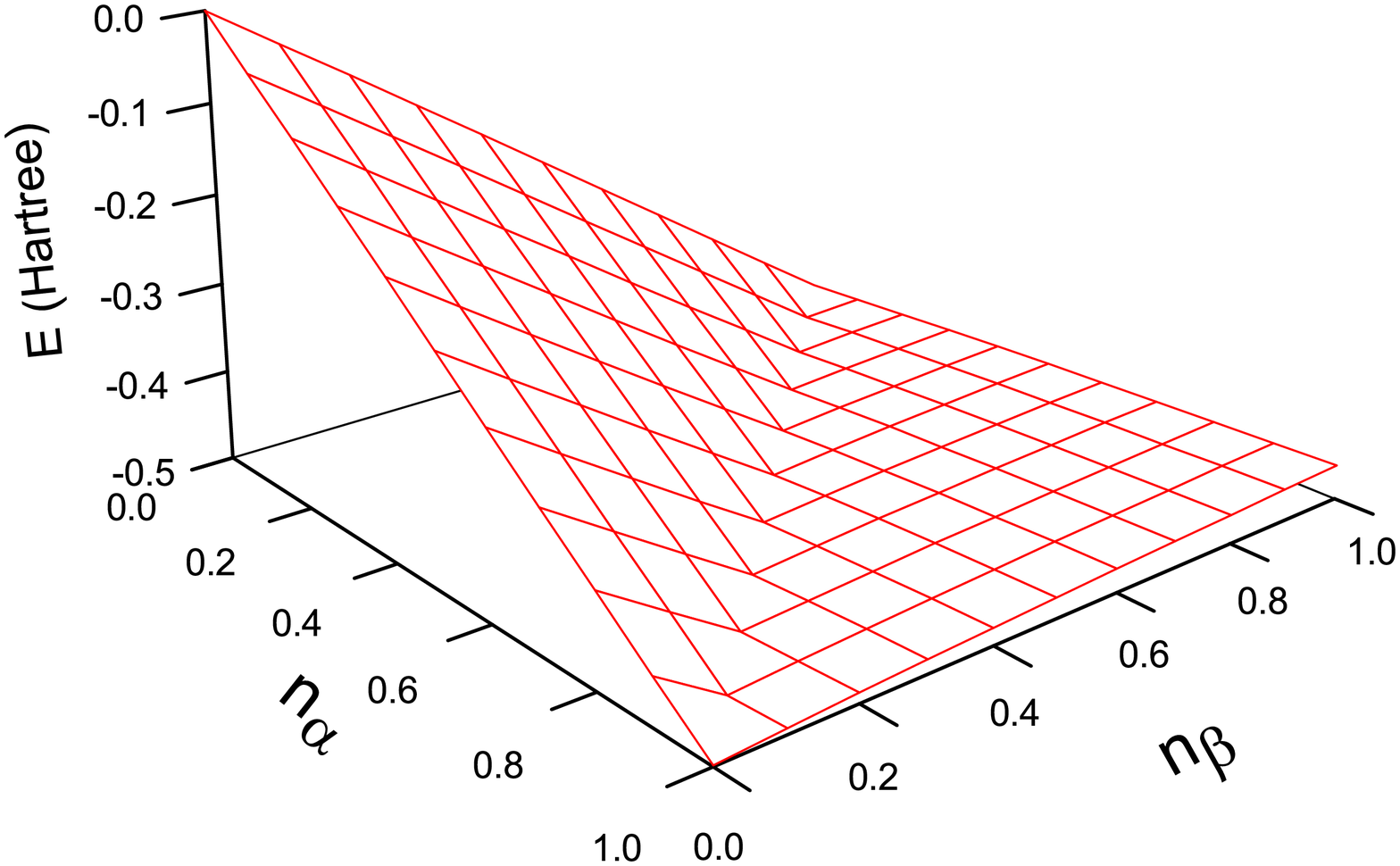} %
\end{minipage}\caption{The ph-RPA energy for the H atom (left) is a nearly constant function
of the fractional spin projection, but is a convex function of the
fractional electron number. The pp-RPA energy (right) is physically
correct: it has a nearly constant function of the fractional spin
projection and a linear function of the fractional electron number.
Like the exact functional, its derivative has a discontinuity at N=1.}
\label{fig:H_phRPA_LDA} 
\end{figure*}

\begin{figure*}
\begin{minipage}[c]{0.45\linewidth}%
\centering \includegraphics[width=0.99\textwidth]{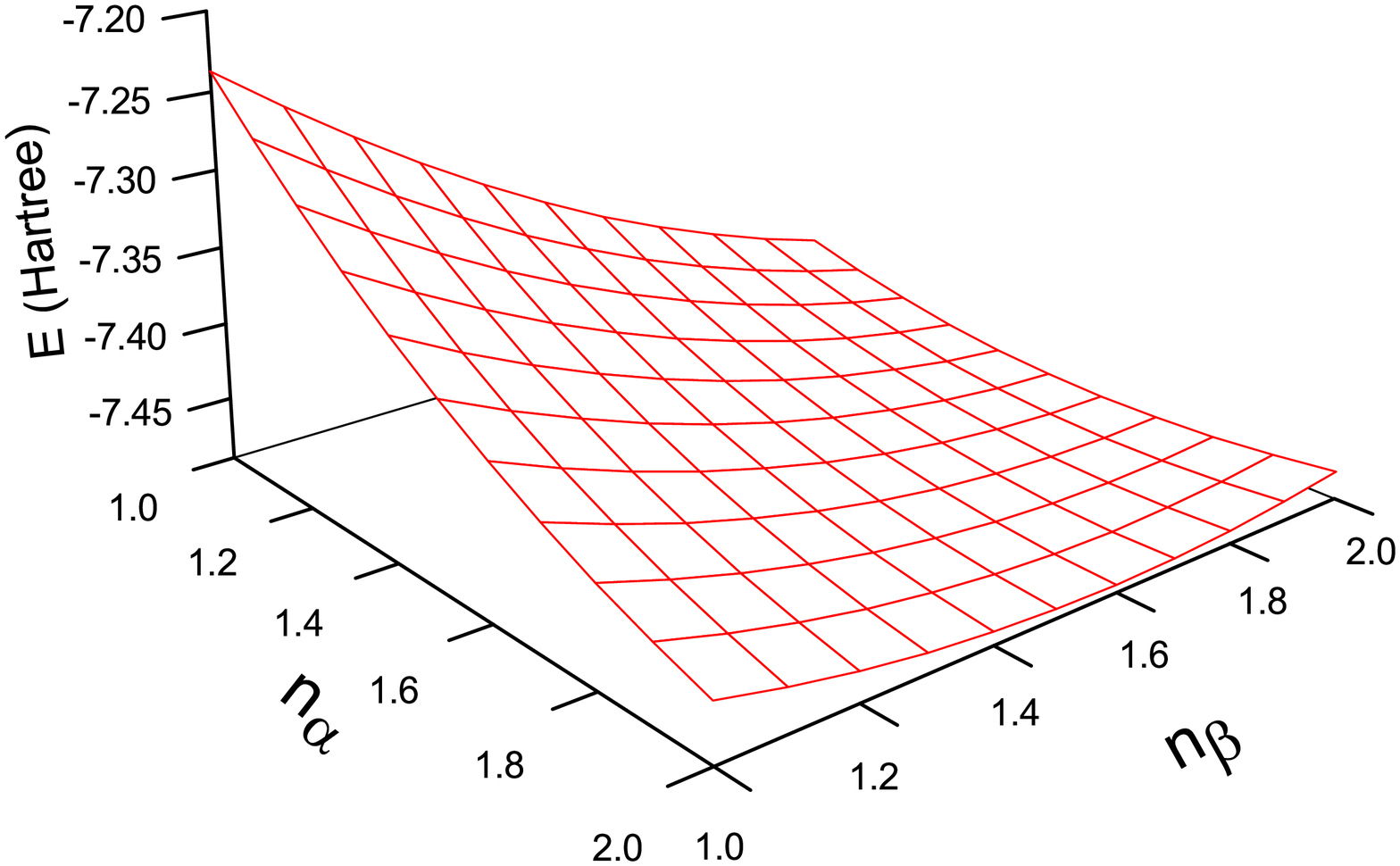} %
\end{minipage}%
\begin{minipage}[c]{0.45\linewidth}%
\includegraphics[width=0.99\textwidth]{Lipp.eps} %
\end{minipage}\caption{The ph-RPA energy for the Li atom (left) is a nearly constant function
of the fractional spin projection, but is a convex function of the
fractional electron number. The pp-RPA energy (right) is a nearly
constant function of the fractional spin projection and a nearly linear
function of the fractional electron number. Like the exact functional,
its derivative has a discontinuity at N=3.}
\label{fig:Li_phRPA_LDA} 
\end{figure*}

\begin{figure*}
\begin{minipage}[c]{0.45\linewidth}%
\centering \includegraphics[width=0.99\textwidth]{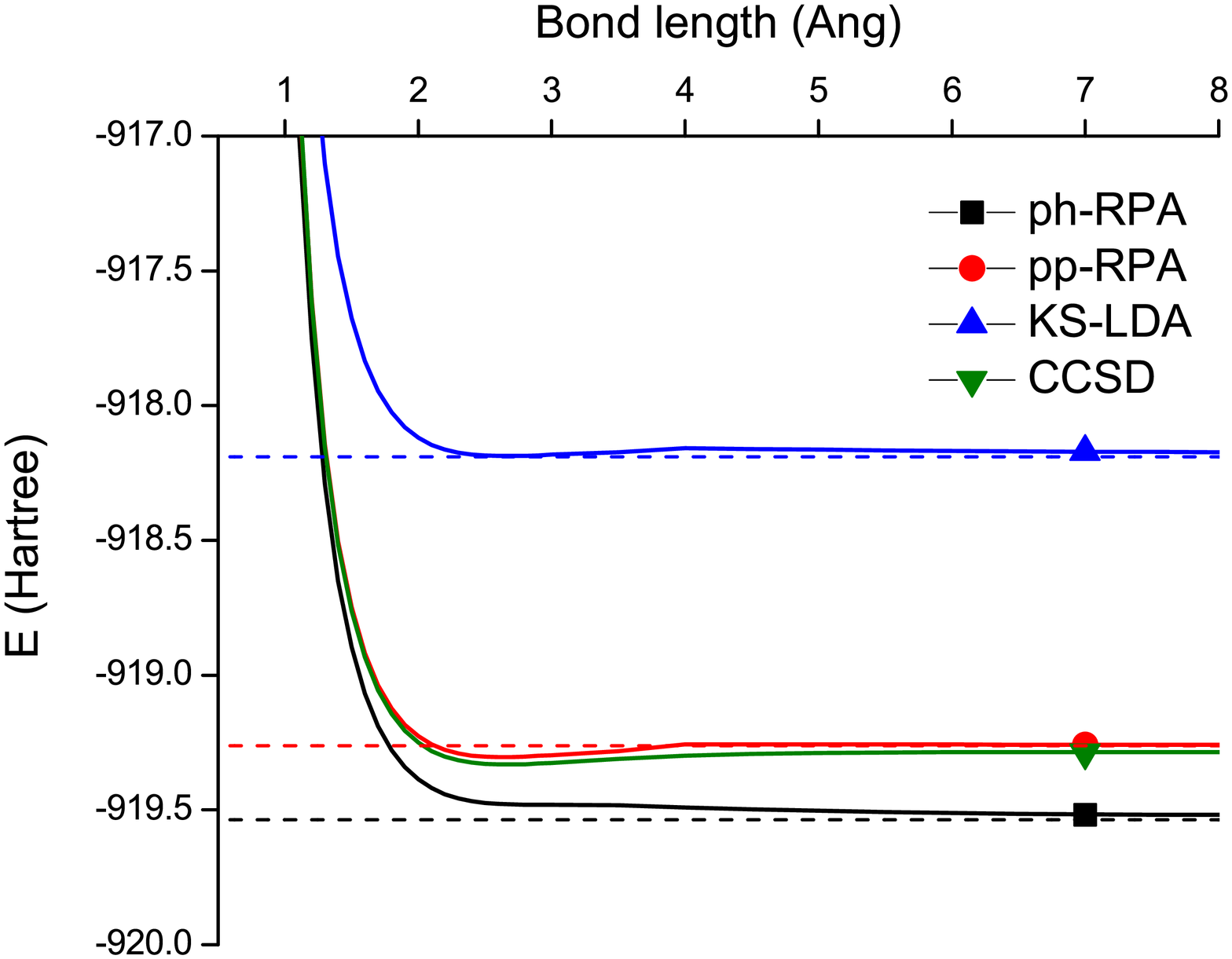} %
\end{minipage}%
\begin{minipage}[c]{0.45\linewidth}%
\includegraphics[width=0.99\textwidth]{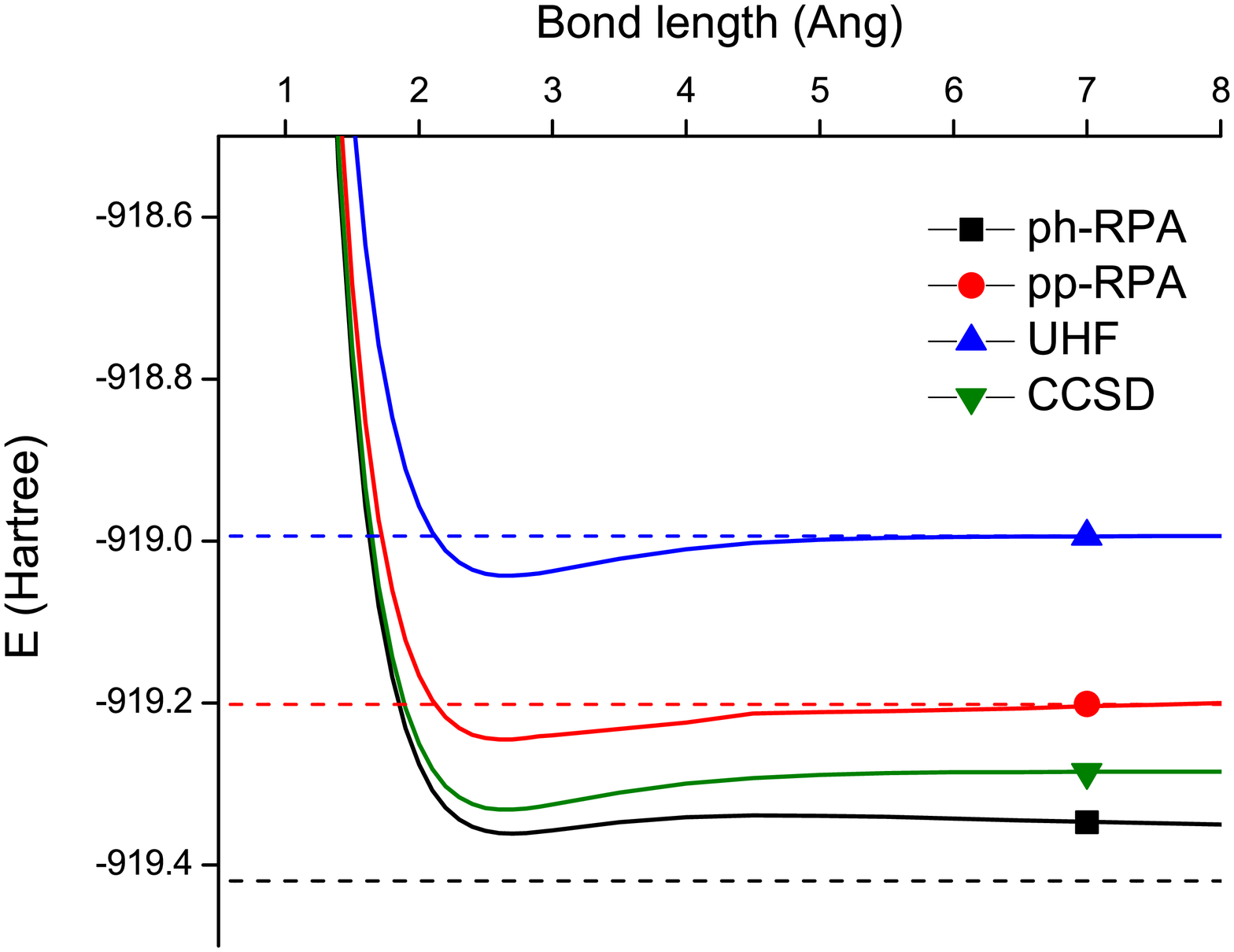} %
\end{minipage}\caption{The pp-RPA also gives a correct energy profile for \ce{Cl2-}, in
contrast to the ph-RPA (left: LDA reference, right: HF reference). The dashed lines indicate the dissociation limit from the fractional analysis of the He atom. }
\end{figure*}

\begin{figure*}
\begin{minipage}[c]{0.45\linewidth}%
\centering \includegraphics[width=0.99\textwidth]{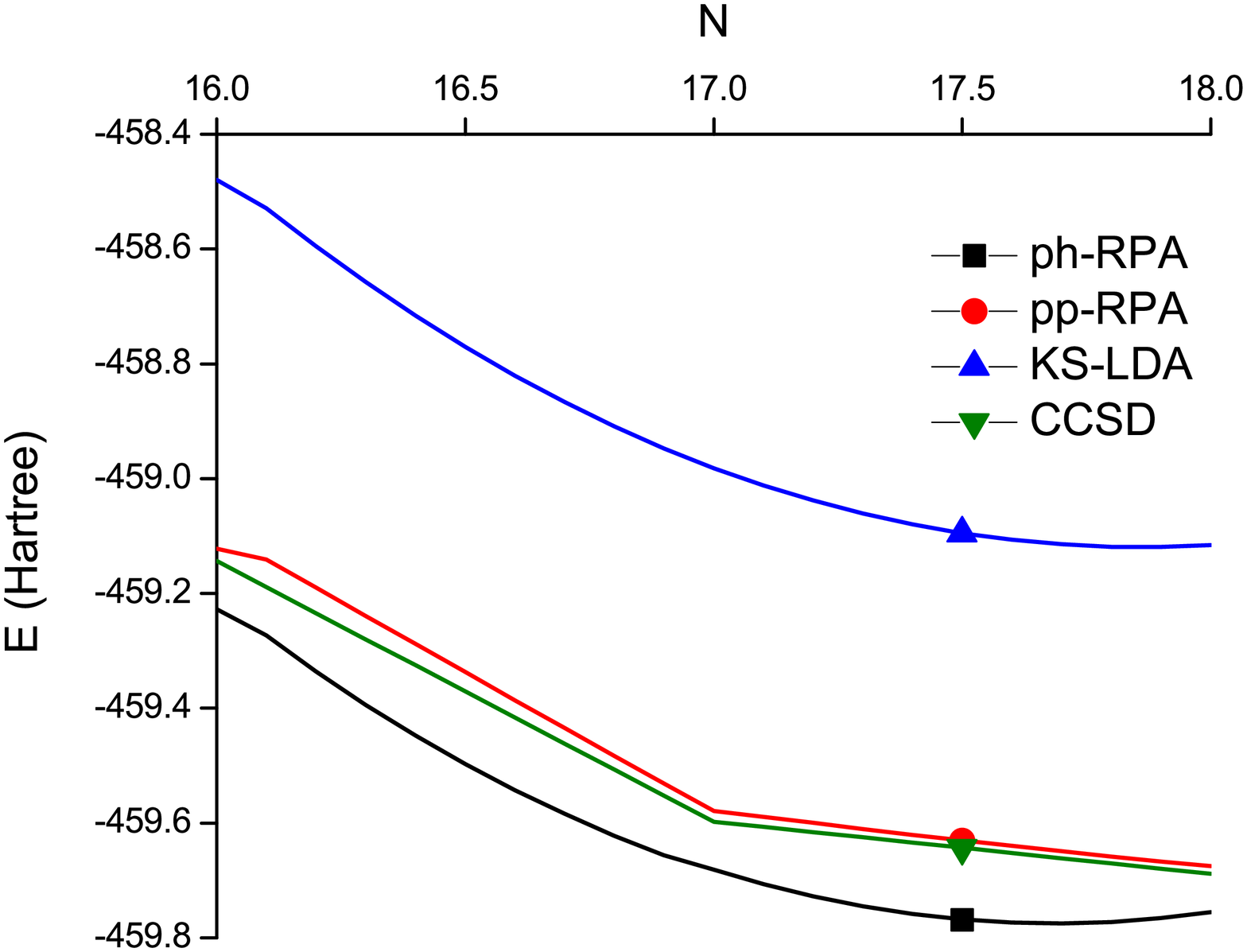} %
\end{minipage}%
\begin{minipage}[c]{0.45\linewidth}%
\includegraphics[width=0.99\textwidth]{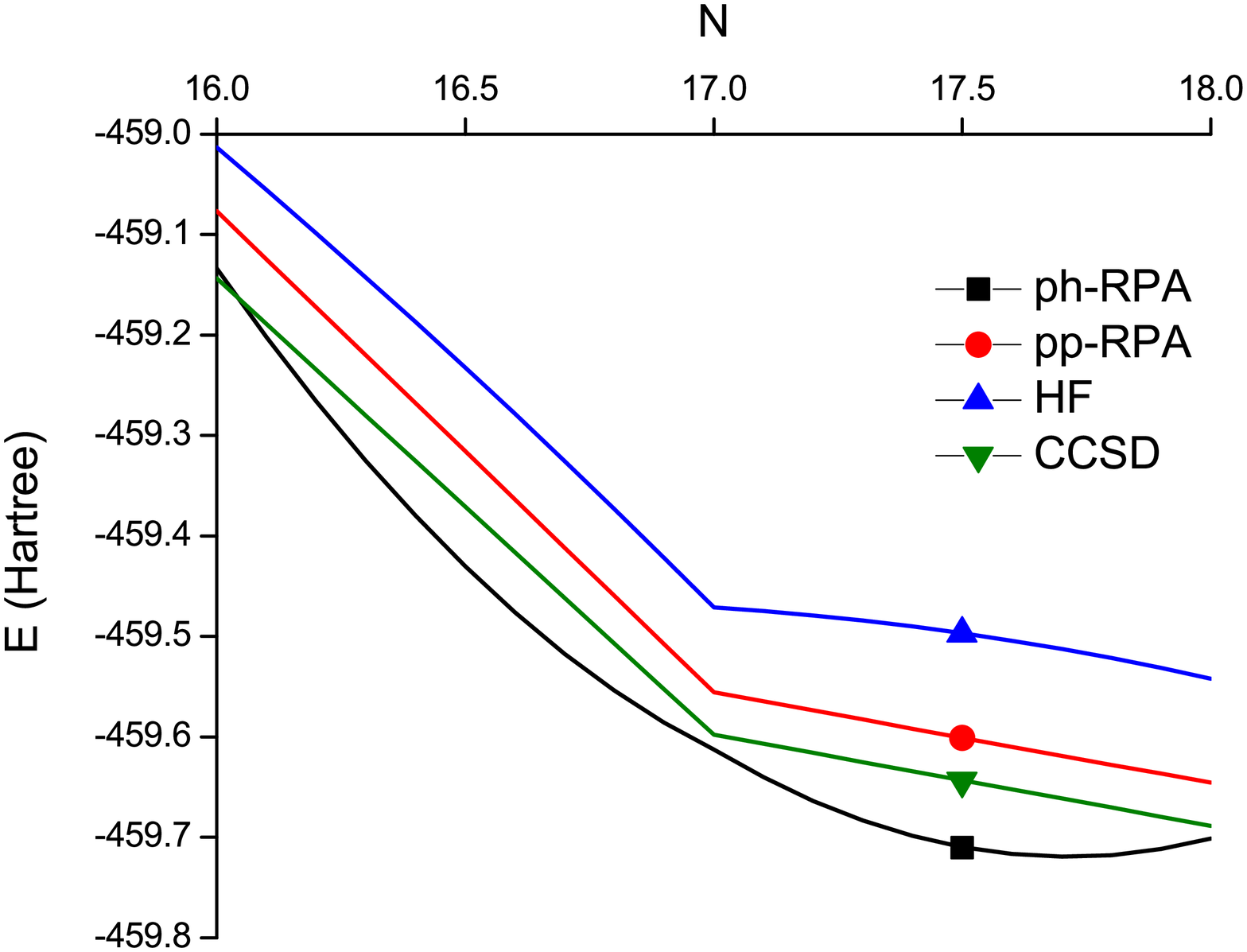} %
\end{minipage}\caption{The pp-RPA energy for the Cl atom is nearly linear in between integer
electron numbers, as opposed to the ph-RPA energy (left: LDA reference, right:
HF reference). The 'accurate' graph consists of line segments between the CCSD
energies for the integer occupations.}
\label{fig:Cl_fracN_LDA} 
\end{figure*}

\begin{figure}
\begin{minipage}[c]{0.45\linewidth}%
\centering \includegraphics[width=0.99\textwidth]{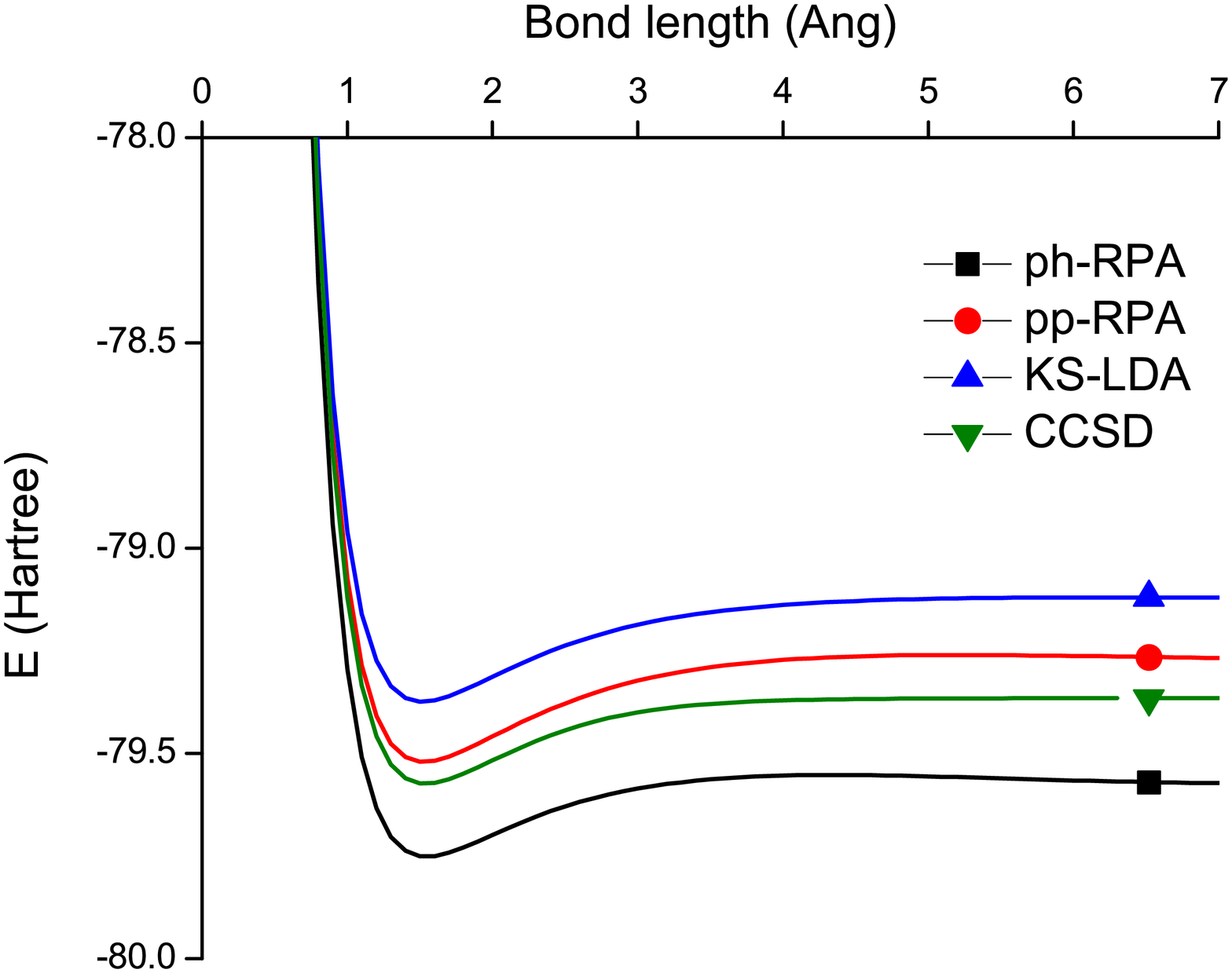} %
\end{minipage}%
\begin{minipage}[c]{0.45\linewidth}%
\includegraphics[width=0.99\textwidth]{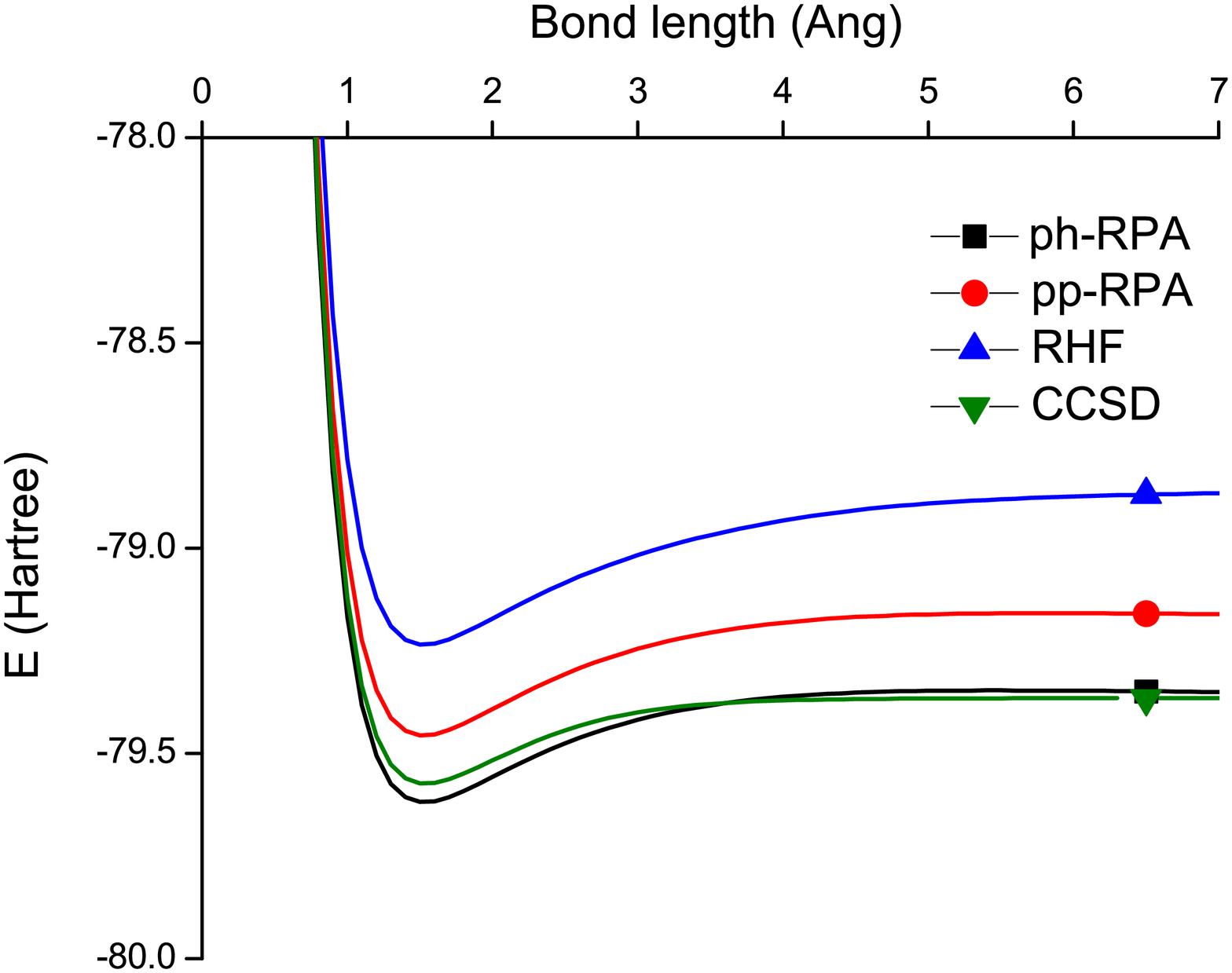} %
\end{minipage}\caption{The pp-RPA describes the stretching of the C-C bond in \ce{C2H6}
correctly (left: restricted LDA reference, right: restricted HF reference). The positions of the H atoms are
kept fixed at their equilibrium position.}
\label{fig:CH3CH3_LDA} 
\end{figure}

\begin{figure*}
\begin{minipage}[c]{0.45\linewidth}%
\centering \includegraphics[width=0.99\textwidth]{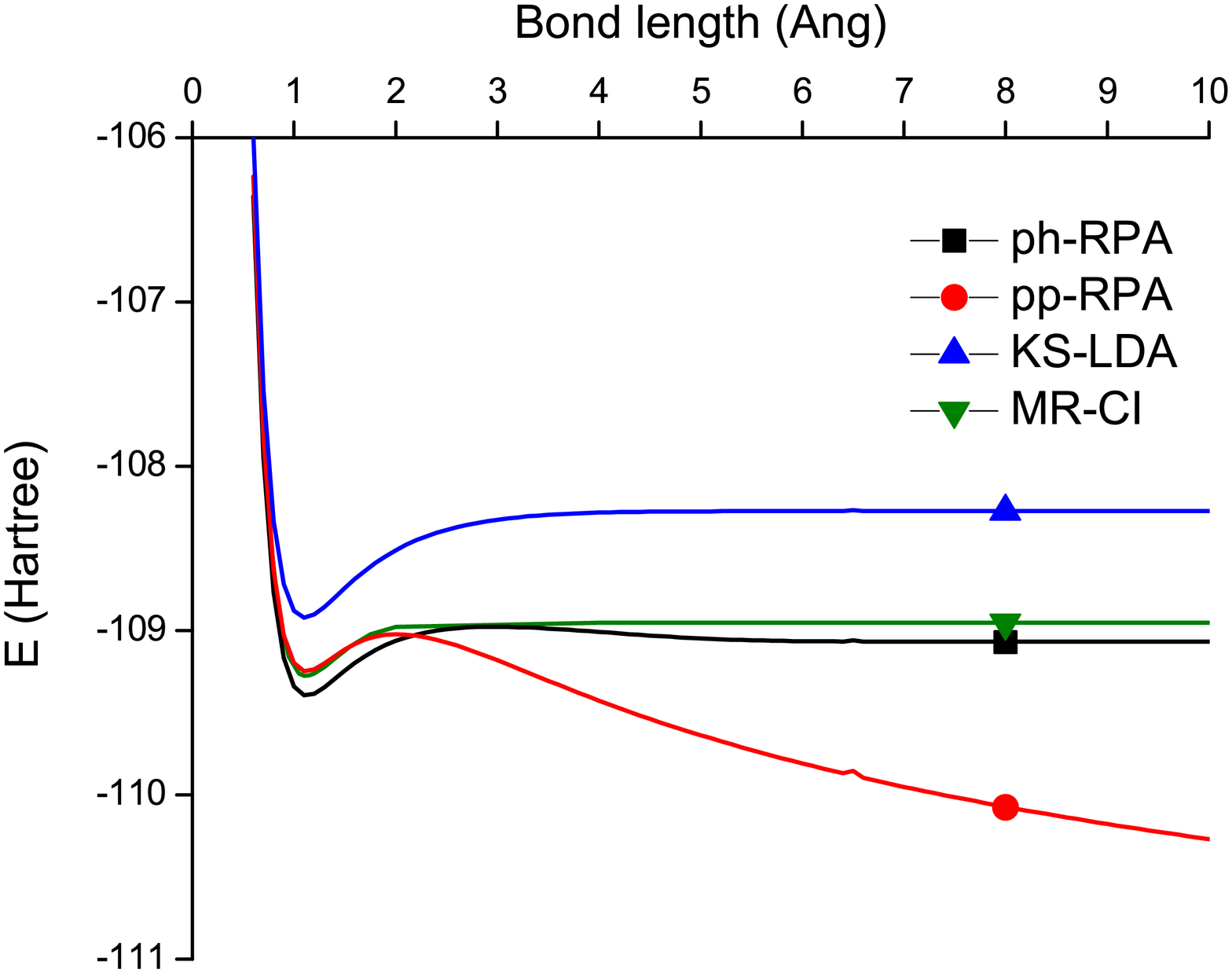} %
\end{minipage}%
\begin{minipage}[c]{0.45\linewidth}%
\includegraphics[width=0.99\textwidth]{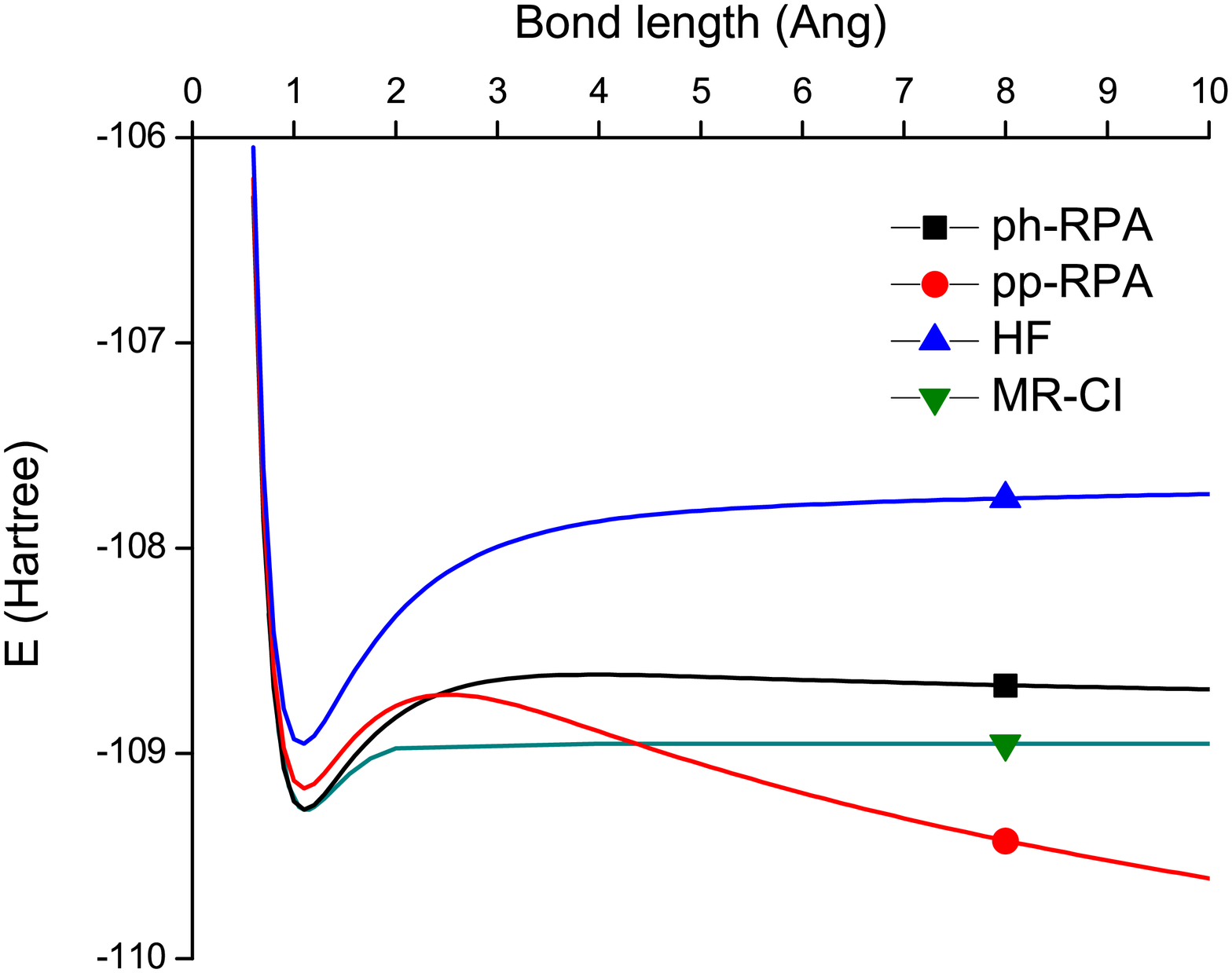} %
\end{minipage}\caption{The pp-RPA leads to a decreasing energy in the dissociation limit
of the triple bond in \ce{N2} (left: restricted LDA reference, right: restricted HF reference).}

\label{fig:NN_LDA} 
\end{figure*}

\begin{figure*}
\begin{minipage}[c]{0.45\linewidth}%
\centering \includegraphics[width=0.99\textwidth]{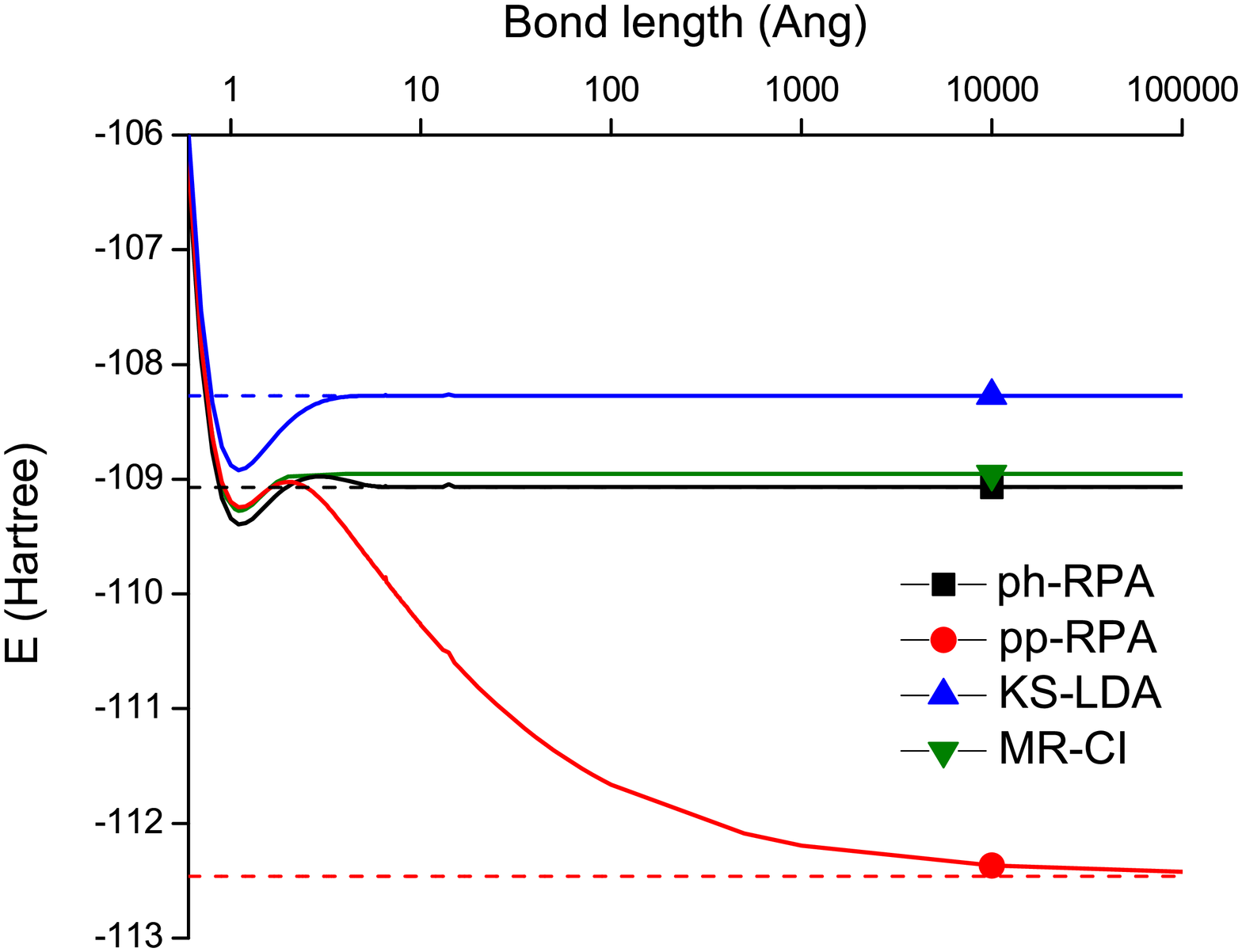} %
\end{minipage}%
\begin{minipage}[c]{0.45\linewidth}%
\includegraphics[width=0.99\textwidth]{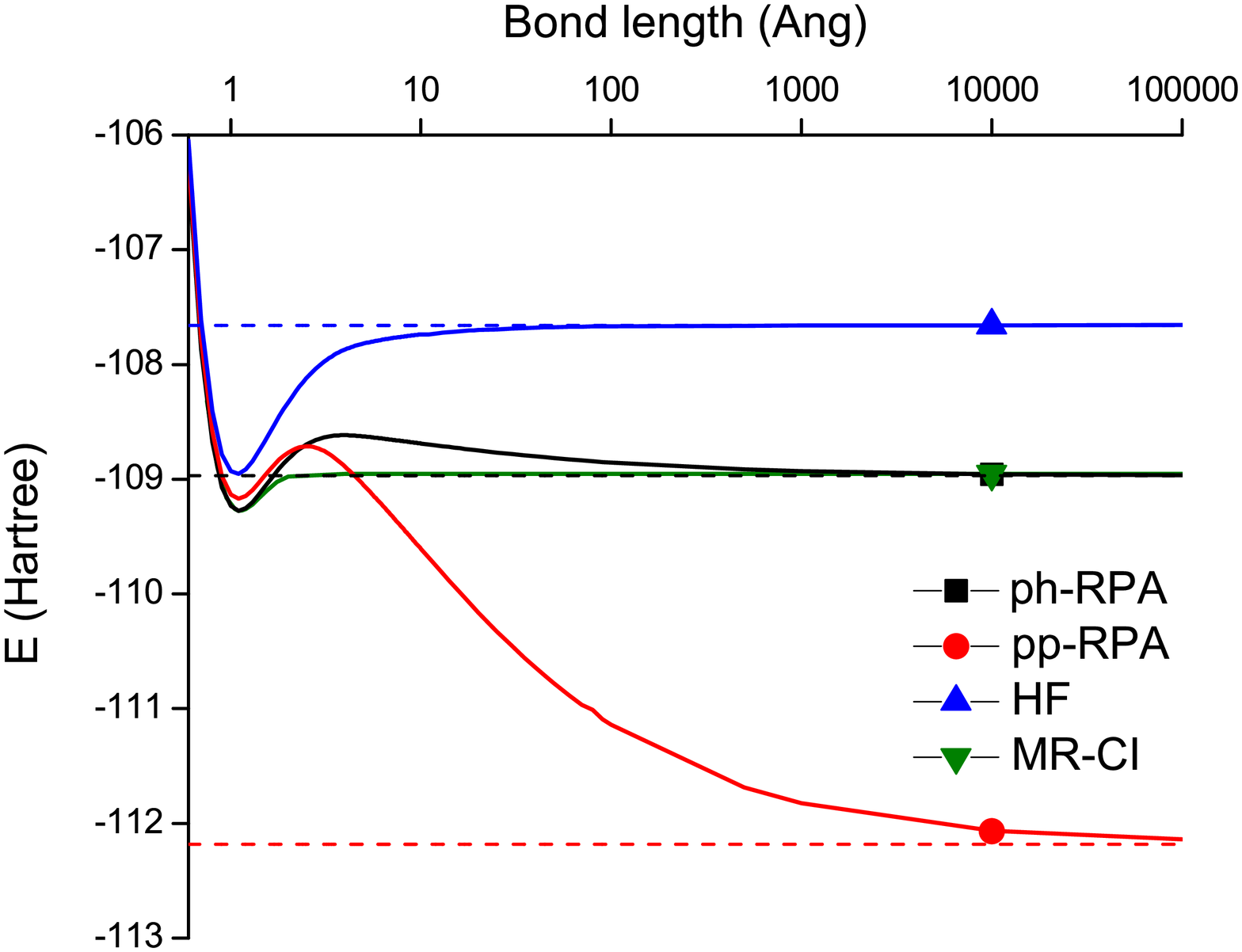} %
\end{minipage}\caption{The dissociation limit of the pp-RPA and ph-RPA energy for \ce{N2}
corresponds to the energy of two spin and angular momentum unpolarized
N atoms, indicated with dashed lines (left: restricted LDA reference, right: restricted HF reference).}
\label{fig:NN_LDA_limit} 
\end{figure*}

\begin{figure*}
\begin{minipage}[c]{0.45\linewidth}%
\centering \includegraphics[width=0.99\textwidth]{ArArLDA.eps} %
\end{minipage}%
\begin{minipage}[c]{0.45\linewidth}%
\includegraphics[width=0.99\textwidth]{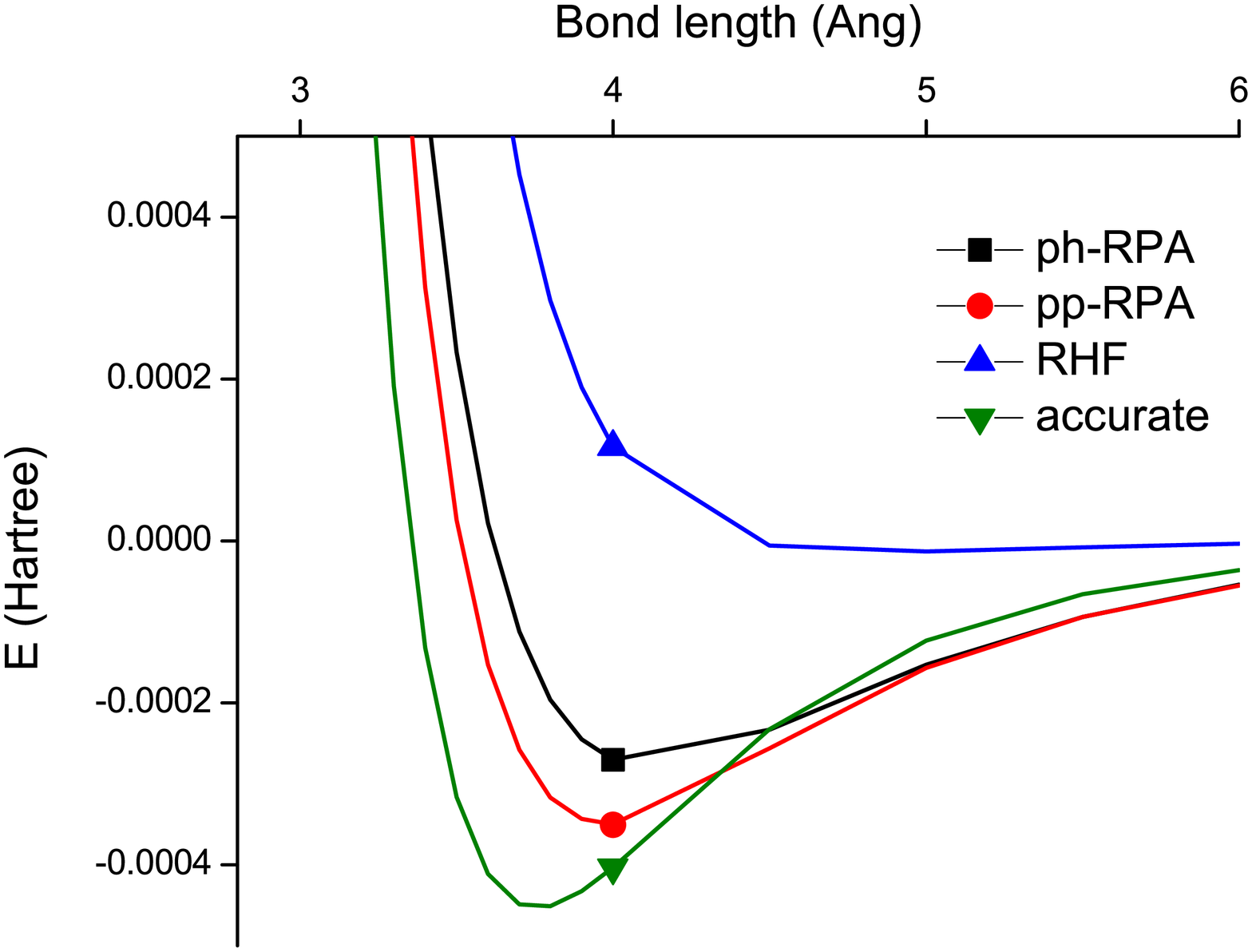} %
\end{minipage}\caption{The ph-RPA and pp-RPA both describe the van der Waals interactions
in the Ar dimer well (left: LDA reference, right: HF reference).}
\end{figure*}

\begin{figure*}
\begin{minipage}[c]{0.45\linewidth}%
\centering \includegraphics[width=0.99\textwidth]{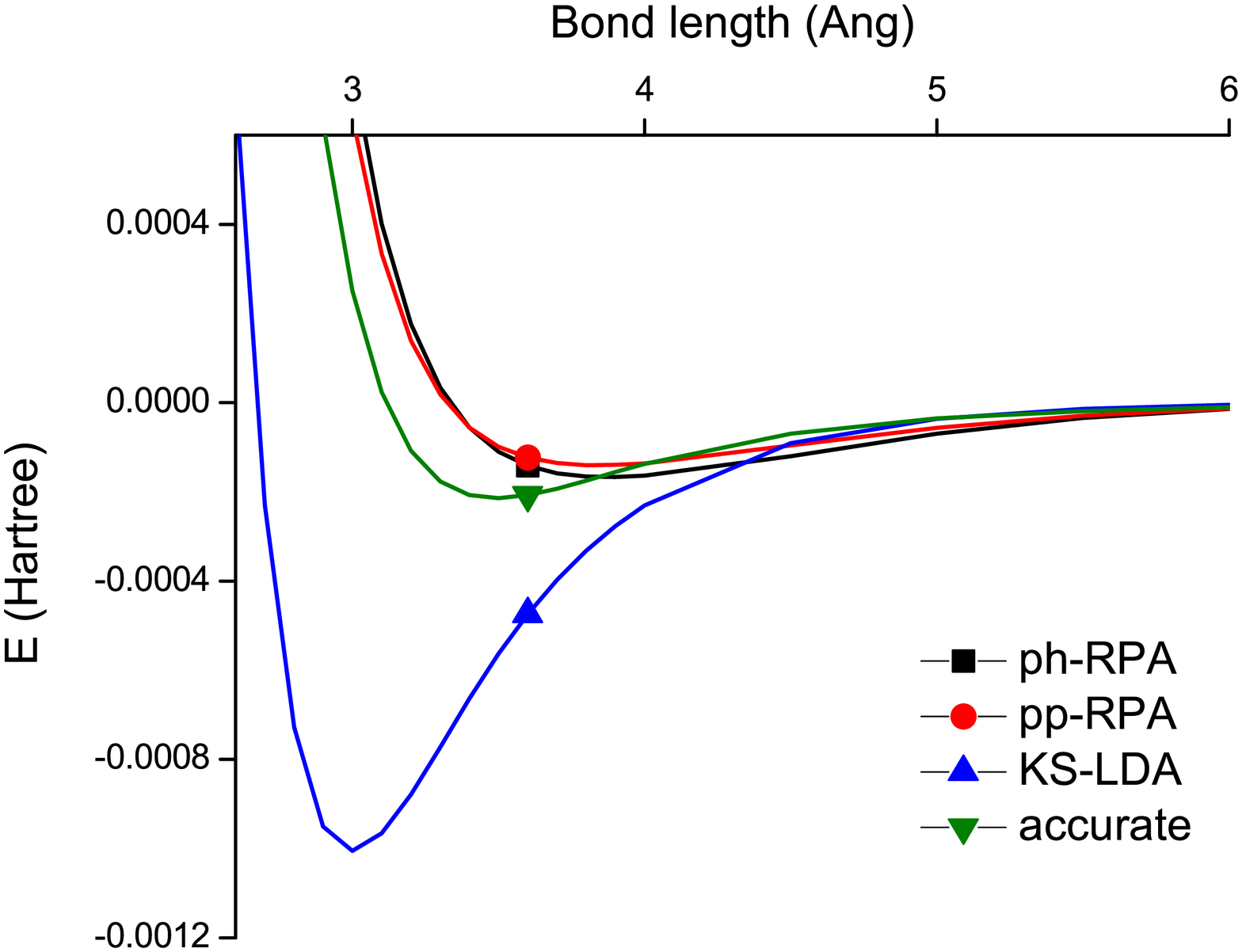} %
\end{minipage}%
\begin{minipage}[c]{0.45\linewidth}%
\includegraphics[width=0.99\textwidth]{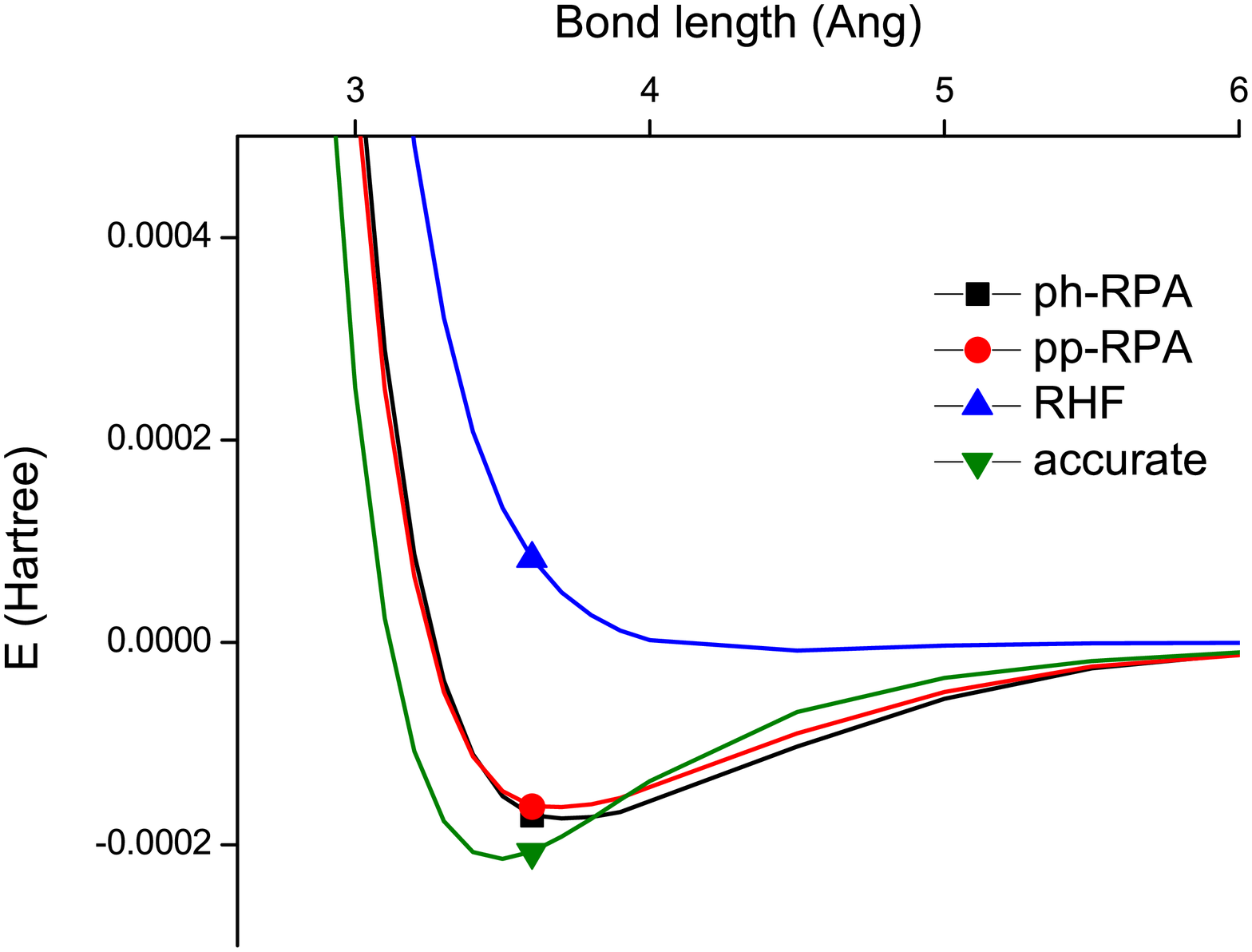} %
\end{minipage}\caption{The pp-RPA also describes the van der Waals interactions in the heteronuclear
\ce{NeAr} well (left: LDA reference, right: HF reference).}
\label{fig:NeAr_LDA} 
\end{figure*}

\begin{figure*}
\begin{minipage}[c]{0.45\linewidth}%
\centering \includegraphics[width=0.99\textwidth]{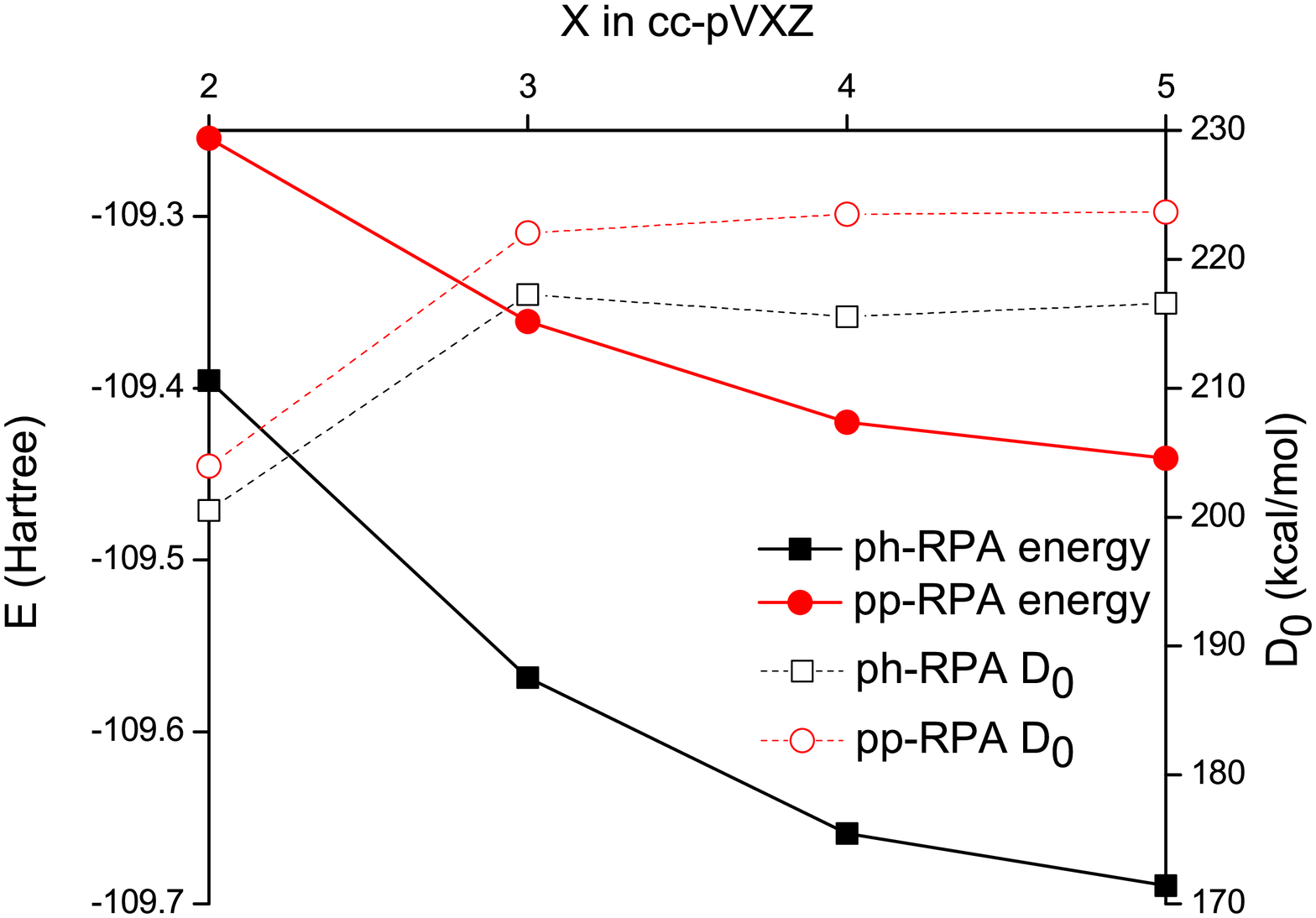} %
\end{minipage}%
\begin{minipage}[c]{0.45\linewidth}%
\includegraphics[width=0.99\textwidth]{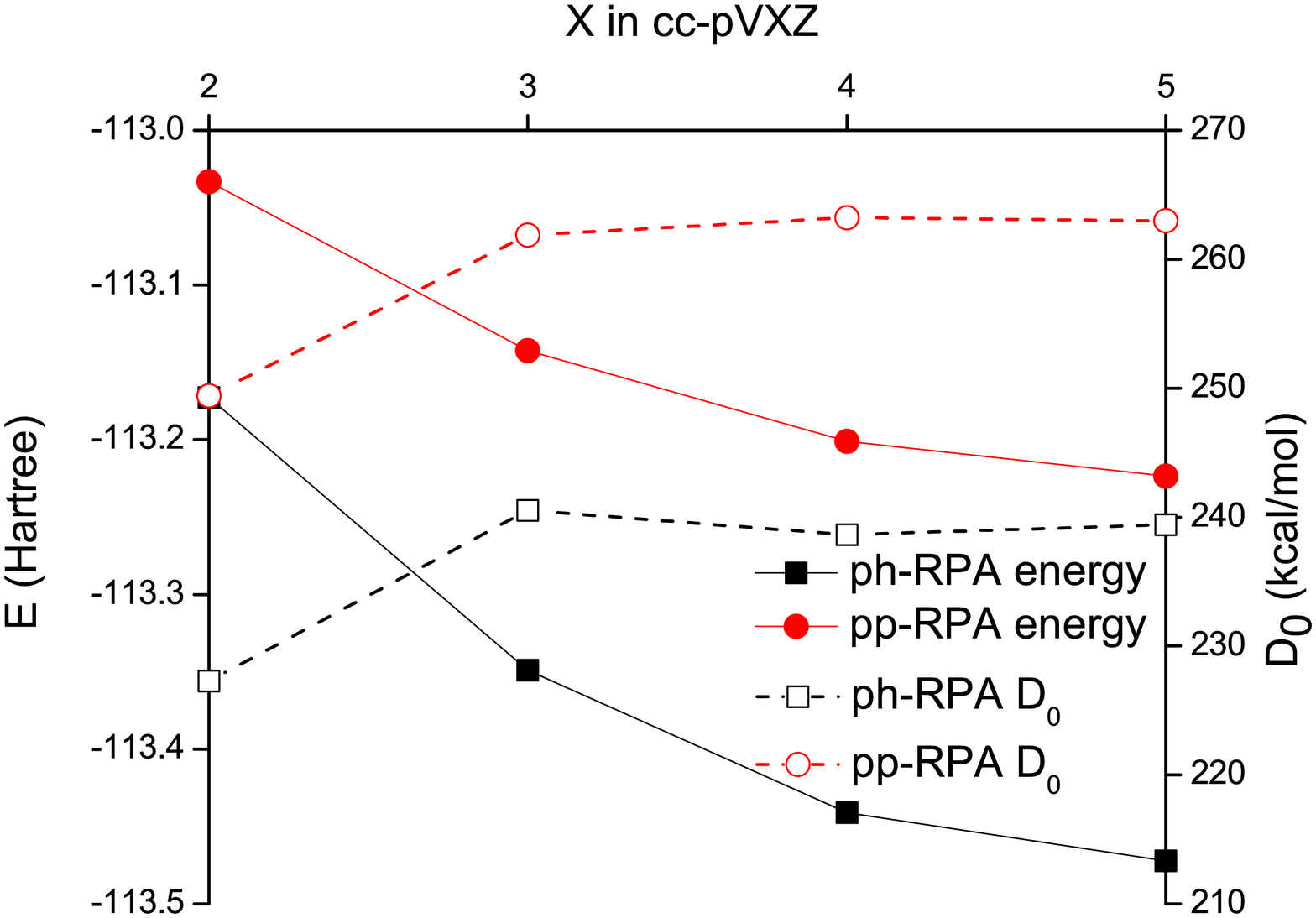} %
\end{minipage}\caption{The basis set convergence of the pp-RPA energy is rather slow, similar
to that of ph-RPA. The atomization energy $D_{0}$ converges
faster to its basis set limit than the absolute energies (left: \ce{N2},
right: \ce{CO}).}
\label{fig:bsc} 
\end{figure*}

\begin{table*}
\caption{The left and right derivatives of the pp-RPA(LDA) and ph-RPA(LDA)
energy in eV, computed by finite difference (with $\Delta=0.001$),
agree well with experiment, especially the derivatives with respect
to the HOMO orbital occupation. }
\begin{tabular}{lccccccccccc}
 & $\left(\frac{\partial E}{\partial n_{f}}\right)_{N-\delta}$  & $\left(\frac{\partial E}{\partial n_{f}}\right)_{N-\delta}$ & $\epsilon_{HOMO} $  & $-I$   &   & $\left(\frac{\partial E}{\partial n_{f}}\right)_{N+\delta}$ & $\left(\frac{\partial E}{\partial n_{f}}\right)_{N+\delta}$ & $\epsilon_{LUMO} $ & $A$ \tabularnewline
\multicolumn{1}{l}{} & pp-RPA(LDA)  & ph-RPA(LDA) & KS-LDA  & \multicolumn{1}{l}{expt.} & & pp-RPA(LDA)  & ph-RPA(LDA) & KS-LDA & expt.  &  & \tabularnewline
Li  & -5.395  & -3.130 & -3.581  & -5.392 &  & 0.125  & -3.013 & -2.169 & -0.618  &  & \tabularnewline
Be  & -8.628  & -5.379 & -6.042  & -9.323 &  & 1.185  & -2.811 & -2.515 & -0.295  &  & \tabularnewline
B  & -8.184  & -3.668 & -4.540  & -8.298&   & 0.772  & -4.010 & -3.812 & -0.280  &  & \tabularnewline
C  & -11.112  & -5.271 & -6.564  & -11.260&  & 0.177  & -4.131 & -5.083 & -1.262  &  & \tabularnewline
N  & -14.281  & -6.636 & -8.849  & -14.534&   & 0.959  & -5.553 & -4.910 & -0.070  &  & \tabularnewline
O  & -15.137  & -8.242 & -9.636  & -13.618&  & -1.395  & -8.299 & -7.709 & -1.461  &  & \tabularnewline
F  & -17.803  & -10.193 & -11.837  & -17.423&   & -4.206  & -11.434 & -10.812 & -3.401  &  & \tabularnewline
\textbf{MAE}  & \textbf{0.445}  & \textbf{5.332} & \textbf{4.114}  &  & & \textbf{0.945}  & \textbf{4.552} & \textbf{4.232} &  &  & \tabularnewline
\end{tabular}
\caption{The left and right derivatives of the pp-RPA(HF) and ph-RPA(LDA) energy
in eV, computed by finite difference (with $\Delta=0.001$) agree
well with experiment.}
\begin{tabular}{lcccccccccc}
 & $\left(\frac{\partial E}{\partial n_{f}}\right)_{N-\delta}$  &  $\left(\frac{\partial E}{\partial n_{f}}\right)_{N-\delta}$ & $\epsilon_{HOMO}$  &  $-I$ &  & $\left(\frac{\partial E}{\partial n_{f}}\right)_{N+\delta}$ &  $\left(\frac{\partial E}{\partial n_{f}}\right)_{N+\delta}$  & $\epsilon_{LUMO}$  & $A$\tabularnewline
 & \multirow{1}{*}{pp-RPA(HF)} & ph-RPA(HF) & HF  & expt.  &  & \multirow{1}{*}{pp-RPA(HF)} & ph-RPA(HF) & HF  & expt.   \tabularnewline
Li  & -5.349  & -2.580 & -5.343  & -5.392 &  & -0.030  & -2.026 & 0.153  & -0.618  & \tabularnewline
Be  & -8.528  & -4.595 & -8.416  & -9.323 &  & 0.336  & -2.054 & 0.396  & -0.295  & \tabularnewline
B  & -8.369  & -3.013 & -8.666  & -8.298  & & 0.424  & -2.821 & 0.795  & -0.280  & \tabularnewline
C  & -11.405  & -4.649 & -11.941  & -11.260 &  & 0.377  & -3.978 & 1.025  & -1.262  & \tabularnewline
N  & -14.696  & -6.639 & -15.531  & -14.534 &  & 1.357  & -3.273 & 2.095  & -0.070  & \tabularnewline
O  & -15.607  & -6.948 & -16.648  & -13.618 & & 0.226  & -5.759 & 1.765  & -1.461  & \tabularnewline
F  & -18.397  & -8.841 & -19.921  & -17.423 &  & -1.787  & -8.885 & 0.967  & -3.401  & \tabularnewline
\textbf{MAE}  & \textbf{0.597}  & \textbf{6.083}  & \textbf{1.219}  &  &  & \textbf{1.184}  & \textbf{3.058} & \textbf{2.083}  &  & \tabularnewline
\end{tabular}
\label{tab:mu_ppRPA} 
\end{table*}

\begin{table}
\caption{The errors in the atomization energies $D_{0}$ and the heats of formation
$\Delta H$ (in kcal/mol) relative to the experimental values $\Delta H_{expt.}$,
computed with pp-RPA in the cc-pVTZ basis set, are significantly better
than those computed with ph-RPA.}
\begin{tabular}{lrrrrr}
 & $D_{0}^{pp-RPA}$  & $D_{0}^{ph-RPA}$  & $\Delta H^{pp-RPA}$  & $\Delta H^{ph-RPA}$  & $\Delta H^{expt}$\tabularnewline
$\mathrm{C_{2}H_{2}}$  & 406.3  & 387.3  & 53.2  & 72.2  & 54.2\tabularnewline
$\mathrm{CH_{4}}$  & 410.6  & 410.8  & -9.3  & -9.6  & -17.9\tabularnewline
$\mathrm{Cl_{2}}$  & 56.6  & 44.2  & 1.4  & 13.7  & 0.0\tabularnewline
$\mathrm{CO}$  & 265.0  & 243.6  & -32.1  & -10.7  & -26.4\tabularnewline
$\mathrm{F}_{2}$  & 37.5  & 27.9  & 1.0  & 10.6  & 0.0\tabularnewline
$\mathrm{H}_{2}$  & 100.4  & 108.3  & 8.8  & 0.9  & 0.0\tabularnewline
$\mathrm{H_{2}O}$  & 225.8  & 218.8  & -51.3  & -44.4  & -57.8\tabularnewline
$\mathrm{HCl}$  & 102.4  & 98.3  & -18.1  & -14.0  & -22.1\tabularnewline
$\mathrm{HF}$  & 139.2  & 128.5  & -63.5  & -52.8  & -65.1\tabularnewline
$\mathrm{HOCl}$  & 161.7  & 148.5  & -15.1  & -1.9  & -17.8\tabularnewline
$\mathrm{HOOH}$  & 262.7  & 250.6  & -26.4  & -14.3  & -32.5\tabularnewline
$\mathrm{LiH}$  & 47.9  & 52.6  & 43.2  & 38.5  & 33.3\tabularnewline
$\mathrm{N_{2}}$  & 225.6  & 221.8  & 3.0  & 6.8  & 0.0\tabularnewline
$\mathrm{NaCl}$  & 94.2  & 82.2  & -39.8  & -27.8  & -43.6\tabularnewline
$\mathrm{NH}$  & 75.7  & 81.3  & 93.0  & 87.3  & 85.2\tabularnewline
$\mathrm{NH_{2}}$  & 170.6  & 177.5  & 56.0  & 49.1  & 45.1\tabularnewline
$\mathrm{NH_{3}}$  & 284.5  & 288.9  & 1.9  & -2.5  & -11.0\tabularnewline
$\mathrm{O}_{2}$  & 129.4  & 111.3  & -8.8  & 9.2  & 0.0\tabularnewline
\textbf{MAE }  &  &  & \textbf{5.8}  & \textbf{10.4}  & \tabularnewline
\textbf{MAX }  &  &  & \textbf{12.9}  & \textbf{18.0}  & \tabularnewline
\end{tabular}
\label{tab:HOF} 
\end{table}

\clearpage{}

\bibliographystyle{plain}